\documentclass[onecolumn]{els-mrw_modified}
\usepackage{amsmath,amssymb,amsfonts,amsthm,makeidx,graphicx}
\usepackage{comment}
\usepackage{amsopn}
\usepackage{amsbsy}
\usepackage{tikz}
\usepackage{tipa}
\usepackage{braket}
\usepackage{verbatim}
\usepackage{epsfig,float}
\usepackage{dcolumn}
\usepackage{bm}
\usepackage{hyperref}
\usepackage{subcaption}
\usepackage{txfonts}
\usepackage{helvet}
\newcommand{\reffig}[1]{\mbox{Fig.~\ref{#1}}}
\newcommand{\refeq}[1]{\mbox{Eq.~(\ref{#1})}}
\newcommand{\refsec}[1]{\mbox{Sec.~\ref{#1}}}

\newcommand{\be}{\begin{equation}}
\newcommand{\ee}{\end{equation}}
\newcommand{\ba}{\begin{eqnarray}}
\newcommand{\ea}{\end{eqnarray}}
\newcommand{\bal}{\begin{align}}
\newcommand{\eal}{\end{align}}
\renewcommand{\Re}{\mathrm{Re}}
\renewcommand{\Im}{\mathrm{Im}}
\newcommand{\T}{${\mathcal T}\,$}

\def\II{\hbox{$1\hskip -1.2pt\vrule depth 0pt height 1.6ex width 0.7pt\vrule depth 0pt height 0.3pt width 0.12em$}}
\usepackage{psfrag}
\usepackage{units}
\usepackage{slashed}
\usepackage{xcolor}
\usepackage{latexsym}
\usepackage{times}
\usepackage{esint}
\usepackage{color}
\usepackage{orcidlink}
\hypersetup{
            unicode=false,          
            pdftoolbar=false,        
            pdfmenubar=true,        
            pdffitwindow=false,     
            pdfstartview={FitH},    
            pdfkeywords={Quantum Chaos} {Quantum Billiards} {Relativistic Quantum Billiards}, 
            pdfnewwindow=true,      
            colorlinks=true,       
            linkcolor=red,          
            citecolor=blue,        
            filecolor=magenta,      
            urlcolor=blue           
        }
\usepackage{xcolor}

\usepackage[T1]{fontenc}
\begin{document}

\chapter{Relativistic Quantum Chaos in Neutrino Billiards}

\author[1,2,3]{Barbara Dietz\orcidlink{0000-0002-8251-6531}}
\address[1]{\orgname{Max-Planck Institute for the Physics of Complex Systems}, \orgaddress{N\"othnitzer Stra\ss e 38, 01187 Dresden, Germany}}
\address[2]{\orgname{TU Dresden}, \orgdiv{Institute of Theoretical Physics}, \orgaddress{01062 Dresden, Germany}}
\address[3]{\orgname{Institute for Basic Science (IBS)}, \orgdiv{Center for Theoretical Physics of Complex Systems}, \orgaddress{Daejeon 34126, Republic of Korea}}

\date{\today}

\articletag{Chapter Article tagline: November 02, 2026}

\maketitle

\begin{glossary}[Keywords]
Quantum Chaos, Relativistic Quantum Chaos, Quantum Billiards, Neutrino Billiards, Graphene Billiards, Microwave Photonic Crystals
\end{glossary}


\begin{abstract}[Abstract]
	Neutrino billiards serve as a model system for the study of aspects of relativistic quantum chaos. These are relativistic quantum billiards consisting of a spin-1/2 particle which is confined to a planar domain by imposing boundary conditions on the spinor components which were proposed in [Berry and Mondragon, {\it Proc. R. Soc.} A {\bf 412}, 53 (1987)]. We review their general features and the properties of neutrino billiards with shapes of billiards with integrable dynamics. Furthermore, we review the features of two neutrino billiards with the shapes of billiards generating a chaotic dynamics, whose nonrelativistic counterpart exhibits particular properties. Finally we briefly discuss possible experimental realizations of relativistic quantum billiards based on graphene billiards, that is, finite-size sheets of graphene.    
\end{abstract}

\section{Introduction}
It has been well established by now that the eigenstates of generic nonrelativistic quantum systems, whose corresponding classical dynamics is chaotic, exhibit universal properties that depend only on the associated symmetry class~\cite{Casati1980,Berry1981a}. This was formulated in the famous Bohigas-Giannoni-Schmit (BGS) conjecture~\cite{Bohigas1984}, stating that they are well described by random-matrix theory (RMT), whereas -- according to the Berry-Tabor (BT) conjecture~\cite{Berry1977a} -- those of generic quantum systems whose corresponding classical dynamics is integrable exhibit Poisson statistics. In the semiclassical limit $\hbar\to 0$, the applicability of RMT was confirmed within the periodic orbit theory~\cite{Heusler2007}, which was pioneered by Gutzwiller~\cite{Gutzwiller1971}. It provides an approximation for the fluctuating part of the spectral density of a quantum system in terms of a sum over the periodic orbits (POs) of the associated classical dynamics. This connection between the eigenvalue spectrum of a quantum system and the classical POs is visible in its Fourier transform from eigenwavenumber to length, which exhibits peaks at the lengths of the POs, and therefore is commonly referred to as length spectrum. The BGS and BT conjectures and applicability of the trace formula have been confirmed in numerous quantum systems~\cite{Guhr1989,LesHouches1989,Haake2018}. 

Besides quantum graphs, which are introduced in the chapter by Gregory Berkolaiko and Sven Gnutzmann~\cite{Berkolaiko2026}, nonrelativistic quantum billiards (QBs) have served for several decades as a paradigm model for the theoretical, numerical and experimental study of aspects of quantum chaos~\cite{Sinai1970,Bunimovich1979,Berry1981}; see also the chapter by Takahisa Harayama~\cite{Harayama2026}, where quantum billiard lasers are introduced. This chapter provides a review of the properties of the eigenstates of relativistic quantum billiards, neutrino billiards (NBs)~\cite{Berry1987}, which became of interest again after the pioneering fabrication of graphene~\cite{Novoselov2004}. They consist of a non-interacting massless or massive spin-1/2 particle, which is confined to the billiard domain by imposing the boundary condition (BC) that the outward current vanishes, and are governed by the Weyl equation~\cite{Weyl1929}, generally referred to as Dirac equation in this context. Massless NBs were proposed by Sir Michael Berry and Raul Mondragon~\cite{Berry1987} as a realization of a quantum system, in which time-reversal (\T) invariance is violated without the presence of a magnetic field, and were not investigated further. It is worth mentioning that they were introduced and given their name before the revelation that neutrinos have mass. About a decade later other relativistic systems, consisting of a charged relativistic spin-1/2 particle in an external electromagnetic field~\cite{Bolte1999}, became of interest in the context of spectral properties and semiclassical approximations~\cite{Bolte1999}. Insights and findings obtained since revival of interest in NBs and a comparison with those of the corresponding nonrelativistic QB are reviewed. A summary of their salient features is provided in~\refsec{NBS}. In distinction to nonrelativistic QBs, NBs do not have a well-defined classical counterpart. The peaks in the length spectra of NBs are at the same lengths as for QBs, except that those with an odd  number of reflections at the billiard boundary are missing. This implies that there is a connection between NBs and the dynamics of the classical hard-wall billiard (CB) of corresponding shape. Thus, the objective of the study of NBs was to understand to what extent the BGS and BT conjectures are applicable to relativistic quantum billiards with shapes of CBs with chaotic, mixed or integrable dynamics, and to develop a semiclassical approach to get information about the semiclassical limit and the transition from the relativistic to the nonrelativistic regime~\cite{Dietz2020}, which is achieved by increasing the mass of the spin-1/2 particle to infinity. In~\refsec{NBS} the methods developed and used for the study of NBs are reviewed. As outlined in~\refsec{Syms}, in distinction to the eigenstates of QBs the eigenspinors of NBs cannot be classified according to mirror symmetries of their shapes, whereas the two spinor components of NBs with a discrete rotational symmetry can be classified, but belong to different symmetry classes~\cite{Zhang2021,Zhang2023}. This has implications concerning the spectral properties of certain NBs. In~\refsec{BIEs} an extension of the boundary-integral equation for massless NBs~\cite{Berry1987,Yupei2019} to massive ones~\cite{Dietz2020,Dietz2022a} is introduced. It originates from the Green theorem, which incorporates the associated BCs and thus provides an exact boundary-integral equation for the eigenfunctions in the interior of a QB or NB in terms of those on the boundary. It is the starting point of the derivation of a semiclassical approximation of the spectral density, which will be reviewed in detail in~\refsec{Trace}; see also the chapter by Martin Sieber and Sebastian M\"uller~\cite{Sieber2026} for a review of the semiclassical periodic orbit theory. To obtain information on the semiclassical limit Husimi functions, which are the topic of the chapter by Martina Hentschel~\cite{Hentschel2026} for nonrelativistic systems, momentum distributions and the associated trace formula are employed~\cite{Dietz2020}. These are introduced in~\refsec{Measures}. 

In~\refsec{Int} and~\refsec{Chaos} results for various exemplary shapes are reviewed. These comprise NBs with the shapes of billiards with integrable classical dynamics in Secs.~\ref{Circle},~\ref{Ellipse} and~\ref{ET}, whose eigenstates can be computed analytically and exhibit the same features as the corresponding QB, and NBs with shapes of sectors of the circle, ellipse and equilateral triangle billiards whose spectral properties differ from those of the corresponding QB~\refsec{Sectors}. Furthermore, in~\refsec{Chaos} NBs exhibiting nongeneric properties are presented. These NBs have the shape of billiards generating a chaotic dynamics. The spectral properties of some of them may deviate from RMT predictions due to the presence of relativistic quantum scars, or more generally, of scarred spinor states. It will be outlined how these can be extracted to achieve the expected agreement with RMT. One prominent example is the stadium billiard, whose classical dynamics is fully chaotic except for neutral-stable and almost POs. More information on the classical and semiclassical dynamics of the can be found in the chapter by Steve Tomsovic~\cite{Tomsovic2026}. These are of measure zero in phase space but lead to a scarring of the wave functions and spinor functions of the corresponding QB and NB, respectively, and therefore affect the spectral properties. Some features of these nonrelativistic and relativistic quantum scars are briefly reviewed in~\refsec{Stadium}. We use a semiclassical approach, Husimi functions and momentum distributions to identify and remove such scarred states. Another approach is provided in Ref.~\cite{HXGL2018}. Importantly, the conformal mapping method introduced there for the computation of the eigenstates of NBs has been proven to be wrong~\cite{Dietz2025}. Other prominent examples are NBs with shapes of constant-width billiards, that are introduced in~\refsec{CW}. Their classical dynamics is unidirectional~\cite{Knill1998,Gutkin2007}, whereas in the corresponding nonrelativistic QB a change of the rotational direction of motion is possible via dynamical tunneling. On the contrary, in the corresponding massless and finite-mass NBs the modes can be separated into clockwise and counterclockwise modes and dynamical tunneling is absent~\cite{Dietz2022a}. More information on tunneling in nonrelativistic systems is provided in the chapter by Akira Shudo~\cite{Shudo2026}.

Section~\ref{Exp} is devoted to the experimental study of the spectral properties of  finite-size honeycomb lattices -- graphene billiards (GBs) constructed by cutting out of a graphene sheet the shape of the billiard -- with superconducting microwave photonic crystals. These are flat resonators containing metallic cylinders arranged on a triangular grid, referred to as Dirac (microwave) billiards~\cite{Dietz2015,Dietz2016}. The density of states of these Dirac billiards comprises two microwave-frequency regions comprising Dirac points, where the conduction and valence band touch each other conically, with adjacent bands including van Hove singularities~\cite{Maimaiti2020}. Within each of these two regions the density of states resembles that of graphene and is well described by a tight-binding model (TBM) for the GB of corresponding shape. The two regions are separated by a narrow region of particularly high resonance density corresponding to a nearly flat band in the band structure, and the density of states comprising all these regions is well captured by a TBM for a lattice composed of a honeycomb and Kagome sublattice. Honeycomb-Kagome and graphene billiards became of high interest because around the Dirac points the dispersion relation is linear and the spectrum is described by the same relativistic Dirac equation for massless spin-1/2 particles as NBs~\cite{Beenakker2008,Neto2009,Polini2013}. However even in that region the spectral properties of the GB agree with those for the nonrelativistic QB of corresponding shape. Recently, GBs subject to the Haldane model~\cite{Haldane1988} were proposed as a suitable model for the study of properties of NBs based on the corresponding TBM and as a possible experimental realization of them~\cite{Nguyen2024}. 

\section{The Dirac equation for NBs\label{NBS}} 
The domains of planar billiards under consideration are covered by a vector $\boldsymbol{r}(r,\gamma)=[x(r,\gamma),y(r,\gamma)]$, $0\leq r\leq r_0$ with $r=r_0$ defining the boundary and $0\leq\gamma < 2\pi$, or in the complex plane by $w(r,\gamma)=x(r,\gamma)+iy(r,\gamma)$, and can be written as a polynomial in $z=f(r)e^{i\gamma}$ with $f(r_0)\ne 0$ and $\frac{df(r)}{dr}|_{r=r_0}\ne 0$,
\be
w(z)=x(r,\gamma)+iy(r,\gamma)=\sum_mc_mz^m,\, z=f(r)e^{i\gamma},
\label{coordinate}
\ee
where $c_m$ are real or complex coefficients. The boundary $\partial\Omega$ is parametrized by the arclength $s\in[0,\mathcal{L}]$ with $\mathcal{L}$ denoting the perimeter, $w(s)=x[r_0,\gamma(s)]+iy[r_0,\gamma(s)]$ or, generally by a parameter $\gamma$, $w(z)=w(\gamma)$, with 
        \be 
        s(\gamma)=\int_0^\gamma\vert w^\prime(\tilde\gamma)\vert d\tilde\gamma,\, ds=\vert w^\prime(\gamma)\vert d\gamma .
        \ee

Planar hard-wall CBs consist of a point particle moving freely in a bounded two-dimensional domain and reflected specularly at the boundary $\partial\Omega$. They consitute a paradigmatic system for the study of aspects of quantum chaos, because their dynamics is solely determined by their shape. The eigenstates of the corresponding nonrelativistic QB are obtained by solving the Schr\"odinger equation for a free particle with mass $m_0$ and imposing the Dirichlet BC~\cite{LesHouches1989},
        \be
	\mathcal{\hat H}_S\psi(\boldsymbol{r})=-\frac{\hbar^2}{2m_0}\Delta\psi(\boldsymbol{r})=E\psi(\boldsymbol{r}),\,
        \psi(\boldsymbol{r})\vert_{\partial\Omega}=0.
        \label{Schr0}
        \ee
Replacing the energy $E$ by the wavenumber $k$, which is related to $E$ as $\frac{2m_0E}{\hbar^2}=k^2$, leads to the Helmholtz equation
\be
        \mathcal{\hat H}_H\psi(\boldsymbol{r})=-\Delta\psi(\boldsymbol{r})=k^2\psi(\boldsymbol{r}),\,
        \psi(\boldsymbol{r})\vert_{\partial\Omega}=0.
        \label{Schr}
        \ee
The two-dimensional Dirac equation for a free spin-1/2 particle with mass $m_0$ and momentum $\boldsymbol{\hat p}=-i\hbar\boldsymbol{\nabla}$ is given by
\be
\mathcal{\hat H}_D\boldsymbol{\psi}(\boldsymbol{r})=\left( c\boldsymbol{\hat\sigma}\cdot\boldsymbol{\hat p}+m_0c^2\boldsymbol{\hat\sigma_z}\right)\boldsymbol{\psi}(\boldsymbol{r})
=E\boldsymbol{\psi},\, \boldsymbol{\psi}(\boldsymbol{r})=
\begin{pmatrix}
\psi_1(\boldsymbol{r}) \\ \psi_2(\boldsymbol{r})
\end{pmatrix},
\label{DE}
\ee
where $\mathcal{\hat H}_D$ denotes the Dirac Hamiltonian and $\boldsymbol{\hat\sigma}=(\boldsymbol{\hat\sigma}_x,\boldsymbol{\hat\sigma}_y)$, $\boldsymbol{\hat\sigma}_{x,y,z}$ are the Pauli matrices, $E=\hbar ck_E=\hbar ck\sqrt{1+\beta^2}$ is the energy, $k$ is the free-space wave vector and $\beta=\frac{m_0c}{\hbar k}$ is the ratio of the rest-energy momentum and free-space momentum. The particle is confined to the billiard domain $\Omega$ by imposing on the solutions $\psi_{1,2}(\boldsymbol{r})$ of~\refeq{DE} along its boundary the BC that the outgoing flow vanishes. This condition ensures that the Hamiltonian of the NB, $\boldsymbol{\hat H}_{NB}$ is self-adjoint (and hence Hermitian) under this BC; cf. Ref.~\cite{Berry1987} for details. Accordingly, the eigenstates of a NB are obtained by imposing the BC that the normal component of the expectation value of the current operator $\boldsymbol{\hat u}=\boldsymbol{\nabla}_{\boldsymbol{p}}\mathcal{\hat H}_D=c\boldsymbol{\hat\sigma}$ -- referred to as local current -- vanishes,
\be
\label{Flux}
\boldsymbol{n}\cdot\left[\boldsymbol{\psi}^\dagger\boldsymbol{\nabla}_{\boldsymbol{p}}\mathcal{\hat H}_{D}\boldsymbol{\psi}\right]=c\boldsymbol{n}\cdot\left[\boldsymbol{\psi}^\dagger\boldsymbol{\hat\sigma}\boldsymbol{\psi}\right]=0,
\ee
leading to the BC~\cite{Berry1987}
\be
\psi_2(\gamma)=i\mu B(\gamma)e^{i\alpha(\gamma)}\psi_1(\gamma).
\label{BC1}
\ee
Here, $\alpha(\gamma)$ is the angle of the outward-pointing normal vector $\boldsymbol{n}(\gamma)=[\cos\alpha(\gamma),\sin\alpha(\gamma)]$ or in the complex plane $n(\gamma)=e^{i\alpha(\gamma)}$ at $w(\gamma)$ with respect to the $x$ axis, and $\mu=\pm 1$ determines the rotational direction of the current at the boundary. We use $\mu =1$ in the following and $B(\gamma)=1$ in accordance with Ref.~\cite{Berry1987}. The normal vector can be expressed in terms of the derivative $w^\prime(\gamma)={d}{w(\gamma)}/{d}\gamma$, writing the tangential vector $\boldsymbol{t}(\gamma)=\left[-\sin\alpha(\gamma),\cos\alpha(\gamma)\right]$ in the complex plane, $t(\gamma)=\frac{w^\prime (\gamma)}{\vert w^\prime (\gamma)\vert}$ and accordingly $n(\gamma)=-i\frac{w^\prime (\gamma)}{\vert w^\prime (\gamma)\vert}$, yielding for the BC~\refeq{BC1}
\be
\psi_2(\gamma)=\frac{w^\prime (\gamma)}{\vert w^\prime (\gamma)\vert}\psi_1(\gamma).
\label{BC2}
\ee
Since Hermiticity is not affected by the additional mass term in~\refeq{DE}, the BC is the same for massive NBs. The BC~\refeq{Flux} is not the only one ensuring self-adjointness of the Dirac Hamiltonian and zero outgoing current. For example, in Ref.~\cite{Gaddah2018} the Dirichlet BC is imposed on one spinor component and the other one is obtained with~\refeq{DE}. In the context of nuclear physics the confinement of relativistic particles to a bounded region is achieved, for example, by employing a 'MIT bag model'; cf. Ref.~\cite{Thomas1984} or pages 108-129 of Ref.~\cite{Greiner1994}.  

Introducing the notations  
\be
\mathcal{K}=\sqrt{\frac{1-\sin\theta_\beta}{1+\sin\theta_\beta}},\,
\cos\theta_\beta=\frac{1}{\sqrt{1+\beta^2}},\, \sin\theta_\beta=\frac{\beta}{\sqrt{1+\beta^2}},
\ee
and a two-component spinor $\boldsymbol{\tilde\psi}(\boldsymbol{r})$ which is obtained by dividing the components $\psi_{1,2}(\boldsymbol{r})$ in~\refeq{DE} by the amplitudes of the corresponding free-space wave-function components~\cite{Dietz2020},
\be
\boldsymbol{\psi}(\boldsymbol{r})=
\begin{pmatrix}
\sqrt{\frac{1+\sin\theta_\beta}{2}}\tilde\psi_1(\boldsymbol{r}) \\
\sqrt{\frac{1-\sin\theta_\beta}{2}}\tilde\psi_2(\boldsymbol{r})
\end{pmatrix},
\label{tildepsi}
\ee
turns~\refeq{DE} into a Dirac equation for massless NBs
\be
k\boldsymbol{\tilde\psi}(\boldsymbol{r})+i\boldsymbol{\hat\sigma}\boldsymbol\cdot{\nabla}\boldsymbol{\tilde\psi}(\boldsymbol{r})=0\label{Hamiltonian},
\ee
with modified  BCs, 
\be
\tilde\psi_2(s)=ie^{i\alpha(s)}\mathcal{K}^{-1}\tilde\psi_1(s).\label{BCm}
\ee
In the following the tilde in $\boldsymbol{\tilde\psi}(\boldsymbol{r})$ will be dropped, keeping in mind that for massive NBs the eigenspinors $\boldsymbol{\psi}(\boldsymbol{r})$ are obtained by multiplying the components with $\theta_\beta$-dependent factors as defined in~\refeq{tildepsi}. 

The BC~\refeq{BCm} can be written as separate ones for each component. This is achieved by introducing a coordinate system $(n,s)$ moving along $\partial\Omega$ with axes normal and tangential to the boundary, yielding for the gradient in the complex plane 
\be
\partial_x\pm i\partial_y=e^{\pm i\alpha(s)}\left(\partial_n\pm i\partial_s\right)
\label{grad}
\ee
in terms of the normal and tangential derivatives $\partial_n=\boldsymbol{n}\cdot\boldsymbol{\nabla}$ and $\partial_s=\boldsymbol{t}\cdot\boldsymbol{\nabla}$, respectively. Then with~\refeq{DE} the BCs are obtained in the form
\ba
\left.\left(\partial_n + i\partial_s\right)\psi_1(n,s)\right\vert_{n\to 0^-}&=-k\mathcal{K}^{-1}\psi_1(s),\,
\label{BC2i}
\left.\left(\partial_n - i\partial_s\right)\psi_2(n,s)\right\vert_{n\to 0^-}&=k\mathcal{K}\psi_2(s),
\ea
where $n\to 0^-$ signifies that the boundary is approached from the interior, $\psi_{1,2}(n=0,s)=\psi_{1,2}(s)$.   

For massless particles $\theta_\beta=0$ and $\mathcal{K}=1$, whereas in the nonrelativistic limit characterized by $E\simeq m_0c^2$, which is reached for $\beta =\frac{m_0c}{\hbar k}\to\infty$ for fixed, nonzero $\hbar k$, or for $\tilde\beta\gg k$ where  $\tilde\beta=\frac{m_0c}{\hbar}$ is the rest-energy momentum in units of $\hbar$, corresponding to $\mathcal{K}\simeq\frac{1}{2\beta}\to 0$ and $\theta_\beta\to\pi /2$. The tangential derivative with respect to $s$ is eliminated by introducing the wave functions
\be
\Phi^\pm_1(n,s)=\left\{{\psi_1(n,s)\pm [-ie^{-i\alpha(s)}\mathcal{K}]\psi_2(n,s)}\right\}/{2},\,
\Phi^\pm_2(n,s)=\left\{{\psi_2(n,s)\pm [ie^{i\alpha(s)}\mathcal{K}^{-1}]\psi_1(n,s)}\right\}/{2}.\label{Phi12}
\ee
Employing $\Phi^+_{1,2}(n,s)\xrightarrow{n\to 0^-}\psi_{1,2}(s)$ and the derivative of the BC~\refeq{BC1} with respect to $s$, $i\partial_s\psi_2(s)=-\kappa(s)\psi_2(s)+ie^{i\alpha (s)}\mathcal{K}^{-1}i\partial_s\psi_1(s)$, with $\kappa(s)=\frac{d\alpha(s)}{ds}$ denoting the curvature of the boundary at $s$, the BCs~\refeq{BC2i} become
\be
\left[\tilde\beta+\frac{1}{2}\kappa(s)\right]\Phi^+_j(s)+\left.\partial_n\Phi^+_j(n,s)\right\vert_{n\to 0^-}=0,\, j=1,2.
\label{BCRobin}
\ee
Hence, $\Phi^+_{1,2}(s)$ obey Robin BCs~\cite{Sieber1995,Berry2008,Yupei2019}, however, they are linked by~\refeq{BC1}.

The reader might wonder why Dirichlet BCs are not simply imposed on both spinor components. The reason is that these do not have common nodal lines. To prove this, it is assumed that $\psi_1(\boldsymbol{r})$ has a nodal line denoted by $\tilde\Gamma$ and a coordinate system $(\tilde n,\tilde t)$ moving along $\tilde\Gamma$ with axes normal and tangential to $\tilde\Gamma$ is introduced. Similar to~\refeq{grad}, with $u(t)$ denoting the angle between the normal vector and the $x$ axis, the gradient becomes
\be
\partial_x\pm i\partial_y=e^{\pm iu(\tilde t)}\left(\partial_{\tilde n}\pm i\partial_{\tilde t}\right).
\label{tn}
\ee
The assumption $\psi_1(\boldsymbol{r})\vert_{\tilde\Gamma}=0$ implies that the first spinor component is constant along the nodal line, that is, $\partial_{\tilde t}\psi_1(\tilde n,\tilde t)\vert_{\tilde\Gamma}=0$. Transforming the Dirac Hamiltonian~\refeq{DE} from Cartesian coordinates to $(\tilde t,\tilde n)$ yields for the second spinor component 
\be
k\psi_2(\tilde n,\tilde t)\vert_{\tilde\Gamma}=-ie^{iu(\tilde t)}\partial_{\tilde n}\psi_1(\tilde n,\tilde t)\vert_{\tilde\Gamma},\,
0=\left(\partial_{\tilde n}-i\partial_{\tilde t}\right)\psi_2(\tilde n,\tilde t)\vert_{\tilde\Gamma}\label{NL}.
\ee
These equations demonstrate that $\psi_1$ and $\psi_2$ cannot exhibit nodal lines simultaneously. Indeed, this would imply that $\partial_{\tilde t}\psi_j(\tilde n,\tilde t)\vert_{\tilde\Gamma}=0$ and $\partial_{\tilde n}\psi_j(\tilde n,\tilde t)\vert_{\tilde\Gamma}=0$ for $j=1,2$ and thus, that they vanish in some region of the billiard area.

The Dirac equation for NBs complies with the nonrelativistic limit $m_0c\gg\hbar k$. Namely, inserting $E\simeq m_0c^2+(\hbar k)^2/(2m_0)$ into the equation for the second wave-function component in~(\ref{DE}) gives~\cite{Dietz2020}
\be
\psi_2(\boldsymbol{r})\simeq -\frac{i}{2\tilde\beta}\left(\partial_x+i\partial_y\right)\psi_1(\boldsymbol{r}), 
\label{psi2nonrel}
\ee
and thus for the first one the Schr\"odinger equation,
\be
\left(\Delta+k^2\right)\psi_1(\boldsymbol{r})=0.
\label{psi1nonrel}
\ee
Along the boundary~\refeq{psi2nonrel} becomes with Eqs.~(\ref{BC1}) and~(\ref{BC2i}), 
\be
\psi_2(s)\simeq -\frac{i}{2\tilde\beta}e^{i\alpha(s)}\left(\partial_n+i\partial_s\right)\psi_1(n,s)\vert_{n\to 0^-}
=\frac{i}{2\beta}e^{i\alpha(s)}\mathcal{K}^{-1}\psi_1(s)
=\frac{\mathcal{K}^{-1}}{2\beta}\psi_2(s),
\label{BCnonrel}
\ee
which is fulfilled in the nonrelativistic limit. Furthermore, in that limit $\psi_1(\boldsymbol{r})$ decouples from $\psi_2(\boldsymbol{r})$ implying that $\Phi^+_j(n,s)\equiv\psi_j(n,s)$ and, as can be seen from~\refeq{tildepsi}, the second wave function component becomes vanishingly small as $\sin\theta_\beta\to 1$. Accordingly, $\psi_1(\boldsymbol{r})$ can be determined by solving the Schr\"odinger equation of a QB subject to the Robin BC~\refeq{BCRobin} which for $\tilde\beta\to\infty$ turns into the Dirichlet BC~\cite{Sieber1995}. Thus, the Dirac equation for NBs with BCs given in~\refeq{BC1} evolves in the nonrelativistic limit into the wave equation~\refeq{Schr} for the corresponding nonrelatistic QB with Dirichlet BCs. This was demonstrated numerically in~\cite{Dietz2020,Zhang2021,Zhang2023}.  

The choice of the coordinate system in Eqs.~(\ref{DE}) and~(\ref{BCm}) depends on the billiard shape. For polygonal shapes cartesian coordinates are appropriate, 
\be
k\boldsymbol{\psi}=
-i\begin{pmatrix}
  0 &\frac{\partial}{\partial x} -i\frac{\partial}{\partial y}\\
  \frac{\partial}{\partial x} +i\frac{\partial}{\partial y} &0
  \end{pmatrix}
\boldsymbol{\psi}.
\label{Dcartesian}
\ee
Its solutions can be written in terms of plane waves~\cite{Berry1987},
\be
\begin{pmatrix}
\psi_1 \\ \psi_2
\end{pmatrix}
=
\frac{1}{\sqrt{2}}e^{i\boldsymbol{k}\cdot\boldsymbol{r}}\begin{pmatrix}
e^{-i\theta_k/2} \\ e^{i\theta_k/2}
\end{pmatrix}
\ee
where $\theta_k$ is the angle between $\boldsymbol{k}=(k_x,k_y)$ and the $x$-axis, $\theta_k=\arctan\left(\frac{k_y}{k_x}\right)$. 

For curved shapes resulting from a conformal mapping of the circle the domain and boundary are defined via~\refeq{coordinate} with $f(r)=r$. The coordinate transformation from cartesian coordinates $(x,y)$ to polar coordinates $(r,\gamma)$ with $z=re^{i\gamma}$ yields with $\left( r\frac{\partial}{\partial r} +i\frac{\partial}{\partial \gamma}\right)z=0$ for the gradient in the complex plane
\be
\label{Grad1}\frac{\partial}{\partial x}+i\frac{\partial}{\partial y}=
\frac{1}{z^\ast\left[w^\prime (z)\right]^\ast}\left( r\frac{\partial}{\partial r} +i\frac{\partial}{\partial \gamma}\right),
\ee
where $^\ast$ denotes complex conjugation. In this coordinate system solutions of the Dirac equation can be written in terms of a plane-wave expansion~\cite{Dietz1993},
\be
\Psi_{1}(r,\gamma) = \sum_ma_{m}(k)i^mJ_m(k\vert w(z)\vert)e^{im\theta(z)},\, \Psi_{2}(r,\gamma)=\sum_ma_{m}(k)i^{m+1}J_{m+1}(k\vert w(z)\vert)e^{i(m+1)\theta(z)}.
\label{Psi1}
\ee
Here $J_m(x)$ denotes the Bessel function of the first kind of order $m$, the coefficients $a_m(k)$ are complex or real and $e^{i\theta(z)}=\frac{w(z)}{\vert w(z)\vert}$; cf. Ref.~\cite{Dietz2025} for more details. To be an eigenstate of the NB with shape $w(z)$, $\Psi_{1,2}(r,\gamma)$ should fulfill at discrete values of $k$ the BC~\refeq{BC2}. Finding such solutions can be a cumbersome task.

An ansatz of that form with $m\geq 0$ might not be possible, like, e.g., in the case of an elliptic boundary. However, there the ansatz~\refeq{coordinate} works, yielding for the gradient
\be
\left(\frac{\partial}{\partial x} +i\frac{\partial}{\partial y}\right)=
\frac{1}{z^\star\left[w^\prime (z)\right]^\star}\left[\left(\frac{df(r)}{dr}\right)^{-1}f(r)\frac{\partial}{\partial r} +i\frac{\partial}{\partial \gamma}\right].
\label{Grad2}\ee
Another commonly used parametrization for convex billiards is $w(r,\theta)=R(r,\theta)e^{i\theta}$, $R(r,\theta)=\vert w(r,\theta)\vert$, and $\frac{\partial R(r,\theta)}{\partial r}\ne 0$,
with
\be
R(r,\theta)\frac{\partial R(r,\theta)}{\partial r}\left(\frac{\partial}{\partial x} +i\frac{\partial}{\partial y}\right)= -i\frac{\partial w(r,\theta)}{\partial\theta}\frac{\partial}{\partial r} +i\frac{\partial w(r,\theta)}{\partial r}\frac{\partial}{\partial\theta}\,  ,
\label{Grad3}\ee
and the boundary again defined by some value $r=r_0$.

\subsection{Symmetries\label{Syms}}
The properties of the eigenstates of QBs and NBs exhibiting a geometric symmetry of their shape differ. Generally, when applying an orthogonal transformation $\mathcal{\hat R}$ to $\boldsymbol{r}$, $\boldsymbol{r^\prime}=\mathcal{\hat R}\boldsymbol{r}$, and the associated unitary transformation $\mathcal{\hat U}$ to a Dirac Hamiltonian $\mathcal{\hat H}_D(\boldsymbol{r})$ with solutions $\boldsymbol{\psi}(\boldsymbol{r})$ of the Dirac equation~\refeq{DE}, then the eigenfunctions of the transformed Dirac Hamiltonian $\mathcal{\hat H}_D(\boldsymbol{r^\prime})=\mathcal{\hat U}^\dagger\mathcal{\hat H}_D(\boldsymbol{r})\mathcal{\hat U}$ are $\boldsymbol{\bar\psi}(\boldsymbol{r^\prime})=\mathcal{\hat U}^\dagger\boldsymbol{\psi}(\boldsymbol{r})$. 

For a billiard with mirror symmetry the coordinate system can be defined such that the symmetry axis coincides either with the $x$ or $y$ axis. For a mirror reflection  of $\boldsymbol{r}$ with respect to these axes the transformations are $\mathcal{\hat R}_{x,y}=\pm{\hat\sigma}_z$ and $\mathcal{\hat U}_x={\hat\sigma}_x$ or $\mathcal{\hat U}_y=i{\hat\sigma}_y$, respectively. The wave functions of a QB exhibiting a mirror symmetry can be separated into symmetric or antisymmetric ones with respect to the symmetry axis, and thus fulfill along it either Neumann or Dirichlet BCs. However, for the corresponding NB the spinors~\cite{Zhang2021} obtained by applying $\mathcal{\hat U}_x$ or $\mathcal{\hat U}_y$ to the eigenspinors of a NB do not fulfill the BC~\refeq{BC1}. Nevertheless, the eigenspinor components of a NB with mirror symmetry with respect to the $y$ or $x$ axis may have the property~\cite{Dietz2019}
\be
\left[\psi_1(-x,y),\psi_2(-x,y)\right]=\pm\left[\psi_1^\ast(x,y),\psi_2^\ast(x,y)\right],
\label{symy}
\ee
or
\be
\left[\psi_1(x,-y),\psi_2(x,-y)\right]=\left[\pm\psi_1^\ast(x,y),\mp\psi_2^\ast(x,y)\right],
\label{symx}
\ee
deduced from the equalities $\mathcal{\hat H}_D(-x,y)=\mathcal{\hat H}^\ast_D(x,y)$ and $\mathcal{\hat H}_D(x,-y)={\hat\sigma_z}\mathcal{\hat H}^\ast_D(x,y){\hat\sigma_z}$. The property that eigenspinors of NBs with a mirror symmetry cannot be assigned to the associated symmetry classes and that the spinor components do not have common nodal lines are directly related. A consequence is that for pairs of NBs with the shapes of isospectral QBs isospectrality fails~\cite{Yupei2020}. Indeed, the question of Kac, whether one can hear the shape of a drum or, equivalently, whether the eigenvalue spectra of billiards with different shapes cannot be identical, has been refuted for pairs of nonisometric nonrelativistic QBs, for which isospectrality was proven by employing the transplantation method~\cite{Giraud2010}. However, it was demonstrated in~\cite{Yupei2020} ifirst, that this method fails for the corresponding pairs of NBs and second, that they cease to be isospectral.   

For a counterclockwise rotation by $\frac{2\pi}{M}$, the orthogonal and unitary transformations are performed by applying 
\be
\label{Rn}
\mathcal{\hat R}_M=\begin{pmatrix}
        \cos\left(\frac{2\pi}{M}\right) & -\sin\left(\frac{2\pi}{M}\right)\\
        \sin\left(\frac{2\pi}{M}\right) &  \, \cos\left(\frac{2\pi}{M}\right)
\end{pmatrix},
\ee
and
\be
\label{Un}
\mathcal{\hat U}_M=
\begin{pmatrix}
e^{i\frac{\pi}{M}} & 0\\
0 &e^{-i\frac{\pi}{M}}
\end{pmatrix},
\ee
respectively. The eigenstates of a QB, whose shape has an $M$-fold rotational symmetry, $w(s^\prime)=e^{i\frac{2\pi}{M}}w(s)$ and $e^{i\alpha(s^\prime)}=e^{i\frac{2\pi}{M}}e^{i\alpha(s)}$, can be separated into $M$ subspaces labeled by $l$ according to the transformation properties of the eigenfunctions under the operator $\mathcal{\hat R}_M$~\cite{Robbins1989,Joyner2012}
        \be
        \psi^{(l)}_m(\mathcal{\hat R}_M\boldsymbol{r})=e^{il\frac{2\pi}{M}}\psi^{(l)}_m(\boldsymbol{r}),\, l=0,\dots ,M-1.
        \label{RotSym}
        \ee
They are real for $l=0$ and for even $M$ also for $l=M/2$, that is, invariant under the time-reversal operator $\hat T$~\cite{Haake2018}, whereas for $l\ne 0,M/2$ they are complex and
\be
\hat T\psi^{(l)}_m(\boldsymbol{r})=\psi^{(M-l)}_m(\boldsymbol{r}).\label{Top} 
\ee
Time-reversal invariance of the QB implies that $\psi_m^{(l)}(\boldsymbol{r})$ and $\psi_m^{(M-l)}(\boldsymbol{r})$ are eigenfunctions with the same eigenvalue $k_m^2$. Accordingly, the spectrum can be separated into nondegenerate eigenvalues (singlets) with $l=0,\frac{M}{2}$ and pairwise degenerate ones (doublets) with labels $l,M-l$. For $l=0$ the eigenfunctions are rotationally invariant. 

For the eigenspinors of the Dirac Hamiltonian Eqs.~(\ref{Rn}) and~(\ref{Un}) yield that $\boldsymbol{\bar\psi}^T_M(\mathcal{\hat R}_M\boldsymbol{r})= [e^{-i\frac{\pi}{M}}\psi_1(\boldsymbol{r}),e^{i\frac{\pi}{M}}\psi_2(\boldsymbol{r})]$. Thus, if the shape of a NB has an $M$-fold rotational symmetry, the BC~\refeq{BC1} is fulfilled for $\boldsymbol{\bar\psi}(\mathcal{\hat R}_M\boldsymbol{r})$ if $\boldsymbol{\psi}(\boldsymbol{r})$ is an eigenspinor of $\mathcal{\hat H}_{NB}$. Furthermore, the spinor components of the eigenstates of the corresponding NB can again be classified according to their transformation properties under a rotation by $\frac{2\pi}{M}$ into $M$ subspaces~\cite{Dietz2021,Zhang2021,Zhang2023}, however, they belong to different ones.  Indeed, for $\psi_{1,m}(\boldsymbol{r})$ belonging to the subspace $l$ it follows,
        \be
        \psi^{(l)}_{1,m}(\mathcal{\hat R}_M\boldsymbol{r})=e^{il\frac{2\pi}{M}}\psi^{(l)}_{1,m}(\boldsymbol{r})\label{sympsi}\Rightarrow
        \psi^{(\tilde l)}_{2,m}(\mathcal{\hat R}_M\boldsymbol{r})=e^{i(l-1)\frac{2\pi}{M}}\psi^{(\tilde l)}_{2,m}(\boldsymbol{r}),
        \ee
implying that $\psi_{2,m}(\boldsymbol{r})$ belongs to the subspace $\tilde l =(l-1)$, where $\tilde l=-1$ corresponds to $l=M-1$, and thus the eigenspinors themselves can not be classified according to their transformation properties under rotation by $\frac{2\pi}{M}$. A direct consequence of this feature is that in contrast to a QB with an $M$-fold symmetry the corresponding NB and sectors of the NB, that are obtained by cutting it along the symmetry lines into fundamental domains, do not have common eigenstates. For NBs the eigenstates are complex for all symmetry classes and there are no degeneracies, because \T invariance is violated. In the following, when referring for NBs to a symmetry class $l$, then that of the first spinor component is meant.  

\subsection{Boundary-integral equations~\label{BIEs}}
In~\cite{Berry1987} a boundary-integral equation (BIE) has been derived for massless NBs, which is also applicable to massive ones, after bringing the Dirac equation~\refeq{DE} to the form~\refeq{Hamiltonian}. It's origin is the Green theorem, which provides an exact integral equation for the computation of the eigenvalues and eigenfunctions of a QB or eigenspinors of a NB within the domain of the billiard in terms of those on the boundary, which are subject to the BCs. The boundary-integral method has the enormous advantage that the eigenvalue problem is reduced from a two-dimensional differential equation to a one-dimensional boundary integral. Furthermore, since the BCs are taken into account, the BIE provides a suitable quantization equation to obtain analytical results for boundary-value problems like the present one, or to derive a trace formula for the spectral density of QBs~\cite{Harayama1992,Sieber1997}. The derivation of these BIEs also involves the matrix equation for the free-space Green function, 
\be
k\boldsymbol{G_0}(\boldsymbol{r},\boldsymbol{r}^\prime)+i\boldsymbol{\hat\sigma}\cdot\boldsymbol{\nabla_r}\boldsymbol{G_0}(\boldsymbol{r},\boldsymbol{r}^\prime)=\delta\left(\boldsymbol{r}-\boldsymbol{r}^\prime\right)\II,\, 
\boldsymbol{G_0}(\boldsymbol{r},\boldsymbol{r}^\prime)
=\left(k+i\epsilon -i\boldsymbol{\hat\sigma}\cdot\boldsymbol{\nabla}\right)\mathcal{G}_0(\boldsymbol{r},\boldsymbol{r}^\prime),
\label{Greeni}
\ee
obtained from the Green operator $\boldsymbol{\hat G}=\left(k+i\epsilon +i\boldsymbol{\hat\sigma}\cdot\boldsymbol{\nabla}\right)^{-1}$ corresponding to the Dirac operator~\refeq{DE}. Here $\II$ stands for the $2\times 2$ identity matrix, $\boldsymbol{G_0}(\boldsymbol{r},\boldsymbol{r^\prime})$ denotes $\boldsymbol{\hat G}$ in the coordinate representation, $\boldsymbol{G_0}(\boldsymbol{r},\boldsymbol{r^\prime})=\langle\boldsymbol{r}\vert\boldsymbol{\hat G}\vert\boldsymbol{r}^\prime\rangle$, and $\mathcal{G}_0(\boldsymbol{r},\boldsymbol{r}^\prime)=-i/4 H^{(1)}_0(k\rho(\boldsymbol{r},\boldsymbol{r}^\prime))$ is the scalar free-space Green function in two-dimensional space, with $\boldsymbol{\rho}(\boldsymbol{r},\boldsymbol{r}^\prime)=\boldsymbol{r}-\boldsymbol{r}^\prime$, $\rho (\boldsymbol{r},\boldsymbol{r}^\prime)=\vert\boldsymbol{\rho}(\boldsymbol{r},\boldsymbol{r}^\prime)\vert$ and $H_{m}^{(1)}(k\rho)=J_{m}(k\rho)+iY_{m}(k\rho)$ denoting the Hankel function of order $m$ of the first kind. 

Multiplying~\refeq{Greeni} from the left with $\boldsymbol{\psi^\dagger}(\boldsymbol{r})$ and~\refeq{Hamiltonian} from the right with $\boldsymbol{G_0}(\boldsymbol{r},\boldsymbol{r}^\prime)$, subtracting the resulting equations from each other and integrating in the billiard domain $\Omega$ over $\boldsymbol{r}$ yields 
\be
i\int\int_\Omega{\rm d}^2\boldsymbol{r}\boldsymbol{\nabla}_{\boldsymbol{r}}\cdot\left[\boldsymbol{\psi}(\boldsymbol{r})^\dagger\boldsymbol{\hat\sigma G_0}(\boldsymbol{r},\boldsymbol{r}^\prime)\right]
=i\oint_{\partial\Omega} {\rm d}s\,\boldsymbol{\hat n}\cdot\left[\boldsymbol{\psi}(\boldsymbol{r})^\dagger\boldsymbol{\hat\sigma G_0}(\boldsymbol{r},\boldsymbol{r}^\prime)\right]
=\left\{
\begin{array}{cll}
\boldsymbol{\psi}^\dagger(\boldsymbol{r}^\prime)&,&\boldsymbol{r}^\prime\in\Omega\backslash\partial\Omega\\
\frac{1}{2}\boldsymbol{\psi}^\dagger(\boldsymbol{r}^\prime)&,&\boldsymbol{r}^\prime\in\partial\Omega \\
0&,&{\rm otherwise}
\end{array}
\right.
\label{Greentheorem}
\ee
Here the Gauss theorem has been employed to convert the two-dimensional integral over the billiard area into a one-dimensional line integral along the billiard boundary $\partial\Omega$. If $\boldsymbol{r}^\prime$ is located at a corner of the boundary, the factor $\frac{1}{2}$ should be replaced by $\frac{\theta_{in}}{2\pi}$ with $\theta_{in}$ denoting the inner angle of the corner~\cite{Baecker2003,Okada2005,Yupei2020}. Introducing in the complex-plane presentation along the boundary the notation $e^{i\xi(\gamma,\gamma^\prime)}=\frac{\boldsymbol{\rho}(\gamma,\gamma^\prime)}{\rho(\gamma,\gamma^\prime)}$, where $\boldsymbol{\rho}(\gamma,\gamma^\prime)=w(\gamma)-w(\gamma^\prime)$, the free-space Green function reads
\be
\boldsymbol{G_0}(\gamma,\gamma^\prime)
=\frac{k}{4}
\begin{pmatrix}
        -iH^{(1)}_0[k\rho (\gamma,\gamma^\prime)] &e^{-i\xi(\gamma,\gamma^\prime)}H^{(1)}_1[k\rho (\gamma,\gamma^\prime)]\\
        e^{i\xi(\gamma,\gamma^\prime)}H^{(1)}_1[k\rho (\gamma,\gamma^\prime)] &-iH^{(1)}_0[k\rho (\gamma,\gamma^\prime)]
\end{pmatrix}\, .
\ee

For $\boldsymbol{r}^\prime\in\partial\Omega$~\refeq{Greentheorem} provides upon application of the BC~\refeq{BCm} for each spinor component an equation for the eigenvalues $k_m$ of NBs,
\begin{align}
\psi_1^\ast(\gamma^\prime)\label{eigenvalpsi1}&=
\frac{ik}{2}\oint_{\partial\Omega}\vert w^\prime(\gamma)\vert d\gamma\psi_1^\ast(\gamma)
\left\{-\mathcal{K}^{-1}H_0^{(1)}[k\rho(\gamma,\gamma^\prime)]+e^{-i\alpha(\gamma)}e^{i\xi(\gamma,\gamma^\prime)}H_1^{(1)}[k\rho(\gamma,\gamma^\prime)]\right\},\\
\psi_2^\ast(\gamma^\prime)\label{eigenvalpsi2}&=
\frac{ik}{2}\oint_{\partial\Omega}\vert w^\prime(\gamma)\vert d\gamma\psi_2^\ast(\gamma)
\left\{\mathcal{K}H_0^{(1)}[k\rho(\gamma,\gamma^\prime)]+e^{i\alpha(\gamma)}e^{-i\xi(\gamma,\gamma^\prime)}H_1^{(1)}[k\rho(\gamma,\gamma^\prime)]\right\}.
\end{align}
Yet, in that form the BIEs cannot be used to compute eigenvalues of a NB accurately, because at $\rho=0$ $H_0^{(1)}(k\rho)$ and $H_1^{(1)}(k\rho)$ have a logarithmic and $1/\rho$ singularity, respectively. For the massless case these are removed by using Eqs.~(\ref{eigenvalpsi1}) and~(\ref{eigenvalpsi2}) to obtain BIEs for $\Phi^+_{1,2}(n,s)$ defined in~\refeq{Phi12}, and thus for $\psi_{1,2}(s)$ in the limit $n\to 0^-$. Using the same procedure for the general case~\cite{Dietz2020} yields
\begin{align}
\label{BIEpsi}
&\cos\theta_\beta\psi_1^\ast(\gamma^\prime)=
\frac{ik}{4}\oint_{\partial\Omega}\vert w^\prime(\gamma)\vert d\gamma\psi_1^\ast(\gamma)\\
&\times\left(\left\{\left[e^{i\left(\alpha(\gamma^\prime)-\alpha(\gamma)\right)}-1\right]-\sin\theta_\beta\left[e^{i\left(\alpha(\gamma^\prime)-\alpha(\gamma)\right)}+1\right]\right\}H_0^{(1)}(k\rho)
	+\cos\theta_\beta\left[e^{i\left(\xi(\gamma,\gamma^\prime)-\alpha(\gamma)\right)}+e^{-i\left(\xi(\gamma,\gamma^\prime)-\alpha(\gamma^\prime)\right)}\right]H_1^{(1)}(k\rho)\right).\nonumber
\end{align}
Here, the argument of $\rho=\rho(\gamma,\gamma^\prime)$ is suppressed. For the massless case, the integrand vanishes in the limit $\gamma\to\gamma^\prime$ and $\rho\to 0$, whereas for nonzero mass all terms vanish except for the one which is proportional to $\sin\theta_\beta H_0^{(1)}[k\rho(\gamma,\gamma^\prime)]$. This term diverges logarithmically for $\delta\gamma^\prime\to 0$, as can be seen when evaluating the different contributions to the integrand for $\gamma=\gamma^\prime+\delta\gamma^\prime,\, \delta\gamma^\prime\to 0$,
\be\label{approx}
e^{i\alpha(\gamma^\prime+\delta\gamma^\prime)}\simeq e^{i\alpha(\gamma^\prime)}\left[1+i\kappa(\gamma^\prime)\vert w^\prime (\gamma^\prime)\vert\delta\gamma^\prime\right],\, e^{i\xi(\gamma^\prime+\delta\gamma^\prime,\gamma^\prime)}
\simeq ie^{i\alpha(\gamma^\prime)}\frac{\delta\gamma^\prime}{\vert\delta\gamma^\prime\vert}\left[1+\frac{i}{2}\kappa(\gamma^\prime)\vert w^\prime(\gamma^\prime)\vert\delta\gamma^\prime\right],\, \rho\simeq\vert w^\prime(\gamma^\prime)\delta\gamma^\prime\vert.
\ee
This singularity is removed by employing a condition for the fulfillment of the BC~\refeq{BCm}, which is obtained based on the BIEs for $\Phi^-_{1,2}(n,s)$ defined in~\refeq{Phi12}, with Eqs.~(\ref{eigenvalpsi1}) and~(\ref{eigenvalpsi2}),
\begin{align}
        &\oint_{\partial\Omega}\vert w^\prime(\gamma)\vert{\rm  d}\gamma\left[e^{i\left(\alpha(\gamma^\prime)-\alpha(\gamma)\right)}+1\right]H_0^{(1)}(k\rho)\psi_1^\ast(\gamma)\label{Eq3}\\
        =&\oint_{\partial\Omega}\vert w^\prime(\gamma)\vert d\gamma
	\left\{\sin\theta_\beta\left[e^{i\left(\alpha(\gamma^\prime)-\alpha(\gamma)\right)}-1\right]H_0^{(1)}(k\rho)
	+\cos\theta_\beta\left[e^{i\left(\xi(\gamma,\gamma^\prime)-\alpha(\gamma)\right)}-e^{-i\left(\xi(\gamma,\gamma^\prime)-\alpha(\gamma^\prime)\right)}\right]H_1^{(1)}(k\rho)\right\}\psi_1^\ast(\gamma).\nonumber
\end{align}
Replacing the corresponding term in~\refeq{BIEpsi} by the right-hand side of this equation leads to the BIE
\begin{align}
\label{BIEpsi1}
        \psi_1^\ast(\gamma^\prime)=&\frac{ik}{4}\oint_{\partial\Omega}\vert w^\prime(\gamma)\vert d\gamma e^{i\frac{\Delta\Phi(\gamma,\gamma^\prime)}{2}}2Q^{(1)}_1(k;\gamma^\prime,\gamma)\psi_1^\ast(\gamma)\\
	\label{Kernelpsi1}	
        Q^{(1)}_1(k;\gamma^\prime,\gamma)=&i\cos\theta_\beta\sin\left[\frac{\Delta\Phi(\gamma,\gamma^\prime)}{2}\right]H_0^{(1)}(k\rho)
        +\left\{i\sin\theta_\beta\sin\left[\Delta\chi(\gamma,\gamma^\prime)\right]
+\cos\left[\Delta\chi(\gamma,\gamma^\prime)\right]\right\}H_1^{(1)}(k\rho),
\end{align}
with $\Delta\Phi(\gamma,\gamma^\prime)=\alpha(\gamma^\prime)-\alpha(\gamma)$ and $\Delta\chi(\gamma,\gamma^\prime)=\frac{\alpha(\gamma^\prime)+\alpha(\gamma)}{2}-\xi(\gamma,\gamma^\prime)$.
In the limit $\gamma=\gamma^\prime+\delta\gamma^\prime\to\gamma^\prime$ all terms except the $\sin\theta_\beta$ term vanish. For $\vert\delta\gamma^\prime\vert\ll\gamma^\prime$ this term becomes with~\refeq{approx} and $H_1^{(1)}(k\rho)\approx -\frac{2i}{k\vert w^\prime(\gamma^\prime)\vert\vert\delta\gamma^\prime\vert}$
\be
\frac{ik}{4}e^{i\frac{\Delta\Phi(\gamma,\gamma^\prime)}{2}}2i\sin\theta_\beta\sin\left[\Delta\chi(\gamma,\gamma^\prime)\right]H_1^{(1)}(k\rho)\vert w^\prime(\gamma)\vert  d\gamma
\approx -\frac{\sin\theta_\beta}{2\pi}\left[\kappa(\gamma^\prime)\vert w^\prime(\gamma^\prime)\vert+\frac{2i}{\delta\gamma^\prime}\right] d\gamma .\label{approx3}
\ee

In the contour integration~\refeq{BIEpsi1} the singularity at $\gamma=\gamma^\prime$ needs to be treated separately. For this the contour is slightly distorted in the region $\gamma\in [\gamma^\prime -\epsilon,\gamma^\prime +\epsilon]$ with $0<\epsilon\ll \gamma^\prime$ into a half circle of radius $\epsilon$ with center at $\gamma^\prime$ such that the singularity lies within the countour. Introducing a shifted integration variable $\tilde\gamma =\gamma-\gamma^\prime$ and taking the limit $\epsilon\to 0$ gives~\cite{Dietz2022a}
\be 
-\frac{2i\sin\theta_\beta}{2\pi}\int_{0}^{\pi} id\phi\psi_1^\ast(\gamma^\prime +\epsilon e^{i\phi})
\xrightarrow{\epsilon\to 0}\sin\theta_\beta\psi_1^\ast(\gamma^\prime),\label{approx4}
\ee
leading to the final result for the BIE,
\be
\left(1-\sin\theta_\beta\right)\psi_1^\ast(\gamma^\prime)=\frac{ik}{4}\fint_{0}^{2\pi}\vert w^\prime(\gamma)\vert d\gamma Q^{(1)}_1(k;\gamma^\prime,\gamma)\psi_1^\ast(\gamma).\label{Eq4}\\
\ee
Here, $\gamma=\gamma^\prime$ is excluded from the integration range. This has been verified numerically and also analytically in the semiclassical limit; cf. e.g.~\cite{Dietz2022a}. The corresponding BIE for $\psi_2^\ast(\gamma^\prime)$ is obtained by multiplying the integrand in~\refeq{BIEpsi1} with $e^{-i\Delta\Phi(\gamma,\gamma^\prime)}$.  The associated eigenspinor components are obtained from~\refeq{Greentheorem} by inserting the eigenvalues and boundary function $\psi^\ast_j(\gamma)$ into the BIE for $\boldsymbol{r}^\prime\in\Omega$. 

A few remarks are appropriate at this point. Namely, in the nonrelativistic limit $\sin\theta_\beta\to 1$ the left hand side approaches zero, and the numerical evaluation of the BIE becomes increasingly cumbersome. Indeed, for $\sin\theta_\beta\simeq 1$ neither this BIE nor Eqs.~(\ref{eigenvalpsi1}) and~(\ref{eigenvalpsi2}) are suitable equations for the determination of the eigenvalues because, as can be deduced from Eqs.~(\ref{tildepsi}) and~(\ref{psi2nonrel})-(\ref{BCnonrel}), then the spinor components decouple, $\psi_2(\boldsymbol{r})$ becomes vanishingly small and the BC becomes $\psi_1(\gamma)=0$. It has been demonstrated in~\cite{Dietz2020} that the BIEs~\refeq{BIEpsi} can be brought to the general form of the BIE for QBs, implying that in the nonrelativistic limit the BIE given in~\cite{Baecker2003} should be applied. 

For a NB with $M$-fold rotational symmetry the numerical effort for the computation can be drastically reduced by separating the BIE into individual ones for each symmetry class. This is achieved by employing the periodicity properties
\be
w\left(\varphi+\lambda\frac{2\pi}{M}\right)=e^{i\lambda\frac{2\pi}{M}}w(\varphi)\label{wl1},\,
w^\prime\left(\varphi+\lambda\frac{2\pi}{M}\right)=e^{i\lambda\frac{2\pi}{M}}w^\prime (\varphi),\,
e^{i\alpha\left(\varphi+\lambda\frac{2\pi}{M}\right)}=e^{i\lambda\frac{2\pi}{M}}e^{i\alpha(\varphi)}, 
\ee
with $\lambda=0,1,2,\dots,M-1$, such that the integration range of $\varphi$ and $\varphi^\prime$ can be restricted to one fundamental domain, $\varphi,\varphi^\prime\in[0,\frac{2\pi}{M})$, yielding the BIEs
\be
\psi_j^{(l)}(\gamma^\prime)=\int_0^{\frac{2\pi}{M}}d\varphi\tilde Q^{(l)}_j(k;\gamma,\gamma^\prime)\psi_j^{(l)}(\gamma)
\label{BIEsym1}
\ee
with
\be
\tilde Q_j^{(l)}(k;\gamma,\gamma^\prime)=\sum_{\lambda=0}^{M-1}e^{il\frac{2\pi}{M}\lambda}
Q^{(1)}_j\left(k;\gamma+\lambda\frac{2\pi}{M},\gamma^\prime\right),
\label{Msym1}
\ee
where $Q^{(1)}_j\left(k;\gamma,\gamma^\prime\right)$ is defined for $j=1$ in~\refeq{Kernelpsi1} and is obtained for $j=2$ by proceeding as explained below~\refeq{Eq4}.

\subsection{A semiclassical trace formula for massive NBs\label{Trace}}
The semiclassical periodic-orbit theory provides a further important approach within the field of quantum chaos.  It was pioneered by Gutzwiller~\cite{Gutzwiller1971}. He derived a semiclassical approximation for the fluctuating part of the spectral density in terms of a sum over POs~\cite{Gutzwiller1971}, the renowned Gutzwiller trace formula. The derivation requires that all POs are isolated, which is the case for systems whose classical dynamics is fully chaotic. This approach made it possible to understand the effect of the classical dynamics on the properties of the corresponding quantum system in terms of purely classical quantities. An analogous trace formula was obtained for integrable systems by Berry and Tabor based on the semiclassical Einstein-Brillouin-Keller quantization~\cite{Berry1977a}. In~\cite{Dietz2019,Dietz2021} trace formulas are derived for NBs. For the derivation the procedure used in~\cite{Harayama1992,Sieber1997} is adapted, which is based on the BIEs. The semiclassical limit is obtained for $\hbar\to 0$ for fixed $E=\hbar ck_E$ and $\hbar k$, i.e., for fixed $\beta$ and thus $m_0c$, which is equivalent to the limit $k_E\to\infty$ or $k\to\infty$ while leaving $\beta$ and $m_0c$ unchanged. Here it is assumed that the NB has the shape of a CB with chaotic classical dynamics. As mentioned at the end of~\refsec{BIEs}, the BIE~\refeq{BIEpsi} or~\refeq{Eq4} are not suitable in the nonrelativistic limit. Thus, to obtain a semiclassical quantization condition which is well-defined for all values $0\leq m_0<\infty$, i.e., $0\leq\theta_\beta\leq \pi/2$ and a trace formula with convergence properties similar to those of Gutzwiller's trace formula, a combination of~\refeq{BIEpsi} and a BIE, which fails in the opposite limit, namely the limit $\tilde\beta\to 0$, has been used in~\cite{Dietz2020}. That BIE is derived from the relation~\refeq{BC2} which comprises normal derivatives of $\Phi_{1,2}(n,s)$. In view of possible jumps accompanying the integrals over normal derivatives~\cite{Kleinman1974},~\refeq{BC2i} is employed to compute tangential derivatives instead,
\be
-\frac{\kappa(s^\prime)}{2}\psi_1^\ast(s^\prime)=
\frac{i\partial_{s^\prime}\psi_1^\ast(s^\prime)-ie^{i\alpha(s^\prime)}\mathcal{K}i\partial_{s^\prime}\psi_2^\ast(s^\prime)}{2}.
\ee
A BIE is then obtained with Eqs.~(\ref{eigenvalpsi1}) and~(\ref{eigenvalpsi2}) and
\ba
\partial_{s^\prime} H^{(1)}_0[k\rho(s,s^\prime)]&=&\sin[\xi(s,s^\prime)-\alpha(s^\prime)]kH^{(1)}_1[k\rho(s,s^\prime)],\\ 
\partial_{s^\prime} H^{(1)}_1[k\rho(s,s^\prime)]&=&\sin[\xi(s,s^\prime)-\alpha(s^\prime)]
k\left(\frac{H^{(1)}_1[k\rho(s,s^\prime)]}{k\rho(s,s^\prime)}-H_0^{(1)}[k\rho(s,s^\prime)]\right),\\
\partial_{s^\prime} e^{i\xi(s,s^\prime)}&=&-i\cos[\xi(s,s^\prime)-\alpha(s^\prime)]\frac{e^{i\xi(s,s^\prime)}}{\rho(s,s^\prime)},
\ea
such that
\begin{align}
&-\frac{\kappa(\gamma^\prime)}{2}\psi_1^\ast(\gamma^\prime)=\frac{ik}{4}\oint_{\partial\Omega}\vert w^\prime(\gamma)\vert{\rm d}\gamma\psi_1^\ast(\gamma)\left\{\mathcal{I}_1+\mathcal{I}_2+\mathcal{I}_3\right\},\label{BIE2}
\end{align}
with
\begin{align}
&\mathcal{I}_1=\left\{\left[\mathcal{K}e^{i[\alpha(\gamma^\prime)-\alpha(\gamma)]}+\mathcal{K}^{-1}\right]kH_1^{(1)}[k\rho(\gamma,\gamma^\prime)]
-\left[e^{i[\xi(\gamma,\gamma^\prime)-\alpha(\gamma)]}+e^{i[\alpha(\gamma^\prime)-\xi(\gamma,\gamma^\prime)]}\right]kH_0^{(1)}[k\rho(\gamma,\gamma^\prime)]\right\}
\cos[\xi(\gamma,\gamma^\prime)-\alpha(\gamma^\prime)]\nonumber\\
&\mathcal{I}_2=\left\{\left[e^{i[\alpha(\gamma^\prime)-\alpha(\gamma)]}+1\right]kH_0^{(1)}[k\rho(\gamma,\gamma^\prime)]
-\left[\mathcal{K}e^{i[\alpha(\gamma^\prime)-\alpha(\gamma)]}+\mathcal{K}^{-1}\right]e^{i[\xi(\gamma,\gamma^\prime)-\alpha(\gamma^\prime)]}kH_1^{(1)}[k\rho(\gamma,\gamma^\prime)]\right\}\nonumber\\
&\mathcal{I}_3=
\left[e^{i[2\xi(\gamma,\gamma^\prime)-\alpha(\gamma^\prime)-\alpha(\gamma)]}+e^{2i[\xi(\gamma,\gamma^\prime)-\alpha(\gamma^\prime)]}\right]\frac{H_1^{(1)}[k\rho(\gamma,\gamma^\prime)]}{\rho(\gamma,\gamma^\prime)}.
\nonumber
\end{align}
With the BIEs~\refeq{eigenvalpsi1} and~\refeq{eigenvalpsi2} the integral over $\mathcal{I}_2$ can be simplified, 
\be
\frac{ik}{4}\oint_{\partial\Omega}\vert w^\prime(\gamma)\vert{\rm d}\gamma\psi_1^\ast(\gamma)\mathcal{I}_2
=k\frac{\left[\mathcal{K}^{-1}-\mathcal{K}\right]}{2}\psi_1^\ast(\gamma)
-2\frac{ik}{4}\oint_{\partial\Omega}\vert w^\prime(\gamma)\vert{\rm d}\gamma\psi_1^\ast(\gamma)\mathcal{K}^{-1}kH_1^{(1)}[k\rho(\gamma,\gamma^\prime)]
\cos[\xi(\gamma,\gamma^\prime)-\alpha(\gamma^\prime)].\nonumber
\ee
Note that the same result is obtained when applying the normal derivatives to Eqs.~(\ref{eigenvalpsi1}) and~(\ref{eigenvalpsi2}). With $\mathcal{K}\cos\theta_\beta=1-\sin\theta_\beta$, $\mathcal{K}^{-1}\cos\theta_\beta=1+\sin\theta_\beta$ and $\frac{\mathcal{K}^{-1}-\mathcal{K}}{2}=\beta=\frac{\sin\theta_\beta}{\cos\theta_\beta}$ the BIE~\refeq{BIE2} takes the form,
\begin{align}
\label{BIEpsi3}
&\left[\sin\theta_\beta+\frac{\kappa(\gamma^\prime)}{2k}\right]\psi_1^\ast(\gamma^\prime)=\frac{ik}{4}\oint_{\partial\Omega}\vert w^\prime(\gamma)\vert d\gamma e^{i\frac{\Delta\Phi(\gamma,\gamma^\prime)}{2}}2Q^{(2)}_1(k;\gamma^\prime,\gamma)\psi_1^\ast(\gamma)\\
&-\cos\theta_\beta\frac{ik}{4}\oint_{\partial\Omega}\vert w^\prime(\gamma)\vert{\rm d}\gamma\psi_1^\ast(\gamma)\label{BIEpsi3a}
\left[e^{i[2\xi(\gamma,\gamma^\prime)-\alpha(\gamma^\prime)-\alpha(\gamma)]}+e^{2i[\alpha(\gamma^\prime)-\xi(\gamma,\gamma^\prime)]}\right]\frac{H_1^{(1)}(k\rho)}{k\rho},\\
&\frac{Q^{(2)}_1(k;\gamma^\prime,\gamma)}{\cos\left[\xi(\gamma,\gamma^\prime)-\alpha(\gamma^\prime)\right]}=
\cos\theta_\beta\cos\left[\Delta\chi(\gamma,\gamma^\prime)\right]H_0^{(1)}(k\rho)
-\left[i\sin\left(\frac{\Delta\Phi(\gamma,\gamma^\prime)}{2}\right)-\sin\theta_\beta\cos\left(\frac{\Delta\Phi(\gamma,\gamma^\prime)}{2}\right)\right]H_1^{(1)}(k\rho)
\end{align}
More details are provided in~\cite{Dietz2020}. Again, the corresponding equation for $\psi_2^\ast(s^\prime)$ is obtained with~\refeq{BCm} by multiplying the integrand with $e^{-i\Delta\Phi(s,s^\prime)}$. 

In the semiclassical limit the term~\refeq{BIEpsi3a} approaches
\be
\frac{H_1^{(1)}[k\rho(s,s^\prime)]}{k\rho(s,s^\prime)}=\frac{H_0^{(1)}[k\rho(s,s^\prime)]+H_2^{(1)}[k\rho(s,s^\prime)]}{2}\xrightarrow{k\to\infty}-i\sqrt{\frac{2}{\pi k\rho}}e^{ik\rho-\frac{i}{4}\pi}\frac{2}{k\rho}
\ee
which is by a factor $\frac{1}{k\rho}$ smaller than $H_0^{(1)}[k\rho(s,s^\prime)]$ and $H_1^{(1)}[k\rho(s,s^\prime)]\simeq -iH_0^{(1)}[k\rho(s,s^\prime)]$ for $k\rho(s,s^\prime)\to\infty$, that is, for $\rho(s,s^\prime)\ne 0$. In the derivation of the semiclassical trace formula only billiards with smoothly varying boundaries are considered and only POs of nonzero length contribute, so that this term is disregarded. For the same reason $\frac{\kappa(s)}{k}$ is negligible in the semiclassical limit $k\to\infty$. The left hand side of~\refeq{BIEpsi} vanishes in the nonrelativistic limit $\theta_\beta=\pi /2$ whereas that of~\refeq{BIEpsi3} is zero in the ultrarelativistic limit $\theta_\beta=0$. In the derivation of a trace formula based on a BIE it is assumed that the left-hand side of the BIE is nonvanishing (see below~\refeq{SecEq}). THerefore, a combination of the BIEs is used, namely $Q^{(1)}_j(k;\gamma^\prime,\gamma)$, $j=1,2$ in~\refeq{Kernelpsi1} is multiplied with  $\cos\theta_\beta$ and $Q^{(2)}_j(k;\gamma^\prime,\gamma)$, $j=1,2$ in~\refeq{BIEpsi3} with $\sin\theta_\beta$ and then their sum, $Q_j(k;\gamma^\prime,\gamma)=\cos\theta_\beta Q^{(1)}_j(k;\gamma^\prime,\gamma)+\sin\theta_\beta Q^{(2)}_j(k;\gamma^\prime,\gamma)$ is taken~\cite{Dietz2020}. The resulting quantization condition has the form~\cite{Sieber1997,Dietz2019a}
\be
\boldsymbol{\psi}^\dagger\left[\boldsymbol{r}(s^\prime)\right]\label{DetEq}
=\oint_{\partial\Omega}{\rm d}s Q\left[\boldsymbol{r}(k;s^\prime),\boldsymbol{r}(s)\right]\boldsymbol{\psi}^\dagger\left[\boldsymbol{r}(s)\right]
=\boldsymbol{\hat Q}(k)\boldsymbol{\psi}^\dagger\left[\boldsymbol{r}(s)\right],
\ee
with $\boldsymbol{\hat Q}_{ij}(k)=\boldsymbol{\hat Q}_{jj}(k)\delta_{ij} $ denoting the integral operator which upon application to $\psi^\ast_j(s)$ yields $\psi^\ast_j(s^\prime)$. This equation has nontrivial solutions at the zeroes of the spectral determinant,
\be
\det\left(\II -\boldsymbol{\hat Q}(k)\right)=0,
\label{SecEq}
\ee
yielding for the fluctuating part of the spectral density (cf.~\cite{Sieber1997,Sieber1998})
\be
\rho^{fluc}(k;\tilde\beta)=-\frac{1}{\pi}\Im\frac{{\rm d}}{{\rm d}k}\ln\det\left[\II -\boldsymbol{\hat Q}(k)\right]
=\frac{1}{\pi}\Im\sum_{p=1}^\infty\frac{1}{p}\frac{{\rm d}}{{\rm d}k}\left[{\rm Tr}\boldsymbol{\hat Q}^p(k)\right],
\label{rhoNB}
\ee
where
\be
{\rm Tr}\boldsymbol{\hat Q}^p(k)=\oint_{\partial\Omega}ds_1\oint_{\partial\Omega}ds_2\cdots\oint_{\partial\Omega}ds_p2\cos\left(\sum_{r=1}^p\frac{\Delta\Phi(s_{r+1},s_r)}{2}\right)\prod_{r=1}^p Q\left[k;\boldsymbol{r}(s_r),\boldsymbol{r}(s_{r+1})\right],
\label{TraceNB}
\ee
with $s_{p+1}=s_1$, $s_0=s_p$. Using approximations for the Hankel functions valid for $k\to\infty$, 
\be
\mathcal{G}_0[k;\boldsymbol{r}(s),\boldsymbol{r}(s^\prime)]\simeq -\frac{i}{4}\sqrt{\frac{2}{\pi\rho}}e^{ik\rho-\frac{i}{4}\pi}~\label{Greenas},\,
\partial_{n^\prime}\mathcal{G}_0[k;\boldsymbol{r}(s),\boldsymbol{r}(s^\prime)]
\nonumber\simeq i\cos\left[\alpha(s^\prime)-\xi(s,s^\prime)\right]\cdot(-k)\cdot \mathcal{G}_0[k;\boldsymbol{r}(s),\boldsymbol{r}(s^\prime)],
\ee
~\refeq{TraceNB} is brought to the form
\be
{\rm Tr}\boldsymbol{\hat Q}^p(k)\simeq\label{Qn1}
(-2)^p\oint_{\partial\Omega}ds_1\cdots\oint_{\partial\Omega}ds_p
\prod_{r=1}^p\partial_{n_r}\mathcal{G}_0\left[k;\boldsymbol{r}(s_r),\boldsymbol{r}(s_{r+1})\right]
\left(\frac{i}{2}\right)^p\frac{\mathcal{\tilde P}_p}{\prod_{r=1}^p\cos\left[\alpha_r-\xi_{r+1,r}\right]},
\ee
with $\alpha_r=\alpha(s_r)$ and $\xi_{r+1,r}=\xi(s_{r+1},s_r)$ denoting the angles of the outward-pointing normal vector at $s_r$ and of the trajectory segment connecting the reflection points at $s_r$ and $s_{r+1}$ with respect to the $x$-axis, respectively, and  $\mathcal{\tilde P}_p$ is obtained from the product in~\refeq{TraceNB}, when replacing the Hankel functions by their asymptotic value for $k\rho\to\infty$, and extracting them~\cite{Dietz2019a}. The only difference between~\refeq{Qn1} and the corresponding expression for nonrelativistic QBs~\cite{Sieber1997} is the fraction $\frac{\mathcal{\tilde P}_p}{\prod_{r=1}^p(\cdots)}$ in the product over $r$. This term does not oscillate rapidly with $k$ and thus does not have any effect on the stationary-phase approximation which is applied to obtain a semiclassical approximation for the $p$ integrals, implying that these steps can be taken from~\cite{Sieber1997}. Consequently, as in the nonrelativistic case, in the semiclassical limit, POs of the classical dynamics of the CB of corresponding shape provide the nonvanishing contributions to ${\rm Tr}\boldsymbol{\hat Q}^p(k)$ and the leading-$k$ contribution to the derivative with respect to $k$ in~\refeq{rhoNB} comes from the phase factor resulting from the integrals over the normal derivatives of $\mathcal{G}_0$, yielding  
\be
\Im\frac{1}{p}\frac{{\rm d}}{{\rm d}k}\left[{\rm Tr}\boldsymbol{\hat Q}^p(k)\right]\label{Trace1}
\simeq\Re\sum_{\gamma_p}\mathcal{A}_{\gamma_p}e^{i\Theta_{\gamma_p}}
\frac{\mathcal{\tilde P}_p^\ast}{\prod_{r=1}^p(-2i)\cos[\alpha_r-\xi_{r+1,r}]},
\ee
where
\be
\mathcal{A}_{\gamma_p}=\frac{l^{(p)}_{\rm PO}}{r_{\rm PO}\sqrt{\vert{\rm Tr}M_{\rm PO}^{(p)}-2\vert}},\,
\Theta_{\gamma_p}=kl_{\rm PO}^{(p)}-\frac{\pi}{2}\mu_{\rm PO}^{(p)},\label{Phase_Ampl}
\ee
with $p$ denoting the number of reflections of the PO $\gamma_p$, $M_{\rm PO}^{(p)}$ the monodromy matrix, $l^{(p)}_{\rm PO}$ the length of the PO, $\mu_{\rm PO}^{(p)}$ the Maslov index and $r_{\rm PO}$ the number of repetitions of the primitive PO which forms the PO. These are the ingredients of Gutzwiller's trace formula for QBs, which is attained when the fraction $\frac{\mathcal{\tilde P}_p^\ast}{\prod_{r=1}^p(\cdots)}$ in~\refeq{Trace1} equals unity,
\be
\rho^{fluc}_{\rm QB}(k)=\frac{1}{\pi}\sum_{p=1}^\infty\sum_{\gamma_p}\mathcal{A}_{\gamma_p}\cos\left(\Theta_{\gamma_p}\right).\label{TraceQB}
\ee
The $^\ast$ in $\mathcal{\tilde P}_p^\ast$ indicates that the product $\mathcal{\tilde P}_p$ should be evaluated along the POs. These are subject to the condition for specular reflection~\cite{Berry1987},
\be
\xi_{r+1,r}-\alpha_r=\alpha_r-\xi_{r,r-1}+\pi =\pi -\chi_r,\, 0\leq\chi_r<\pi/2
\label{Refl}
\ee
with $\chi_r$ corresponding to the angle of the segment $\xi_{r,r-1}$, which connects the reflection points $s_{r-1}$ and $s_r$ of the PO, with respect to the normal vector at $s_r$. With these notations and
\be
\frac{\alpha_{r+1}+\alpha_r}{2}-\xi_{r+1,r}=(\chi_r-\chi_{r+1})/2-\pi/2,\, \alpha_{r+1}-\alpha_r=\pi-(\chi_r+\chi_{r+1}),\, \Phi_{\gamma_p}=\sum_{r=1}^p\chi_r.
\label{Delt}
\ee
the product reads
\begin{align}
&\frac{\mathcal{\tilde P}_p}{\prod_{r=1}^p\cos\left[\alpha_r-\xi_{r+1,r}\right]} =2\cos\left(\Phi_{\gamma_p}-p\frac{\pi}{2}\right)\label{M}\\
	&\times\prod_{r=1}^p\left(\frac{2}{i}\right)\left[-\frac{\cos\theta_\beta}{\cos\chi_r}+i\sin\theta_\beta\right]
\left[\cos\left(\frac{\chi_r+\chi_{r+1}}{2}\right)-i\sin\theta_\beta\sin\left(\frac{\chi_r+\chi_{r+1}}{2}\right)
+\cos\theta_\beta\sin\left(\frac{\chi_r-\chi_{r+1}}{2}\right)\right].\nonumber
\end{align}
With~\refeq{Refl}, $\Phi_{\gamma_p}$ can be further evaluated,
\be
\Phi_{\gamma_p}-p\frac{\pi}{2}=\sum_{j=1}^p[\xi_{j,j-1}-\alpha_j]-p\frac{\pi}{2}
=\frac{\xi_{1,0}-\xi_{p+1,p}}{2},
\label{Delt1}
\ee
where the last term, $\xi_{p+1,p}-\xi_{1,0}=N_p2\pi$, is the total phase accumulated after looping the PO and thus an integer multiple,  $N_p$, of $2\pi$~\cite{Berry1987}.
Separating the second line of~\refeq{M} into modulus $\mathcal{B}^{\tilde\beta}_{\gamma_p}$ and phase $e^{i\Gamma^{\tilde\beta}_{\gamma_p}}$ brings~\refeq{Trace1} to the form  
\be
\Im\frac{1}{p}\frac{{\rm d}}{{\rm d}k}\left[{\rm Tr}\boldsymbol{\hat Q}^p(k)\right]\label{Trace2}
=\Re\sum_{\gamma_p}2\cos\left(\Phi_{\gamma_p}-p\frac{\pi}{2}\right)\mathcal{B}^{\tilde\beta}_{\gamma_p}e^{i\Gamma^{\tilde\beta}_{\gamma_p}}\mathcal{A}_{\gamma_p}e^{i\Theta_{\gamma_p}},
\ee
where the sum is over clockwise and counterclockwise propagating POs. Due to \T invariance their contributions are equivalent for nonrelativistic QBs, whereas reversing the rotational direction of the PO corresponds for NBs to swapping the sign of $\chi_r$ in~\refeq{M} and to a change of sign of $\Phi_{\gamma_p}$, leading to a modification of $\mathcal{B}^{\tilde\beta}_{\gamma_p}e^{i\Gamma^{\tilde\beta}_{\gamma_p}}$. Consequently, the contributions of the clockwise and counterclockwise propagating orbits are not the same and thus NBs exhibit the chirality property. The real part of the summands in~\refeq{Trace2} becomes
\be
2\mathcal{P}^{\tilde\beta}_{\gamma_p}=\cos\left(\Phi_{\gamma_p}-p\frac{\pi}{2}\right)
        \left[\mathcal{B}^{\tilde\beta}_{\gamma_p}\cos\left(\Theta_{\gamma_p}+\Gamma^{\tilde\beta}_{\gamma_p}\right)
        +(-1)^p\mathcal{\tilde B}^{\tilde\beta}_{\gamma_p}\cos\left(\Theta_{\gamma_p}+\tilde\Gamma^{\tilde\beta}_{\gamma_p}\right)\right],\nonumber
\ee
where $\mathcal{\tilde B}^{\tilde\beta}_{\gamma_p}e^{i\tilde\Gamma^{\tilde\beta}_{\gamma_p}}$ is deduced from the last term in~\refeq{Trace1} by reversing the rotational direction, yielding for the trace formula of massive neutrino billiards
\be
\rho_{mass}^{fluc}(k;\tilde\beta)=\frac{1}{\pi}\sum_{p}\sum_{\gamma_{p}}\mathcal{A}_{\gamma_{p}}\mathcal{P}^{\tilde\beta}_{\gamma_p}.
\label{TraceNBf}
\ee
Note, that corresponding trace formulas for NBs with the shapes of billiards with an integrable dynamics are obtained from this expression, by inserting the amplitudes and phases entering that for the corresponding QB in~\refeq{TraceQB}.

It has been shown in~\cite{Dietz2020} that in the nonrelativistic limit the trace formula for a massive NB~\refeq{TraceNBf} turns into Gutzwiller's trace formula for the corresponding QB given in~\refeq{TraceQB}, whereas in the ultrarelativistic limit $\beta\to 0$ the trace formula turns into
\be
\rho^{fluc}_{\rm NB}(k)=\frac{1}{\pi}\sum_p\sum_{\gamma_p}(-1)^p\cos\Phi_{\gamma_p}\cos\left(p\frac{\pi}{2}\right)\label{rhoNB0} 
\mathcal{A}_{\gamma_p}\cos\Theta_{\gamma_p}.
\ee
Thus, for massless NBs POs with an odd number $p$ of reflections at the boundary do not contribute, because then $\cos\left(p\frac{\pi}{2}\right)=0$~\cite{Bolte1999,Wurm2011}. This feature originates from the chirality property and the additional spin degree of freedom. Yet, for nonzero mass all terms in~\refeq{M} are nonvanishing implying that also POs with an odd number of reflections contribute to the trace formula. 

Concerning convergence problems of the trace formula it should be noted that these arise due to the product over $\cos(\alpha_r-\xi_{r+1,r})=-\cos\chi_r$ in the denominator of~\refeq{Trace1}, which is cancelled out by terms in $\mathcal{\tilde P}_p^\ast$ in the ultrarelativistic and nonrelativistic limits. Thus, for these cases the convergence properties are similar to those of Gutzwiller's trace formula. However, in the intermediate region $0<\theta_\beta<\pi/2$ convergence problems might arise due to POs like, for example, whispering gallery orbits, which are nearly tangential to the boundary in some parts, i.e., for which $\alpha_r-\xi_{r+1,r}\approx \pi/2$.  

Starting from the BIEs~\refeq{Msym1} and the trace formula~\refeq{TraceNBf} in~\cite{Zhang2021,Zhang2023} symmetry-projected trace formulas have been derived for each symmetry class for massive NBs with a discrete rotational symmetry. These comprise  in addition to POs of the full system pseudo orbits, that are POs in the symmetry-projected fundamental domains. In the full systems they are not POs, however, their initial and final points are related via the symmetry operations of the underlying irreducible representation~\cite{Robbins1989,Joyner2012}. Furthermore, it has been shown in~\cite{Zhang2021,Zhang2023} that, similar to the POs, pseudo orbits of massless NBs with an odd number of reflections cancel each other. 

\begin{figure}[!h]
\centering
\begin{subfigure}[b]{0.45\textwidth}
        \centering
        \subcaption{} 
        \includegraphics[width=\textwidth]{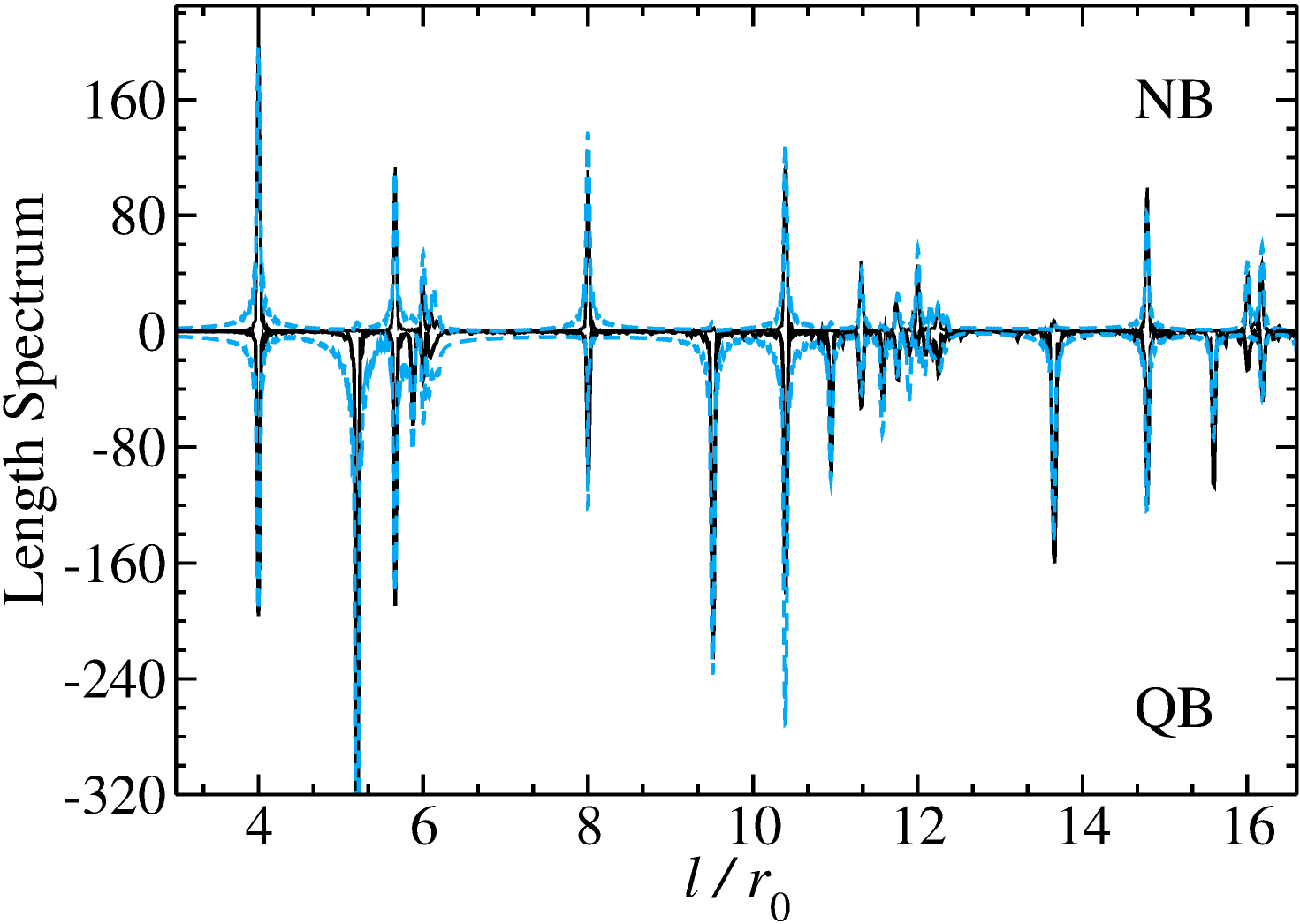}
    \end{subfigure}
\begin{subfigure}[b]{0.485\textwidth}
        \centering
        \subcaption{} 
        \includegraphics[width=\textwidth]{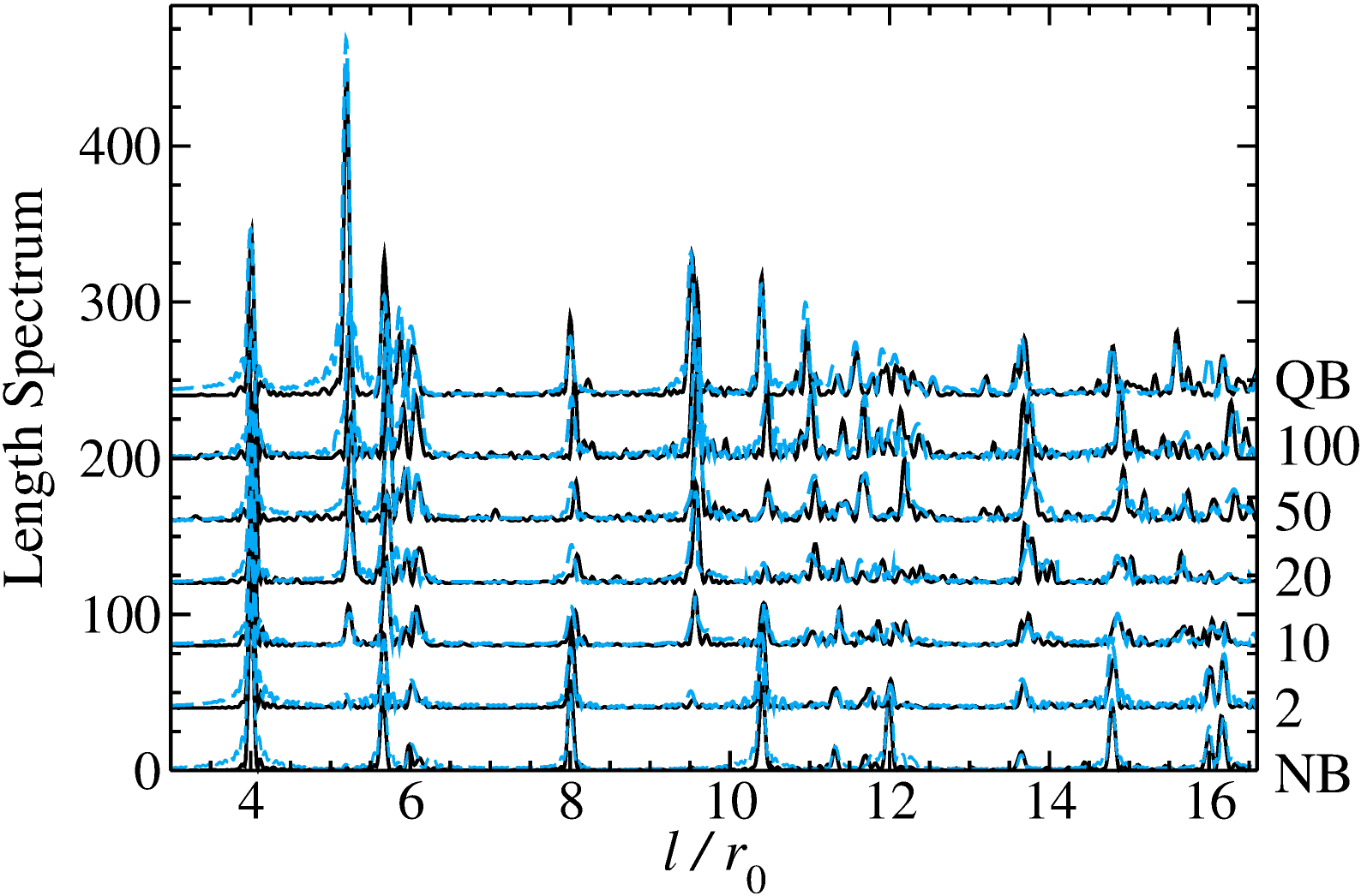}
    \end{subfigure}
\caption{
	(a) Length spectra of the massless semicircle NB (upper curves) and QB (lower curves) (black solid lines). They are compared to the length spectra deduced from the trace formulas for the massless NB and QB (blue dashed lines). (b) Length spectra of massive semicircle NBs for, from bottom to top, $\tilde\beta r_0 =0,2,10,20,50,100$ and the QB (black solid lines). They are compared to the length spectra deduced from the trace formula (blue dashed lines).
        }
\label{Fig_Semicircle}
\end{figure}

The direct link between the spectral density and classical POs is best visualized by length spectra, that is, the Fourier transform of the spectral density from wavenumber to length. Namely, length spectra of QBs and NBs exhibit peaks at the lengths of POs, where in those of massless NBs peaks at the lengths of POs with an odd number of reflections at the boundary are missing. This is illustrated in~\reffig{Fig_Semicircle} for the semicircle NB. In the left part the length spectrum of the massless NB (upper part) is compared to that of the QB (lower part). They exhibit peaks at the same lengths for POs with even number of bounces at the billiard boundary, whereas no peaks appear for the NB in cases with an odd number. The right part exhibits length spectra for the massive NB for various values of the mass $m_0$, that is, of $\tilde\beta r_0$. With increasing mass these missing peaks appear and for $\tilde\beta r_0\gtrsim 100$ the length spectrum becomes similar to that of the QB.

\subsection{Statistical measures for the fluctuation properties of the eigenstates\label{Measures}}
For single-particle quantum systems with a well-defined classical limit the field of quantum chaos entails the search for signatures of the classical dynamics in the spectral properties and the properties of the wave functions of the corresponding quantum system. According to the BT conjecture~\cite{Berry1977a}, the level statistics of generic integrable systems is Poissonian. The BGS conjecture~\cite{Bohigas1984} states that the fluctuations in the eigenvalue spectra of generic quantum systems, whose classical analogue is chaotic, are well described by those of Hermitian random matrices from one of the Wigner-Dyson ensembles associated with Dyson's threefold way~\cite{Dyson1962,Mehta2004}. For integer-spin systems these are the Gaussian orthogonal ensemble (GOE) if time-reversal invariance is preserved and the Gaussian unitary ensemble (GUE) if it is violated. A paradigm example for an untypical system with integrable dynamics is the harmonic oscillator~\cite{Berry1977a}. Examples of billiards with chaotic classical dynamics, that do not comply with BGS are billiards whose shape exhibits a discrete $M$-fold rotational symmetry~\cite{Robbins1989} (cf.~\refsec{Syms}) or billiards with a unidirectional classical dynamics~\cite{Knill1998,Gutkin2007}, cf.~\refsec{CW}.  

Neutrino billiards do not have a well-defined classical limit~\cite{Berry1987}, however, as demonstrated in~\refsec{Trace}, in the semiclassical limit they are connected to the POs of the CB of corresponding shape~\cite{Dietz2020,Yupei2022,Zhang2021}. Numerous numerical studies demonstrated that if the NB has the shapes of a chaotic CB, its spectral properties are well described by RMT, if this is the case for the corresponding QB. To be explicit, if the billiard has no geometrical symmetries, then they agree well with those of random matrices from the GUE, otherwise from the GOE. Yet, for billiards with integrable shape, discrepancies have been found.

For the study of spectral properties, and the comparison with those of other systems, the ordered eigenvalues $k_m$ with $k_1\leq k_2\leq\dots$ need to be unfolded to a uniform mean spacing unity. A commonly used procedure is to replace the eigenvalues by the smooth part of the integrated spectral density  $\mathcal{N}(k_m)$, $\epsilon_m=\mathcal{N}^{smooth}(k_m)$, which for QBs with Dirichlet BCs is given by Weyl's formula~\cite{Weyl1912}, 
\be
N^{Weyl}(k_m)=\frac{\mathcal{A}}{4\pi}k_m^2 -\frac{\mathcal{L}}{4\pi}k_m+C_0,
\ee
with $\mathcal{A}$ and $\mathcal{L}$ denoting the area and perimeter of the billiard, respectively. For massless NBs the perimeter contribution cancels out~\cite{Berry1987}. Accordingly, for QBs and NBs $\mathcal{N}^{smooth}(k_m)$ is determined by fitting a second-order polynomial to $\mathcal{N}(k_m)$~\cite{Sieber1995}. Comparison of the thereby obtained coefficients with $\mathcal{A}$ and $\mathcal{L}$ provides a check for missing levels. An even more sensitive test is the analysis of the fluctuating part of the integrated spectral density, $\mathcal{N}^{fluc}(k_m)=\mathcal{N}(k_m)-\mathcal{N}^{smooth}(k_m)$, which exhibits jumps at $k$ values corresponding to missing or spurious eigenvalues. 

Information on short-range correlations in the spectra are obtained from the distribution $P(s)$ of nearest-neighbor spacings $s_i=\epsilon_{i+1}-\epsilon_i$ and its cumulant, $I(s)=\int_0^sds^\prime P(s^\prime)$, which has the advantage that it does not depend on any binning. Another measure is the distribution of the ratios~\cite{Oganesyan2007,Atas2013} of consecutive spacings between next-nearest neighbors, $r_j=\frac{\epsilon_{j+1}-\epsilon_{j}}{\epsilon_{j}-\epsilon_{j-1}}$, which can take values between zero and infinity. Generally, therefore, instead of the $r_j$ the ratios $\tilde r_j=\min\left(r_j,\frac{1}{r_j}\right)$~\cite{Oganesyan2007} are considered, which take values between zero and unity. Ratios have the advantage that they are dimensionless so that, if the spectral density varies smoothly with $k_m$ no unfolding is needed. Information on long-range correlations is provided by the variance
\be
\Sigma^2(L)=\left\langle\left[\mathcal{N}(L)-\langle\mathcal{N}(L)\rangle\right]^2\right\rangle
\ee
of the number of unfolded eigenvalues in an interval of length $L$, and the rigidity~\cite{Mehta2004}
\be
\Delta_3(L)=\left\langle\min_{a,b}\int_{\epsilon -L/2}^{\epsilon +L/2}d\epsilon\left[N(\epsilon )-a-b\epsilon\right]^2\right\rangle ,
\ee
where $\langle\cdot\rangle$ denotes an average over an ensemble or spectral windows; correct unfolding gives $\langle\mathcal{N}(L)\rangle =L$ . Another commonly used characteristics is the spectral form factor~\cite{Mehta2004,Altland2025}, given by
\be
K(t)=\left\langle\left\lvert \frac{1}{N_{\rm max}}\small\sum_{n}^{N_{\rm max}}e^{2\pi i \epsilon_{n}t}\right\rvert^2\right\rangle.\label{eq:SFF}
\ee
Analytical results for these measures can be found for the Gaussian ensembles in~\cite{Mehta2004,Dietz1990,Atas2013}.

 An example for a typical billiard system with chaotic counterpart is the Africa billiard~\cite{Berry1986}, one which exhibits Poissonian statistics for a sufficiently large number of eigenvalues is the circle billiard. These are members of a family of billiards whose domain and boundary are generated by a conformal map of the type~\refeq{coordinate} with $f(r)=r$~\cite{Berry1986,Berry1987},
\be
\label{AFShape}
w^{AF}(z;\omega)=\frac{z+\omega z^2+\omega e^{i\pi/3}z^3}{\sqrt{1+5\omega^2}},\, r_0=1,\, \omega\lesssim 0.23.
\ee
The shape is circular for $\omega=0$ , whereas with increasing $\omega$ the classical dynamics changes from regular to mixed regular-chaotic, and then to chaotic. For $\omega=0.2$ the shape of the Africa billiard is attained, whose classical dynamics is fully chaotic. The spectral properties agree well with those predicted by the BGS after extracting scarred states~\cite{Nguyen2024,Dietz2025}. The angle $\omega(s)$ of the normal vector with respect to the $x$ axis changes nearly linearly with $s$ and the curvature is close to unity like in a circle billiard, except in the regions around the two bulges.
\begin{figure}[!h]
\centering
\begin{subfigure}[b]{0.45\textwidth}
        \centering
        \subcaption{} 
        \includegraphics[width=\textwidth]{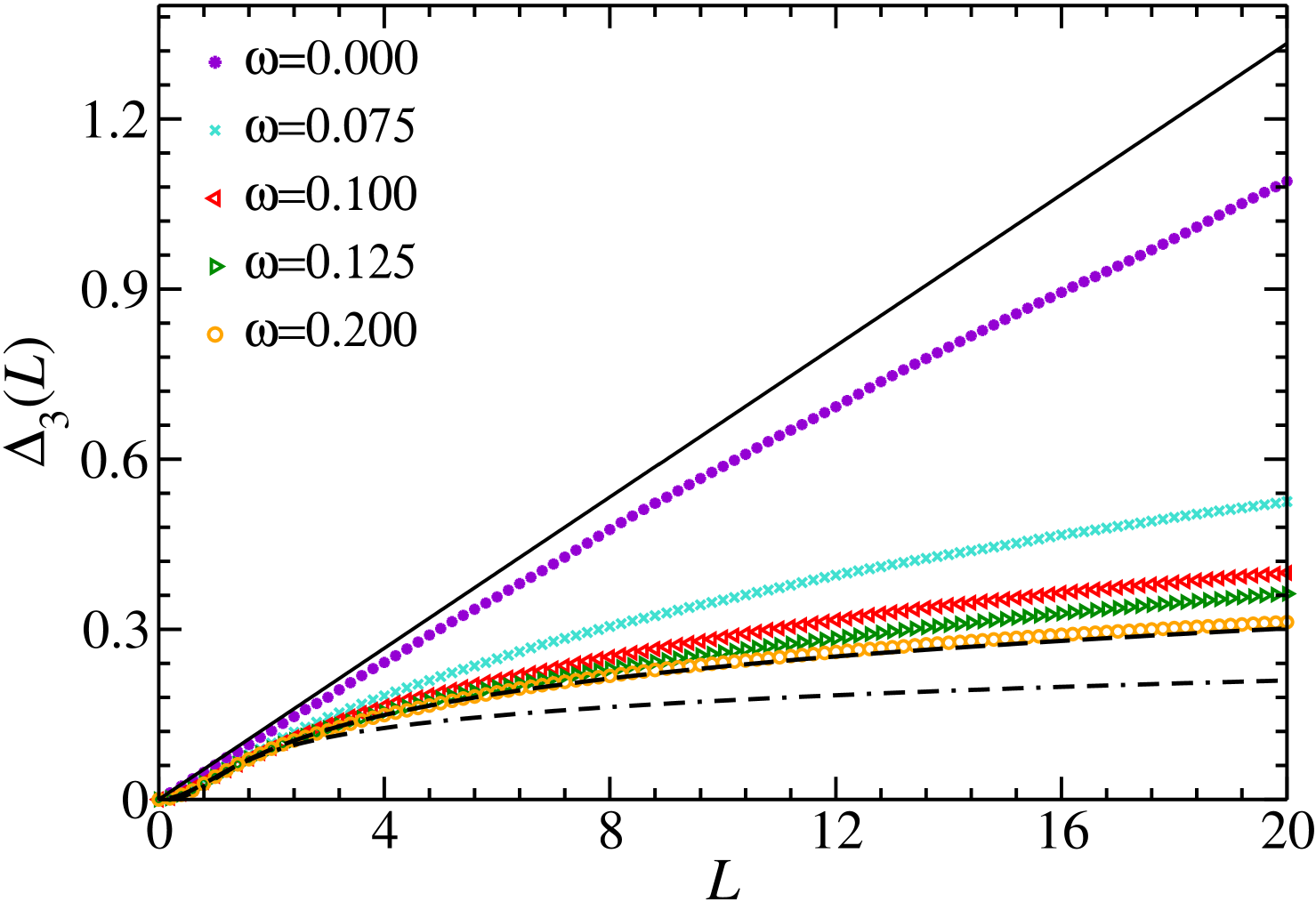}
    \end{subfigure}
\begin{subfigure}[b]{0.45\textwidth}
        \centering
        \subcaption{} 
        \includegraphics[width=\textwidth]{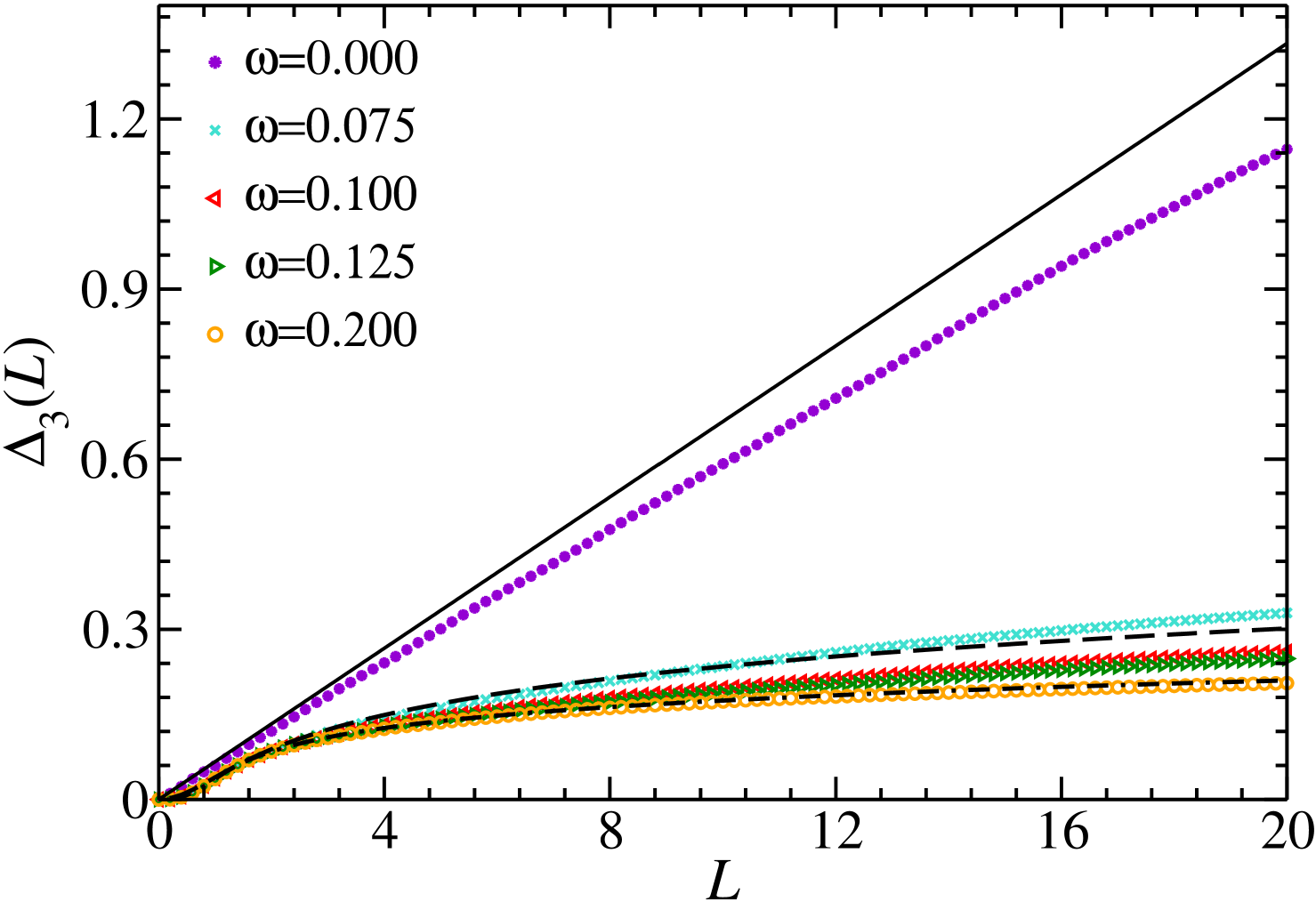}
    \end{subfigure}
	\caption{(a) Spectral rigidity $\Delta_3(L)$ for QBs with shapes~\refeq{AFShape} for various values of $\omega$. The curves for Poissonian, GOE and GUE statistics are plotted as solid, dashed and dash-dotted lines, respectively. (b) Same as panel (a) for the NBs.}
\label{Fig_Africa}
\end{figure}
In~\reffig{Fig_Africa} the spectral rigidity is depicted for QBs (a) and NBs (b) for various values of $\omega$. With increasing $\omega$ the rigidity and, more generally the spectral properties undergo a transition from Poisson to GOE statistics for the QBs and GUE statistics for the NBs.

The wave functions of typical quantum systems with a chaotic classical dynamics are uniformly spread over the whole available billiard area and phase space, and according to a Berry conjecture~\cite{Berry1977} well described by a superposition of random plane waves with fixed wavenumber $k=k_m$, but random directions $\theta_k$. On the contrary, those of a typical quantum system with integrable classical dynamics are well approximated by a superposition of plane waves with a finite number of well-defined directions. A suitable measure to determine the dominantly contributing directions, $\theta_k=\arctan(k_{m,y}/k_{m,x})$ of the plane waves that form an eigenmode~\cite{Bogomolny2021a} is the momentum distribution~\cite{Baecker1999}, that is the Fourier transforms of the wave functions from coordinate space $(x,y)$ to momentum space $(k_x,k_y)$. It is localized on the energy shell, that is, at $k=k_m=\sqrt{k_{m,x}^2+k_{m,y}^2}$, so that it is sufficient to compute the distribution for that value of $k$, 
        \be
        \label{Mom_Distr}
	\mathcal{M}_m(\theta_k)=\int_\Omega\Psi_m(x,y)e^{-ik_m\left(x\cos\theta_k+y\sin\theta_k\right)}dxdy.
        \nonumber\ee
The localization properties of the momentum distribution of wave functions that are localized in momentum space, e.g., on a quantum scar~\cite{McDonald1979,Heller1984}, or generally scarred by classical orbits, can be identified with help of the inverse of the participation number, also commonly referred to as inverse participation ratio~\cite{Wegner1980},
        \be
	\mathcal{I}_2=\frac{\sum_iz_i^2}{\left(\sum_iz_i\right)^2},\, z_i=\vert\mathcal{M}_m\left(\theta_{k,i}\right)\vert^2,\label{IPR}
        \ee
The participation number provides an estimate of the effective number of angular-direction bins $i$ carrying weights $z_i$. However, differences in the localization properties of the momentum distributions are clearer visible in its inverse. Since the momentum distribution is not necessarily normalized to unity, the denominator is included. Both for QBs and NBs, the larger the value of $\mathcal{I}_2$ is, the more localized is the corresponding eigenstate in momentum space, and the smaller is the number of plane waves needed to reconstruct the associated wave function or spinor component.  

 The momentum distribution provides information on dominant directions of the plane waves from which a wave function is constructed. Information on localization properties in phase space can be obtained from  Husimi functions~\cite{Baecker2004}. They are defined in classical phase space~\cite{Husimi1940,LesHouches1989} and are often referred to as quantum Poincar\'e surface of section for nonrelativistic QBs. For NBs Husimi functions provide similar information~\cite{Yupei2022,Dietz2022a,Dietz2023} and thus have been employed to obtain information on their semiclassical limit. Because the classical dynamics of a billiard is determined by its shape, and also the eigenstates of the corresponding QB or NB are fully determined by the boundary functions (cf.~\refsec{BIEs}), it is sufficient to restrict phase space to a Poincar\'e surface of section defined on $\partial\Omega$. Following Ref.~\cite{Baecker2004}, Husimi functions are defined in terms of Poincar\'e-Birkhoff coordinates. These are the arclength parameter $s$ and $p=\sin\chi(s)$ at the point of impact, where $\chi(s)$ is defined in~\refeq{Refl}. For QBs they are obtained from the projection of the normal derivative of the boundary wave function onto a coherent state~\cite{Baecker2004} which is localized at $\partial\Omega$ and periodic with period $\mathcal{L}$,
\ba
&&{\rm H}_m(s,p)=\label{Hus}
\frac{1}{2\pi k_m}\frac{1}{\int_0^\mathcal{L} ds^\prime\left\vert\partial_{n^\prime}\psi_m(s^\prime)\right\vert^2}
\left\vert\int_0^\mathcal{L} ds^\prime\partial_{n^\prime}\psi_m(s^\prime) C^\delta_{(s^\prime,p)}(s^\prime;k_m)\right\vert^2,\\
&&C^\delta_{(s,p)}(s^\prime;k_m)=\nonumber
\left(\frac{k_m}{\pi\delta^2}\right)^{1/4}\sum_{\tilde m=-\infty}^\infty\exp\left(ipk_m\left(s^\prime-s+\tilde m\mathcal{L}\right)-\frac{k_m}{2\delta^2}\left(s^\prime-s+\tilde m\mathcal{L}\right)^2\right),
\ea
with $n^\prime=n(s^\prime)$ the normal vector at $s^\prime$ and $\delta$ is the width of the Gaussian. For NBs $\psi_m(s)$ is replaced by the first or the second eigenspinor component. The Husimi functions of NBs are asymmetric with respect to the $p=0$ axis, the origin being the BC~\refeq{BC2}, which induces a unidirectional current along the boundary and thus differs for clockwise and counterclockwise modes. 
\section{Neutrino billiards with shapes of integrable billiards\label{Int}} 
\subsection{Circle NB\label{Circle}}
Circular billiards belong to the U(1) symmetry class, which comprises all $M$-fold rotational symmetries with $M\geq 2$, implying that the angular momentum is conserved and the classical dynamics is integrable. The domain and boundary are given in polar coordinates, $0\leq r\leq r_0$ and $0\leq\gamma\leq 2\pi$, $w(z)=z$ with $z=re^{i\gamma}$. For the QB the eigenfunctions read 
\be
\psi_{m,\nu}(r,\gamma)=J_m(k_{m,\nu}r)g_m(\gamma),\label{efcirc}
\ee
where $m=0,1,2,\cdots$ and $g_m(\gamma)=\cos(m\gamma)$ for solutions which are symmetric with respect to the $x$ axis and $g_m(\gamma)=\sin(m\gamma)$ for $m\ne 0$ for the antisymmetric ones. The eigenwavenumbers $k_{m,\nu}$ are the same for both symmetry classes and obtained from the quantization condition,
\be
J_m(k_{m,\nu}r_0)=0,\label{ewcirc}
\ee
where $\nu$ counts the zeroes for a given quantum number $m$. The eigenvalues correspondng to $m$ and $-m$ fulfill this equation for the same value of $k$ and thus are degenerate. Restricting to one mirror symmmetry class, e.g. to $m\geq 0$, the spectral properties coincide with those of uncorrelated Poissonian random numbers, as predicted for generic quantum systems with an integrable classical dynamics. For NBs with circular shape the Dirac equation~\refeq{Hamiltonian} with the BC~\refeq{BCm} can also be solved analytically~\cite{Berry1987}, yielding in the ultrarelativistic, massless case for the spinor components
\be
\label{WF_Circle}
\psi_{1,m}(r,\gamma)\propto i^mJ_m(k_{m,\nu}r)e^{im\gamma},\, \psi_{2,m}(r,\gamma)\propto i^{m+1}J_{m+1}(k_{m,\nu}r)e^{i(m+1)\gamma},
\ee 
and for the eigenwavenumbers the quantization condition~\cite{Berry1987}
\be
J_{m+1}(k_{m,\nu}r_0)=J_m(k_{m,\nu}r_0),\label{QCC}
\ee
implying that the eigenvalues for $m$ and $-m-1$ corresponding to equal and opposite angular momenta are not degenerate.

In the following we restrict to the first spinor component. Solutions for the second one can be deduced from the Dirac equation~\refeq{DE} and the BCs~\refeq{BCm}. For the massive case the BIE~\refeq{Eq4} and the corresponding one for the second spinor component can be solved analytically~\cite{Dietz2022a}, which with the notation $\tilde\gamma =\gamma-\gamma^\prime$ reads
\be
\left(1-\sin\theta_\beta\right)\psi_1^\ast(\gamma^\prime)=\frac{i}{4}k\fint_0^{2\pi} d\gamma\label{BIEcircle}\left\{\cos\theta_\beta\left[e^{-i\tilde\gamma}-1\right]H_0^{(1)}(k\rho)+2\sin\theta_\beta\frac{e^{-i\tilde\gamma}-1}{\vert e^{-i\tilde\gamma}-1\vert}H_1^{(1)}(k\rho)\right\}\psi_1^\ast(\gamma).
\ee
Choosing for the spinor components the ansatz~\refeq{Psi1}, the BIE turns into
\begin{align}
&\left(1-\sin\theta_\beta\right)\sum_ma_mi^mJ_m(kr_0)e^{-im\gamma^\prime}=\sum_ma_mi^mJ_m(kr_0)e^{-im\gamma^\prime}\label{circle}\\
\times &\frac{ik}{4}\fint_{-\pi}^{\pi} d\tilde\gamma
e^{-im\tilde\gamma}\left\{\cos\theta_\beta\left[e^{-i\tilde\gamma}-1\right]H_0^{(1)}(k\rho)+2\sin\theta_\beta\frac{e^{-i\tilde\gamma}-1}{\rho}H_1^{(1)}(k\rho)\right\}.\nonumber
\end{align}
Employing the addition theorems for the Hankel functions,
\be
\frac{e^{-i\tilde\gamma}-1}{\rho}H_1^{(1)}(k\rho)=-\sum_{m=-\infty}^{\infty}H_{m+1}(kr_0)J_m(kr_0)e^{im\tilde\gamma},\,
H_0^{(1)}(k\rho)=\sum_{m=-\infty}^{\infty}H_{m}(kr_0)J_m(kr_0)e^{im\tilde\gamma}.
\ee
leads to the equation~\cite{Dietz2022a}
\begin{align}
&\sum_ma_mi^mJ_m(kr_0)e^{-im\gamma^\prime}=\sum_ma_mi^mJ_m(kr_0)e^{-im\gamma^\prime}\label{circle1}\\
\times&\left\{\frac{ik\pi}{2}\left[\cos\theta_\beta\left(H_{m+1}(kr_0)J_{m+1}(kr_0)-H_m(kr_0)J_m(kr_0)\right)-2\sin\theta_\beta H_{m+1}(kr_0)J_m(kr_0)\right]+\sin\theta_\beta\right\}.
\end{align}
Multiplication of both sides of the equation with $e^{im\gamma^\prime}$ and integration over $\gamma^\prime$ and separating the Hankel functions into their real and imaginary parts [see below~\refeq{Greeni}], leads to the quantization condition
\begin{align}
1-\sin\theta_\beta\label{Qeq}
&=\frac{i\pi k}{2}\left\{\cos\theta_\beta\left[J_{m+1}^2(kr_0)-J_m^2(kr_0)\right]-2\sin\theta_\beta J_{m+1}(kr_0)J_m(kr_0)\right\}\\
&-\frac{\pi k}{2}\left\{\cos\theta_\beta\left[Y_{m+1}(kr_0)J_{m+1}(kr_0)-Y_m(kr_0)J_m(kr_0)\right]-2\sin\theta_\beta Y_{m+1}(kr_0)J_m(kr_0)\right\}.
\end{align}
The imaginary part on the right hand side of the equation has to vanish. The corresponding solutions are easily found after dividing it by $J_{m+1}^2(r_0)$ and defining the variable $x=J_{m}^2(r_0)/J_{m+1}^2(r_0)$. Then, the solutions are given by the zeros of a quadratic polynomial in $x$. These are inserted into the equation for the real part and the Wronskian $W\left\{J_m(kr_0),Y_m(kr_0)\right\}=\frac{2}{\pi kr_0}$ is used to solve it. The resulting conditional equations for the imaginary and real parts of~\refeq{Qeq} are given by 
\be
J_{m+1}(kr_0)=\frac{\sin\theta_\beta\pm 1}{\cos\theta_\beta}J_m(kr_0),\,
1-\sin\theta_\beta=\pm\left(1\mp\sin\theta_\beta\right).
\ee
The equation for the real part is fulfilled for the upper sign, leading to the quantization equation for massive NBs,
\be
J_{m+1}(kr_0)=\mathcal{K}^{-1}J_m(kr_0)\label{QCCm},
\ee
which can be obtained directly from~\refeq{Hamiltonian} with the BC~\refeq{BCm} and~\refeq{QCC}, thus confirming the results of Ref.~\cite{Dietz2020}. The spectral properties of the circle QB, and the massless and massive circle NB coincide with those of uncorrelated Poissonian random numbers. The same is the case for the eigenvalues of their symmetry-projected eigenstates.

The trace formula introduced in~\refsec{Trace} has been validated for massless~\cite{Dietz2019} and massive circle NBs~\cite{Dietz2020}. For circle billiards the reflection angles $\chi_r$ of the POs are all equivalent. For a PO consisting of $p$ reflections at the boundary and looping the circle center with winding number $m_\gamma$ it equals 
\be
\chi_r=\left(\frac{{\rm sgn}(m_\gamma)}{2}-\frac{m_\gamma}{p}\right)\pi,\, r=1,\dots,p\label{ChiCirc}, 
\ee
where ${\rm sgn}(m_\gamma)=\pm 1$ for clockwise and counterclockwise POs, respectively, yielding for the contributions~\refeq{Trace2} to the trace formula,
\be
\mathcal{B}^{\tilde\beta}_{\gamma_p}e^{i\Gamma^{\tilde\beta}_{\gamma_p}} =\label{Amplitude}
\left[-\cos\theta_\beta +i\sin\theta_\beta\sin\left(\pi\frac{\vert m_\gamma\vert}{p}\right)\right]^p
\left[1\pm i\sin\theta_\beta\cot\left(\pi\frac{\vert m_\gamma\vert}{p}\right)\right]^p,
\cos\left(\Phi_{\gamma_p}-p\frac{\pi}{2}\right)=\cos\left(\left[m_\gamma-\frac{p}{2}\pm\frac{p}{2}\right]\pi\right).
\ee
and
\be
\mathcal{A}_{\gamma_p}e^{i\Theta_{\gamma_p}}=\sqrt{\frac{k}{\pi}}\left(2-\delta_{p,2m_\gamma}\right)\label{C3}
\frac{\left(\sin\frac{m_{\gamma}}{p}\pi\right)^{3/2}}{\sqrt{p}}
e^{i\left[2kr_0p\sin\left(\frac{m_{\gamma}}{p}\pi\right)-\frac{3\pi}{2}p+\frac{\pi}{4}\right]}.
\ee
Due to the presence of the $\sin\theta_\beta\cot\left(\pi\frac{\vert m_\gamma\vert}{p}\right)$ term in $\mathcal{B}^{\tilde\beta}_{\gamma_p}e^{i\Gamma^{\tilde\beta}_{\gamma_p}}$ in~\refeq{Amplitude}, its contribution to the trace formula becomes large for nonzero, finite mass and $\frac{m_\gamma}{p}\simeq 0$. This is the case for whispering gallery modes with lengths close to $2\pi r_0$ after one loop of the PO, entailing convergence problems in the numerical analysis beyond a certain value of $p$ in~\refeq{ChiCirc}. The value depends on that of $\theta_\beta$ in situations, where such terms do not interfere destructively with other POs, or the corresponding clockwise and counterclockwise POs do not cancel each other. Convergence can, e.g., be controlled by disregarding the contributions of POs with reflection angles above a certain value, or with more sophisticated approaches~\cite{Keating2007}.

\subsection{Ellipse NB\label{Ellipse}}
The ellipse billiard has a twofold symmetry and mirror symmetries with respect to the $x$ and $y$ axes and a second constant of motion, which is the product of the angular momenta $L_1$ and $L_2$ with respect to the two focal points~\cite{Berry1981}. The domain of an ellipse with boundary at $\mu = \mu_0$, semiminor axis $b=\sinh\mu_0$, semimajor axis $a=\cosh\mu_0$ and eccentricity $\epsilon =\frac{1}{\cosh\mu_0}$ are defined in terms of elliptical coordinates, 
\be
w(z)=\frac{1}{2}\left(e^\mu e^{i\gamma}+e^{-\mu}e^{-i\gamma}\right),\, zw^\prime(z)=\frac{1}{2}\left(e^\mu e^{i\gamma}-e^{-\mu}e^{-i\gamma}\right),\, \mu\leq\mu_0,\, z=e^\mu e^{i\gamma}.
\ee
The orbits of the ellipse CB and, accordingly, the eigenmodes of the ellipse QB, are either of librational or of rotational type. Rotational orbits surround both focal points, whereas librational ones bounce back and forth at the billiard boundary between the focal points. With decreasing eccentricity the ellipse is transformed into a circle and the librational modes turn into the diameter orbit, whereas with increasing eccentricity the modes resemble those reflected back and forth at the longer sides in a rectangular billiard.

The Schr\"odinger equation is separable into equations for $\gamma$ and $\mu$, which, in distinction to that for the circle QB, are linked by a common $k$ dependent quantity $h=h(k)$,
\ba
\frac{\partial^2}{\partial\gamma^2}\Phi(\gamma)+\left(h(k)-k^2\cos^2\gamma\right)\Phi(\gamma)&=&0\\
-\frac{\partial^2}{\partial\mu^2}M(\mu)+\left(h(k)-k^2\cosh^2\mu\right)M(\mu)&=&0.
\ea
The eigenstates of the ellipse QB are given by products of the radial and the periodic Mathieu functions~\cite{McLachlan1947,Dietz1993,Waalkens1997},
\be
\psi(k_{m,\nu};\mu,\phi)\propto M_m(k_{m,\nu};\mu)\Phi_m(k;\phi), \label{efell}
\ee
where the eigenvalues are obtained from the quantization condition
\be
M_m(k_{m,\nu}; \mu_0)=0.\label{ewell}
\ee
The periodic Mathieu functions $\Phi_m(\phi)$ belong to four symmetry classes according to the reflection symmetry of $\psi(k_{m,\nu};\mu,\phi)$ with respect to the major and minor axes of the elliptical billiard. The spectral properties of the ellipse QB again coincide with those of uncorrelated Poissonian random numbers, also those of the fully desymmetrized elliptical QB which corresponds to a quarter-ellipse QB with Dirichlet boundary conditions, and the semi-ellipse QBs obtained by cutting the full ellipse along either of its mirror-symmetry axes. Note, that these are the only ellipse-sector QBs with integrable classical dynamics~\cite{Sieber1997}, whereas all other sectors exhibit a mixed one.

The Dirac equation reads in elliptic coordinates
\be
k\boldsymbol{\psi}\label{Dellipse}=
\begin{pmatrix}
0 &\frac{-i}{z\left[w^\prime (z)\right]}\left(\frac{\partial}{\partial\mu} -i\frac{\partial}{\partial\gamma}\right)\\
\frac{-i}{z^\ast\left[w^\prime (z)\right]^\ast}\left(\frac{\partial}{\partial\mu} +i\frac{\partial}{\partial\gamma}\right) &0
\end{pmatrix}
\boldsymbol{\psi}.
\ee
and its solutions for the corresponding ellipse NB can be written in the form~\refeq{Psi1} with $\vert w(z)\vert=\sqrt{\cosh^2\mu-\sin^2\gamma}$. In Ref.~\cite{Dietz2019} an equation for the determination of the eigenstates is derived in terms of Mathieu functions. For this the Bessel functions entering these equations are expanded in sums of products of the radial and periodic Mathieu functions~\cite{McLachlan1947} and the resulting expansions are separated into the symmetry classes associated with the twofold  symmetry  of the eigenstates of the ellipse NB. Furthermore, trace formulas are derived in~\cite{Dietz2019} for the full, semi and quarter ellipse for the ultrarelativistic case based on those for the corresponding ellipse QB~\cite{Sieber1997}, which are quite cumbersome. These have been generalized to massive ellipse NBs in~\cite{Yupei2022} by proceeding as for the circle NB~\cite{Dietz2020}. The spectral properties of the ellipse QB and NB agree well with those of uncorrelated Poissonian random numbers, as expected for generic quantum systems with integrable classical dynamics.

\subsection{Equilateral-triangle NB\label{ET}}
The equilateral-triangle NB is another example for a NB with integrable classical dynamics and Poissonian spectral statistics, whose eigenstates have been derived analytically in~\cite{Dietz2021}. Equilateral triangles have a threefold rotational symmetry and mirror symmetries with respect to the three main axes, that is, $C_{3v}$ symmetry. Thus the equilateral-triangle CB can be divided into six fundamental domains~\cite{Lame1852}i, while the eigenstates of the QB are classified by the irreducible representations of $C_{3v}$. Those of the corresponding NB can be assigned to three symmetry classes according to their transformation properties under rotation by $\frac{2\pi}{3}$. The side lengths of the triangle are set to unity, the origin of the coordinate system is at its center and the boundary of the first fundamental subdomain is chosen parallel to the $y$ axis. The boundaries of the second and third subdomains, $\boldsymbol{r}_{0,j}$, are generated by rotating $\boldsymbol{r}_0$ by $j\frac{2\pi}{3},\, j=1,2$,
\be
\boldsymbol{r}_0=
\begin{pmatrix}
\frac{1}{2\sqrt{3}} \\ \frac{y}{2}
\end{pmatrix},
\, \boldsymbol{r}_{0,j}=\mathcal{R}_3^j\boldsymbol{r}_0,
\ee
with $y\in [-1,1]$ and $\mathcal{R}_3$ is defined in~\refeq{Rn}. In the complex plane the vectors $\boldsymbol{r}_0$ and $\boldsymbol{r}_{0,j}$ correspond to $w_0(s)=\frac{1}{2\sqrt{3}}+i\frac{y}{2}$ and $w_j(s)=e^{ij\frac{2\pi}{3}}w_0(s)$, respectively. The notations
\be
\boldsymbol{k}_j=\left(\mathcal{R}_3^\dagger\right)^j\boldsymbol{k}_0
=k\begin{pmatrix}
\cos\left(\theta_k-j\frac{2\pi}{3}\right) \\ \sin\left(\theta_k-j\frac{2\pi}{3}\right)
\end{pmatrix}\label{kj},\,
\boldsymbol{\kappa}_j=\left(\mathcal{R}_3^\dagger\right)^j\boldsymbol{k}^\ast_0
=k\begin{pmatrix}
\,\,\,\, \cos\left(\theta_k+j\frac{2\pi}{3}\right) \\ -\sin\left(\theta_k+j\frac{2\pi}{3}\right)
\end{pmatrix}
\ee
are introduced, where $\boldsymbol{\kappa}_1=\boldsymbol{k}^\ast_2$ and $\boldsymbol{\kappa}_2=\boldsymbol{k}^\ast_1$. Let $\boldsymbol{r}$ be a vector in the first fundamental domain defined by $\boldsymbol{r}_0$ and $\boldsymbol{r}_j=\mathcal{R}_3^j\boldsymbol{r},\, j=1,2$ vectors in the other two fundamental domains,
\be
\boldsymbol{k}\cdot\boldsymbol{r}_j=\boldsymbol{k}\cdot\mathcal{R}_3^j\boldsymbol{r}=\left(\mathcal{R}_3^\dagger\right)^j\boldsymbol{k}\cdot\boldsymbol{r}=\boldsymbol{k}_j\cdot\boldsymbol{r}.
\ee
An ansatz for the first component of $\boldsymbol{\psi}(\boldsymbol{r})$ with symmetry class $l$ is given by
\be
\psi_1^{(l)}(\boldsymbol{r})=a(k)\sum_{j=0}^2e^{ilj\frac{2\pi}{3}}\left[e^{i\boldsymbol{k}_j\boldsymbol{r}}+c(k)e^{-i\boldsymbol{\kappa}_j\boldsymbol{r}}\right],\label{psi1}
\ee
 yielding for the second one with~\refeq{Hamiltonian}
\be
\psi_2^{(l)}(\boldsymbol{r})=ik\left[\frac{\partial}{\partial x} +i\frac{\partial}{\partial y}\right]\psi_1^{(l)}(\boldsymbol{r})
\label{psi2}
=a(k)\sum_{j=0}^2e^{i(l-1)j\frac{2\pi}{3}}\left[e^{i\theta_k}e^{i\boldsymbol{k}_j\boldsymbol{r}}-c(k)e^{-i\theta_k}e^{-i\boldsymbol{\kappa}_j\boldsymbol{r}}\right],\nonumber
\ee
and when imposing the BC~\refeq{BCm} the quantization condition
\be
\sum_{j=0}^2e^{ilj\frac{2\pi}{3}}e^{i\boldsymbol{k}_j\boldsymbol{r}_0}\left[1+\mathcal{K}e^{i\left(\theta_k-j\frac{2\pi}{3}+\frac{\pi}{2}\right)}\right]\label{BCTr}
=-c(k)\sum_{j=0}^2e^{ilj\frac{2\pi}{3}}e^{-i\boldsymbol{\kappa}_j\boldsymbol{r}_0}\left[1+\mathcal{K}e^{-i\left(\theta_k+j\frac{2\pi}{3}+\frac{\pi}{2}\right)}\right].
\ee
By construction of the plane-wave ansatz the BC is fulfilled at all sides of the triangle if it is at $\boldsymbol{r}_0$ and~\refeq{BCTr} holds for all values of $y\in [-1,1]$ if
\be
c(k)=-e^{i2k_{x,0}\rho_0}e^{i2\Phi_0}\label{Eq0},\,
e^{ik_{x,1}\rho_0}e^{-il\frac{\pi}{3}}e^{i\Phi_1}
=e^{ik_{x,0}\rho_0}e^{i\Phi_0}e^{-in_1\pi},\,
e^{ik_{x,2}\rho_0}e^{il\frac{\pi}{3}}e^{i\Phi_2}
=e^{ik_{x,0}\rho_0}e^{i\Phi_0}e^{-in_2\pi},
\ee
with $\rho_0=\frac{1}{2\sqrt{3}}$. Furthermore,
\be
e^{i\Phi_j}=\sqrt{\frac{\left[1+\mathcal{K}e^{i\left(\theta_k-j\frac{2\pi}{3}+\frac{\pi}{2}\right)}\right]}{\left[1+\mathcal{K}e^{-i\left(\theta_k-j\frac{2\pi}{3}+\frac{\pi}{2}\right)}\right]}}
=\frac{k+i\mathcal{K}\left[k_{x,j}+ik_{y,j}\right]}{\sqrt{k^2\left(1+\mathcal{K}^2\right)-2k\mathcal{K}k_{y,j}}},
\ee
for $j=0,1,2$, where with Eq.~(\ref{Eq0}) the components of the rotated wave vectors $\boldsymbol{k}_j$ are related as
\be
k_{x,1}+k_{x,2}=-k_{x,0}=-k_x,\,\left(k_{x,1}-k_{x,2}\right)\rho_0=\frac{k_{y,0}}{2}=\frac{k_y}{2}.
\ee
These conditional equations lead to quantization conditions for $k_x$ and $k_y$ in terms of transcendental equations,
\ba
&&\tan\left(\frac{\sqrt{3}}{2}k_{x}-\tilde n\pi\right)\label{Quant1}
=3\frac{\mathcal{K}\left(\mathcal{K}^2-1\right)\cos\theta_k}{1+\mathcal{K}^4-4\mathcal{K}^2-\mathcal{K}\left(1+\mathcal{K}^2\right)\sin\theta_k+4\mathcal{K}^2\sin^2\theta_k}\\
&&\tan\left(\frac{k_{y}}{2}-l\frac{2\pi}{3}-\tilde m\pi\right)
=\nonumber\sqrt{3}\frac{\frac{\mathcal{K}^2}{2}+\mathcal{K}\sin\theta_k
}{1-\frac{\mathcal{K}^2}{2}+\mathcal{K}\sin\theta_k},
\ea
with $\tilde n=n_1+n_2$ and $\tilde m=n_2-n_1$. Because $\Phi_a(\theta_k)=-\Phi_a(\pi-\theta_k)$ and $\Phi_b(\theta_k)=\Phi_b(\pi-\theta_k)$,  $(-k_x,k_y)$ is a solution of~\refeq{Quant1} if $(k_x,k_y)$ is one.

In the nonrelativistic limit $\tilde\beta\to\infty$ the phases approach zero, $\Phi_{a,b}\to 0$, implying that for $l=0$ all possible combinations of $\pm k_x$ and $\pm k_y$ are solutions if $(k_x,k_y)$ is one, that the spinor eigenfunction is real for $l=0$, and that
\be
\label{Quantization}\frac{\sqrt{3}}{2\pi}k_{x}=(n_1+n_2),\, \frac{3}{2\pi}k_{y}=2l+3(n_2-n_1),
\ee
where, for $l=0$ the cases $n_1=0$, $n_2=0$, $2n_1=n_2$ and $2n_2=n_1$ need to be excluded, and similarly for $l=1$ the cases $2n_1=n_2-1$ and $2n_2=n_1+1$ and for $l=2$ the cases $2n_1=n_2+2$ and $2n_2=n_1-2$, because the corresponding wave functions vanish identically. In the nonrelativistic limit the eigenvalues coincide with those of the QB~\cite{Richens1981}.

In the ultrarelativistic limit $m_0=0$ and $\mathcal{K}=1$, implying that $\Phi_j=\theta_k+\frac{\pi}{2}-j\frac{2\pi}{3},\, j=0,1,2$ and the quantization conditions become
\be
\label{Quantization1}\frac{\sqrt{3}}{2\pi}k_{x}=(n_1+n_2),\, \frac{3}{2\pi}k_{y}=2(l+1)+3(n_2-n_1).
\ee
Here, for the same reasons as above for $l=0$ the case $2n_1=n_2+1$ and for $l=1$ the case $2n_2=n_1+2$ need to be excluded. Thus, the eigenvalues of the massless NB with symmetry class $l$ coincide with those of the QB with symmetry class $\tilde l=l+1$, where $\tilde l=3$ corresponds to $\tilde l=0$. Accordingly, since the eigenvalues of the QB with $l=1,2$ are degenerate, in the ultrarelativistic limit those of the NB corresponding to $l=0,1$ coincide, and the fluctuation properties in the eigenvalue spectra of the QB corresponding to $l=1,2$ and $l=0$ coincide with those of the massless NB with $l=0,1$, respectively, $l=2$. In the nonrelativistic and ultrarelativistic limit the nearest-neighbor spacing and ratio distribution exhibit nongeneric gaps, whereas the long-range correlations are close to semi-Poisson.

\subsection{Sectors of NBs with a discrete rotational symmetry\label{Sectors}}
\begin{figure}[!h]
\centering
 \begin{tabular}{@{}c@{}}
        \begin{tabular}{@{}c@{}}
	\large (a)
        \includegraphics[width=.49\linewidth]{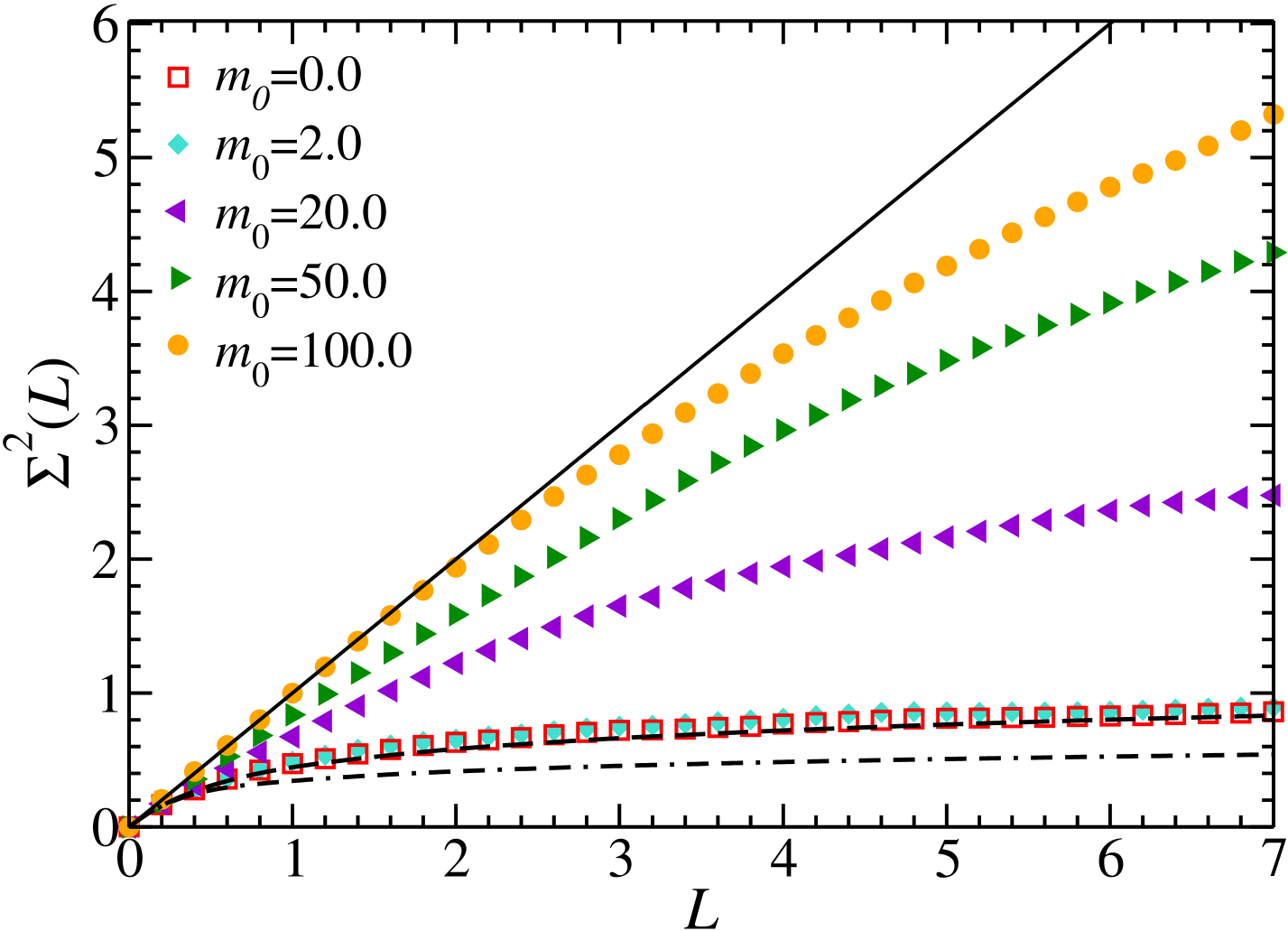}
        \end{tabular}
        \end{tabular}
        \begin{tabular}{@{}c@{}}
        \begin{tabular}{@{}c@{}}
	\large (b)
        \includegraphics[width=.2\linewidth]{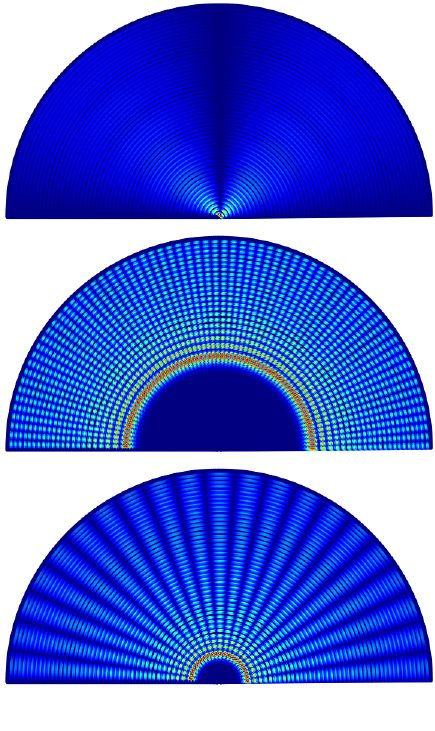}
	\includegraphics[width=.245\linewidth]{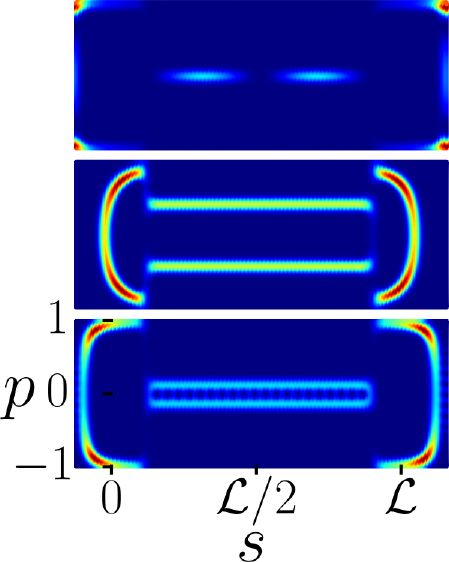}
        \end{tabular}
        \begin{tabular}{@{}c@{}}
	\large (c)
        \includegraphics[width=.2\linewidth]{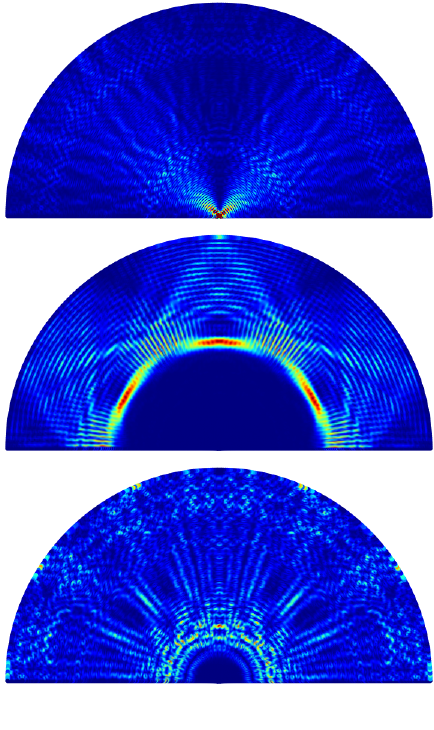}
        \includegraphics[width=.245\linewidth]{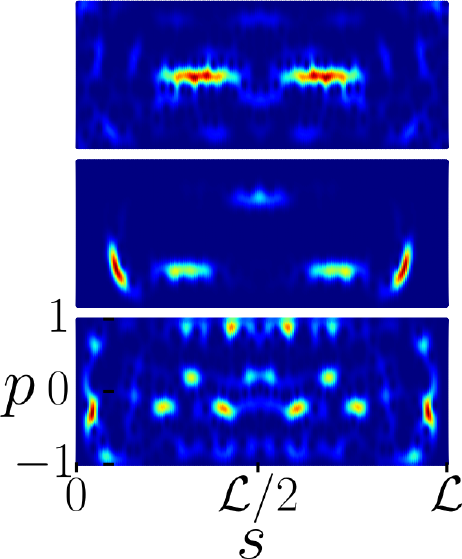}
        \end{tabular}
        \begin{tabular}{@{}c@{}}
        \includegraphics[width=.05\linewidth]{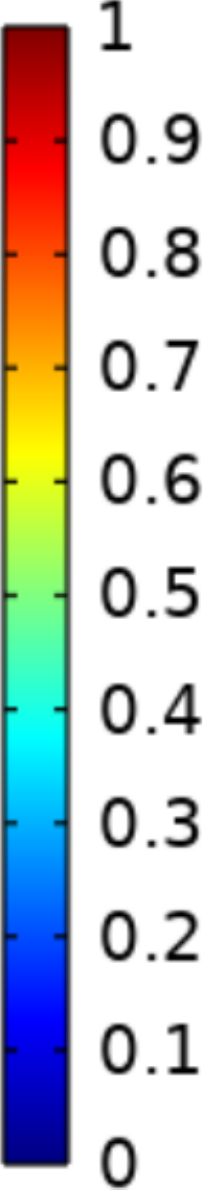} \\[\abovecaptionskip]
        \end{tabular}
        \end{tabular}
	\caption{(a) Number variance $\Sigma^2()$ for the semicircle NB for various masses. The curves for Poissonian, GOE and GUE statistics are plotted as solid, dashed and dash-dotted lines, respectively. To obtain good agreement with GOE statistics for the massless NB, contributions from the diameter orbit were extracted by employing the corresponding trace formula. (b) Intensity distributions of the wave functions (left) and Husimi functions (right) of the QB for eigenstate numbers from top to bottom $m=3241, 3247, 3255$. (c) Intensity distributions of the local currents (left) and Husimi functions (right) of the NB for eigenstate numbers from top to bottom $m=3242, 3245, 3246$, respectively. Here, the arclength parameter $s$ was chosen such that it takes the value $s=0$ at the center of the circle.}
\label{Semicircle_Delta}
\end{figure}
The billiards considered in Ref.~\cite{Yupei2022} have the shapes of a sector of the circle, ellipse or the equilateral triangle. Their classical dynamics is integrable, and the spectral properties of the corresponding QBs agree with those of the full one, that is, they exhibit Poisson statistics. Since the circle billiard belongs to the U(1) symmetry class, which comprises all $M$-fold rotational symmetries with $M\geq 2$, the sectors can have any inner angle $2\pi /M$ with $1<M<\infty$. In~\cite{Yupei2022} values of $M\geq 2$ have been considered and the spectral properties are similar for all $M$. The ellipse billiard has mirror symmetries with respect to its semiminor and semimajor axes, and the only sectors with integrable classical dynamics are the half and quarter ellipses. The sector of the equilateral billiard is obtained by cutting it along a mirror-symmetry axis. The eigenstates of the full and symmetry-projected QBs and NBs are known analytically~\cite{Berry1987,Dietz2019,Dietz2021,Gaddah2018} for all three shapes. The components of the symmetry-projected eigenspinors belong to different symmetry classes as outlined in~\refsec{Syms}. This entails that the sector NBs do not have any common eigenstates with the full and symmetry-projected ones and that these cannot be computed analytically. 

It has been demonstrated~\cite{Yupei2022} that NBs with shapes of the circle- and ellipse-sectors may exhibit spectral properties that are similar to those of QBs with a chaotic dynamics. In~\reffig{Semicircle_Delta} (a) is shown the number variance $\Sigma^2(L)$ for the semicircle NB for various masses.  After extraction of contributions to the spectral properties originating from eigenstates with vanishing support at the corners, that is, from libration-like modes which are localized on the diameter orbit, the spectral properties of the massless semicircle NB agree well with those of random matrices from the GOE, whereas with increasing mass a transition to Poisson statistics takes place. For $m_0\simeq 100$ the spectral properties are similar to those of the semicircle QB. Contributions of librational modes to the spectral properties lead to slow oscillations in the integrated spectral density, which become visible after removing its smooth part. A semiclassical description of the slow oscillations is obtained from the contributions of librational orbits to the associated trace formula. The effect of librational modes on the spectral properties are removed by proceeding as for the bouncing-ball orbits of the stadium billiard~\cite{Sieber1993}, that is, taking into account in addition to the Weyl formula the slow oscillations~\cite{Yupei2022,Dietz2022a} in the unfolding procedure; cf.~\refsec{Stadium} for an explicit example. 

In the left parts of~\reffig{Semicircle_Delta} (b) and (c) are exhibited examples for wave functions of the QB and local currents for the NB, respectively. In the first row ($m=3241$ for the QB and $m=3242$ for the NB) are shown examples for librational modes localized along a diameter orbit, whereas in the second row ($m=3247$ for the QB and $m=3245$ for the NB) examples are exhibited for which they are strongly localized on a rotational mode. For $m=3246$ the local current of the NB (third row) seems to be spread over the whole billiard area, except for a semicircular part around the circle center, whose radius is determined by the associated angular momentum value (see the example for the QB with $m=3255$). Generally, for high-lying states the local current is localized in regions where the corresponding wave functions of the QB are nonvanishing. Thus, it does not spread over the whole billiard area as expected for QBs with a chaotic classical counterpart. In the right parts of~\reffig{Semicircle_Delta} (b) and (c) are shown the corresponding Husimi functions. For the NB they are localized at values of the $s$-parameter, at which the corresponding one of the QB is nonvanishing. However, they spread over a larger range of $p$ values in the third example. Thus, contrary to those of QBs with classically chaotic counterpart, the Husimi functions of the sector NBs do not extend over the whole Poincar\'e surface of section.  

Small-mass ellipse-sector NBs with a large eccentricity $\epsilon$ exhibit spectral properties similar to those of small-mass circle-sector NBs, that is, they are close to GOE after extracting eigenstates associated with the librational modes, which vanish at the corners. For small eccentricity the extraction of librational modes is not possible because their number is too large~\cite{Yupei2022,Dietz2023}. Then, for the semi-ellipse NB the spectral properties are close to those of Poissonian random numbers. Generally, the spectral properties of small-mass circle- and ellipse-sector NBs that predominantly comprise eigenstates with nonvanishing support at the corners, are GOE-like. The right-angled triangle NB, which is obtained by halving the equilateral triangle along a mirror-symmetry line, exhibits Poisson statistics for small masses. For large masses the spectral properties are similar to those of the corresponding QB whose nearest-neighbor spacing and ratio distributions exhibit gaps and thus are nontypical, whereas the long-range correlations approach Poisson statistics with increasing number of included eigenvalues. These results confirm the assumption that the GOE-like behavior is induced by the corners connecting curved and straight boundary parts and the discontinuity caused by the BCs. 

The effect of the corners is reminiscent of that of diffractive orbits in pseudointegrable, almost integrable or singular QBs~\cite{Sieber1997a,Bogomolny1999}. Yet, in distinction to the spectral properties of the sector NBs these exhibit intermediate statistics. The GOE-like behavior of the NB can be attributed to the complexity originating from the different symmetry classes of the spinor components which are mixed when cutting a NB with rotational symmetries into its fundamental domain. Indeed, besides the BCs~\refeq{BCm} an additional requirement is the continuity of the boundary wave functions or spinor components at the corners. Similarly, in polygonal QBs, the occurrence of diffractive orbits ensures continuity of the wave dynamics around diffractive corners. As a result, the wave functions and Husimi functions of the low-mass sector NBs may exhibit complex structures, however, they do not extend over the whole available space, which is expected for generic QBs with a classically chaotic counterpart. Thus, similar to pseudointegrable, almost integrable or singular QBs that comprise diffractive orbits leading to level repulsion and intermediate statistics, the eigenstates of the circle-sector and semi- and quarter-ellipse NBs do not exhibit quantum-chaotic behavior, even though the spectral properties feature quantum signatures of classical chaos. 

 In~\cite{Dietz2023} the properties of symmetry-projected eigenstates of rectangular NBs whose side lengths are either commensurable or incommensurable were investigated. Rectangular billiards have a twofold rotational symmetry. In distinction to the eigenvalues of rectangular QBs, those of the corresponding NBs are nondegenerate and exhibit Poisson statistics. Furthermore, independently of the choice of the ratio of the side-lengths, the spectral properties of their symmetry-projected eigenstates agree with semi-Poisson statistics. Those of the corresponding QBs exhibit nontypical short-range correlations for rational side-length ratios and for the long-range correlations Poisson statistics for a sufficiently long eigenvalue sequence~\cite{Marklof1998}. The spectral properties of the right-angled triangle QBs, that are constructed by cutting a rectangular billiard along the diagonal, may exhibit Poisson, semi-Poisson, intermediate or GOE statistics depending on the ratio of the side lengths~\cite{Artuso1997,Losej2022}. It has been found in~\cite{Dietz2023} that, if they agree with GOE statistics, those of the corresponding NB exhibit GUE behavior. Generally, the eigenvalues of a right-angled triangle NB exhibit stronger level repulsion than the corresponding QB, if the spectral statistics of the latter is intermediate between Poisson and GOE statistics. 
\section{Neutrino billiards with shapes of chaotic billiards exhibiting particular features\label{Chaos}}
\subsection{Scarred quantum states in stadium NBs\label{Stadium}}
\begin{figure}[!h]
\centering
\begin{tabular}{@{}c@{}}
        \begin{tabular}{@{}c@{}}
        \large (a)
        \includegraphics[width=.3\linewidth]{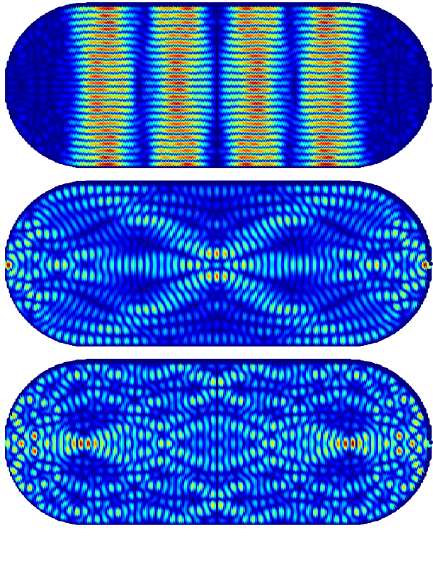}\\[\abovecaptionskip]
        \end{tabular}
        \begin{tabular}{@{}c@{}}
        \large (b)
        \includegraphics[width=.28\linewidth]{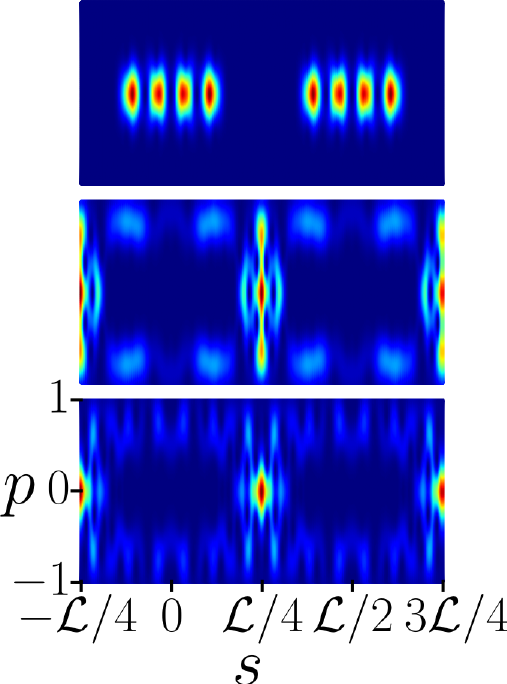}
        \end{tabular}
        \end{tabular}
        \begin{tabular}{@{}c@{}}
        \begin{tabular}{@{}c@{}}
        \large (c)
        \includegraphics[width=.27\linewidth]{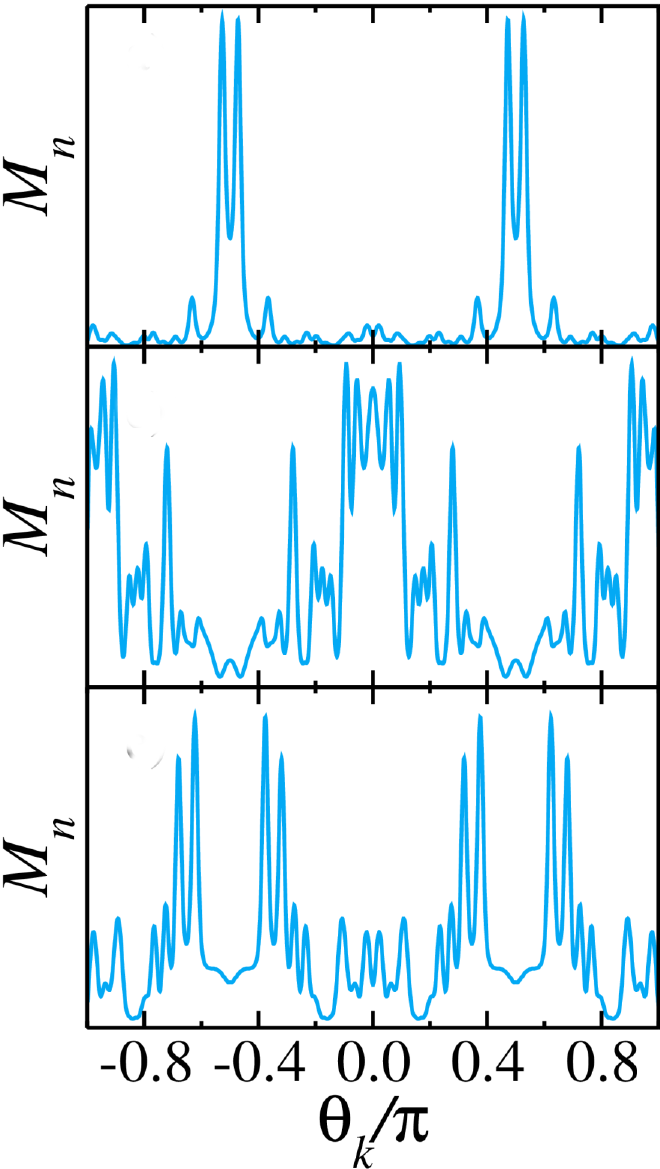}
        \end{tabular}
        \begin{tabular}{@{}c@{}}
        \large (d)
        \includegraphics[width=.7\linewidth]{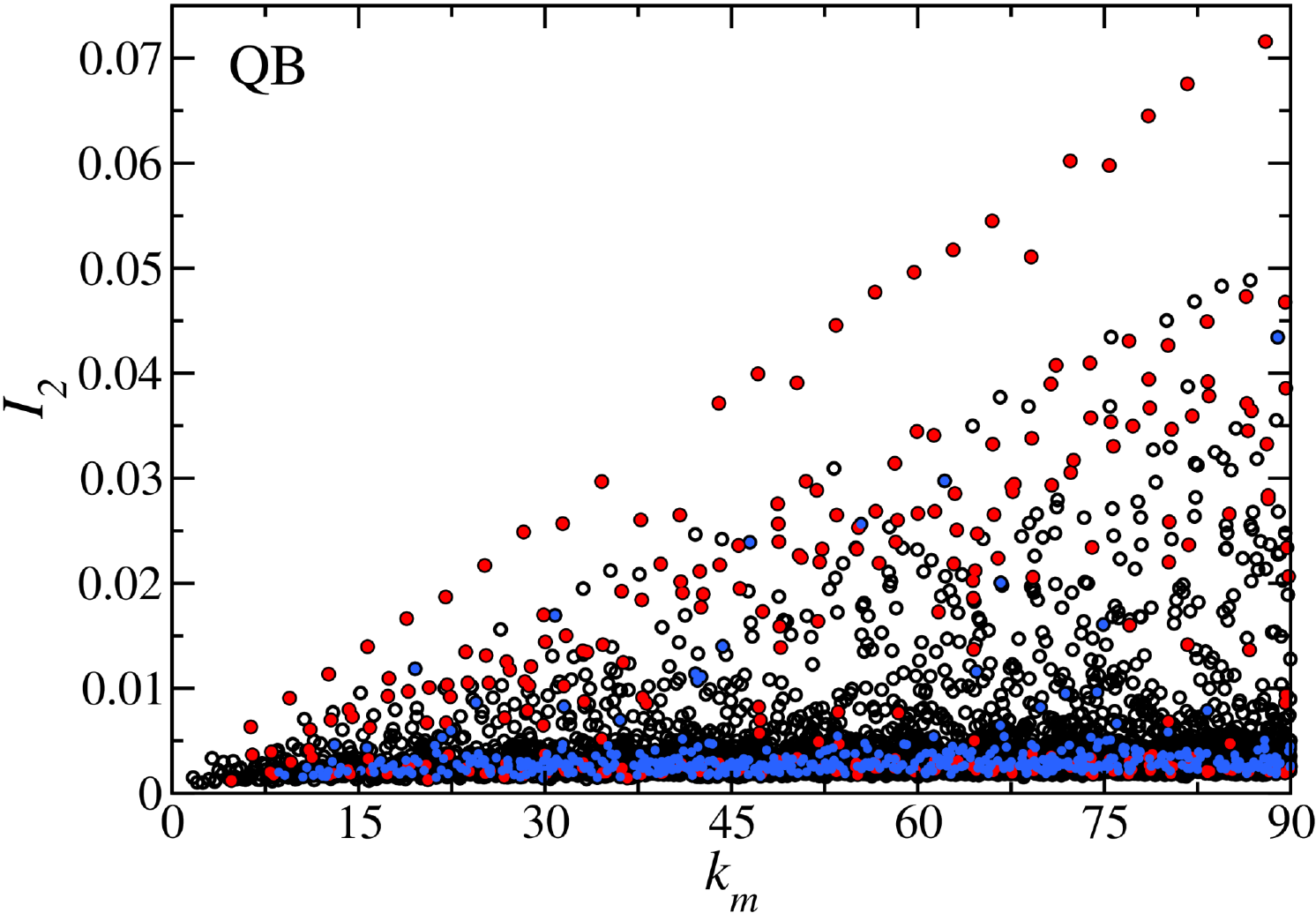}
        \end{tabular}
\end{tabular}
	\caption{(a) Intensity distributions of the wave functions, (b) Husimi functions, and (c) momentum distributions of the stadium QB for the eigenstates with numbers $m=1048,1102,1020$. In (b) the values $s=0,\mathcal{L}/4,\mathcal{L}/2,3\mathcal{L}/4$ ($-\mathcal{L}/4$) correspond to the centers of the lower straight part, the right semicircular part, the upper straight part and the left semicircular part, respectively. (d) Inverse participation number of the on-shell momentum distribution of wave functions $\Psi_m(\boldsymbol{r})$ of the QB. Red dots mark BBOs. Blue dots mark edge states localized along whispering gallery modes of the semicircular boundary parts and either along the straight parts or along paths reflected at the joining corners. 
	}
\label{WFs_QB_Hus_Stadium}
\end{figure}
\begin{figure}[!h]
\centering
\begin{tabular}{@{}c@{}}
        \begin{tabular}{@{}c@{}}
        \large (a)
        \includegraphics[width=.3\linewidth]{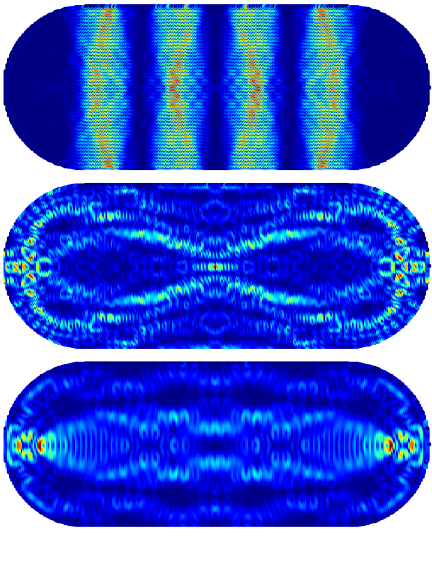}\\[\abovecaptionskip]
        \end{tabular}
        \begin{tabular}{@{}c@{}}
	\large (b)
        \includegraphics[width=.28\linewidth]{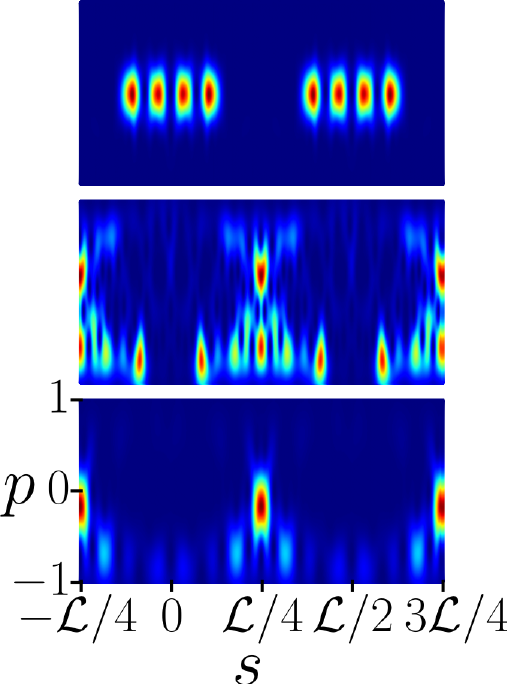}
	\end{tabular}
	\end{tabular}
        \begin{tabular}{@{}c@{}}
        \begin{tabular}{@{}c@{}}
        \large (c)
        \includegraphics[width=.27\linewidth]{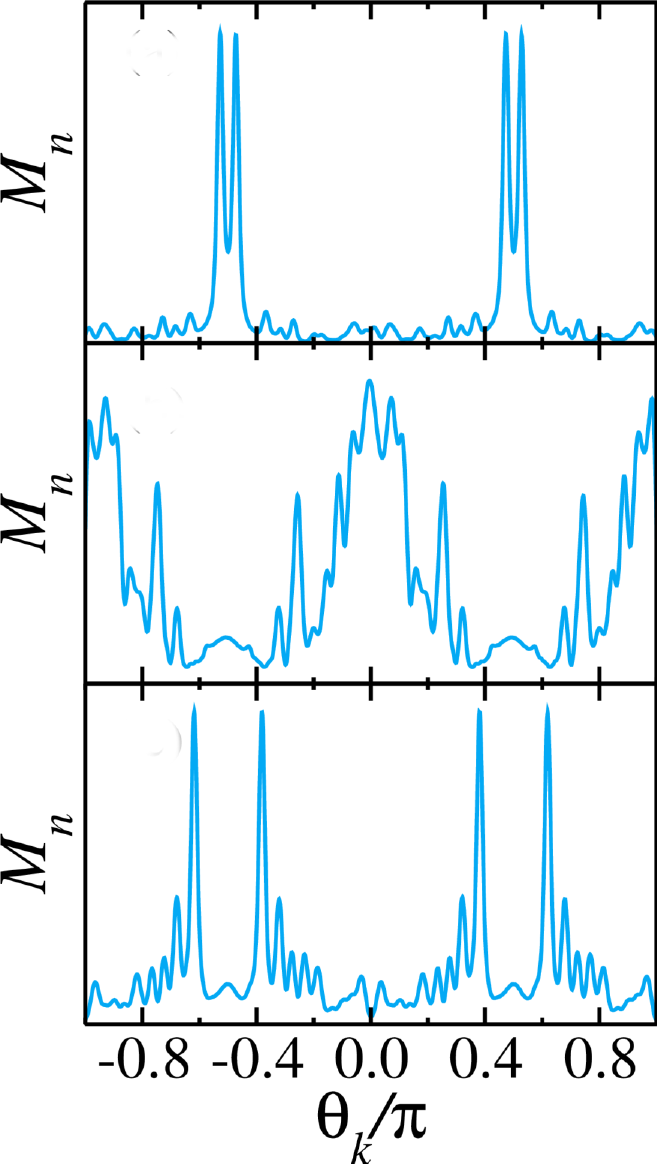}
        \end{tabular}
        \begin{tabular}{@{}c@{}}
        \large (d)
	\includegraphics[width=.7\linewidth]{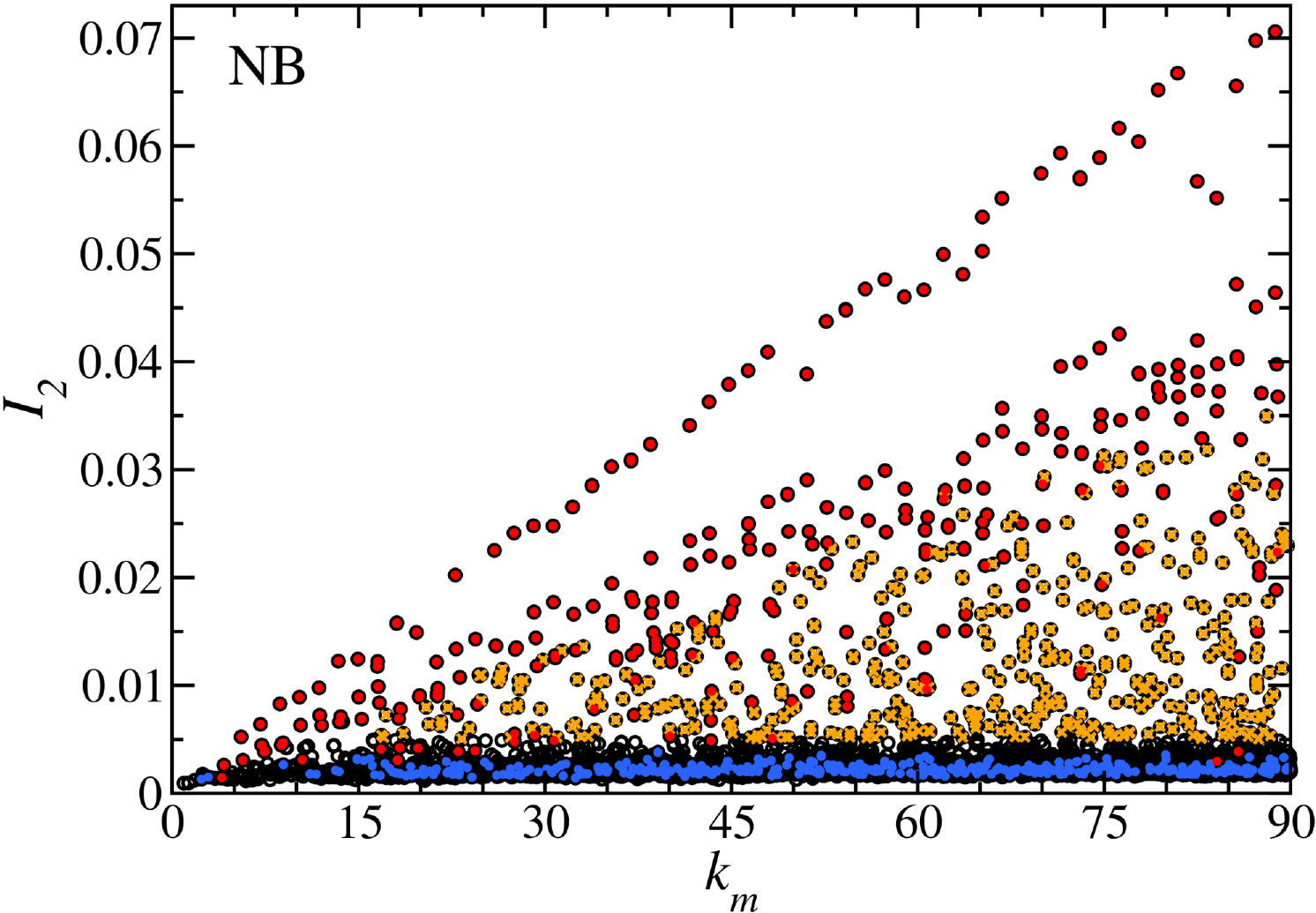}
        \end{tabular}
\end{tabular}
	\caption{(a) Local current, (b) Husimi functions, and (c) momentum distributions of the stadium NB for the eigenstates with numbers $m=1136,482,1001$. In (b) the values $s=0,\mathcal{L}/4,\mathcal{L}/2,3\mathcal{L}/4$ ($-\mathcal{L}/4$) correspond to the centers of the lower straight part, the right semicircular part, the upper straight part and the left semicircular part, respectively. (d) Inverse participation number of the on-shell momentum distribution of the first eigenspinor component $\Psi_{1,m}(\boldsymbol{r})$ of the NB. Red dots mark BBOs. Blue dots mark edge states localized along whispering gallery modes of the semicircular boundary parts and either along the straight parts or along paths reflected at the joining corners. Orange crosses result from wave functions that are either localized along almost BBOs that are reflected at an angle close to $90^\circ$ along the straight part and leak into the semicircular parts or along orbits from the family of the bow-tie orbit, which are predominantly localized in the rectangular part.
        }
\label{WFs_NB_Hus_Stadium}
\end{figure}
\begin{figure}[!h]
\centering
        \includegraphics[width=.49\linewidth]{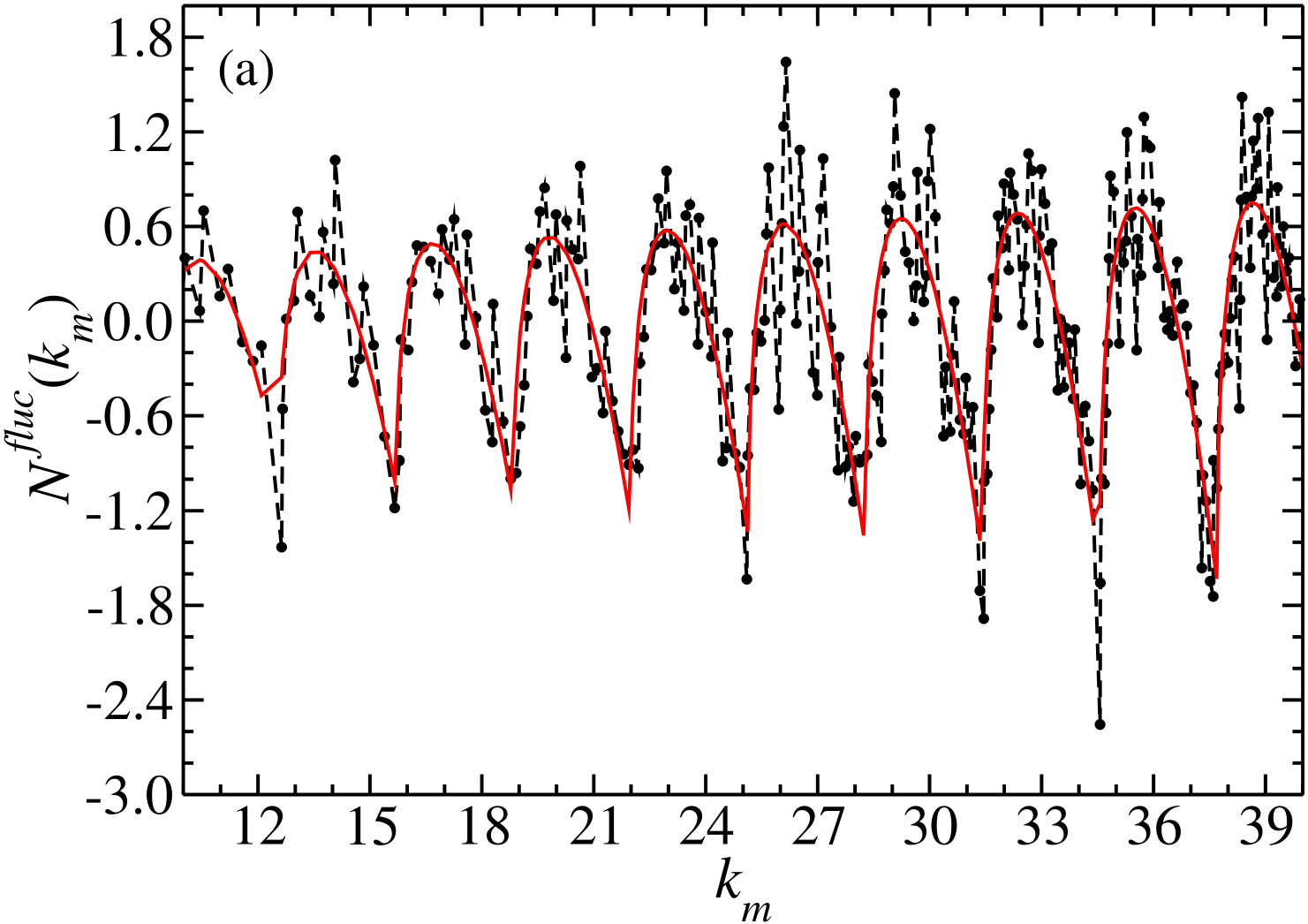}
        \includegraphics[width=.49\linewidth]{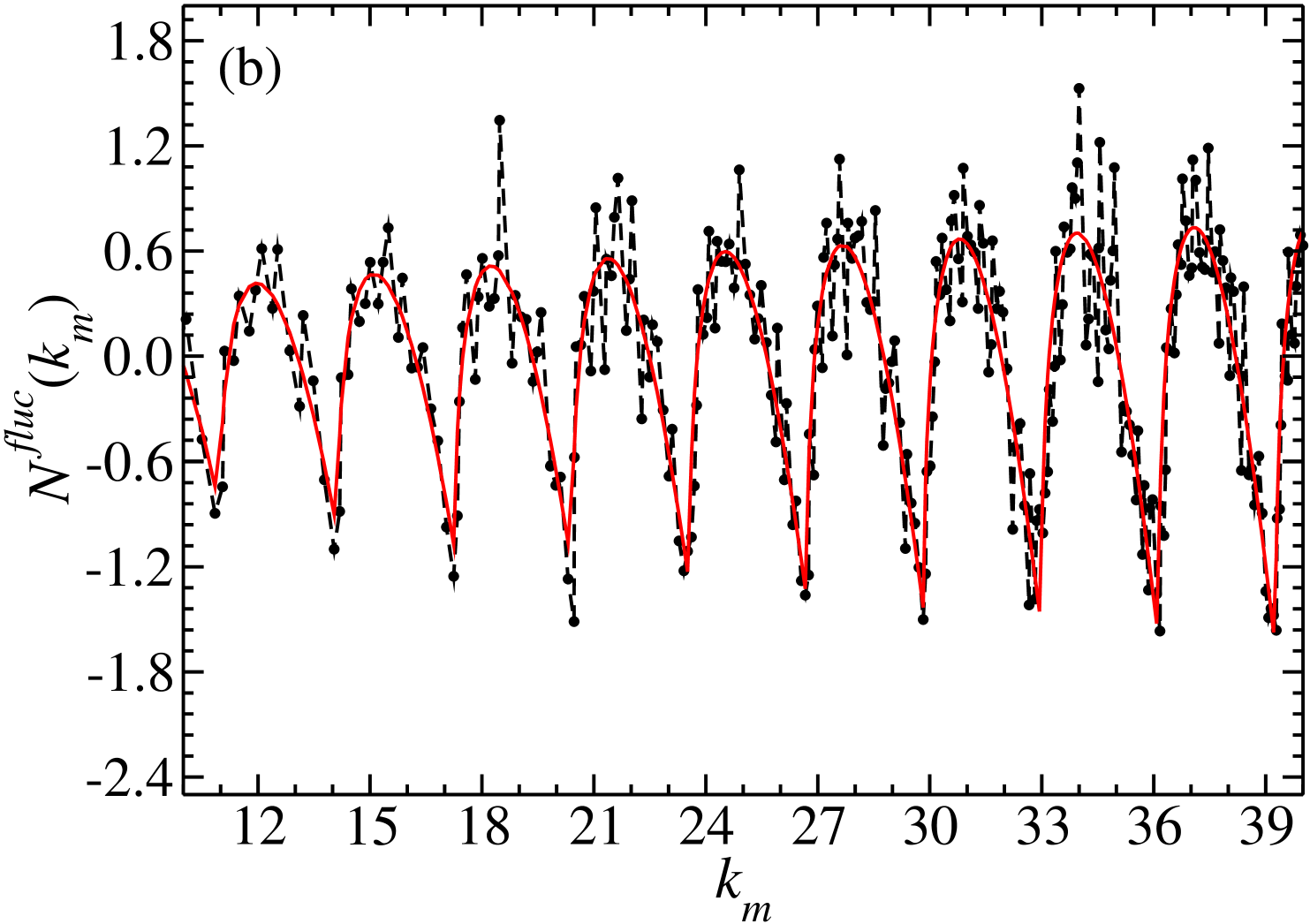}
        \includegraphics[width=.49\linewidth]{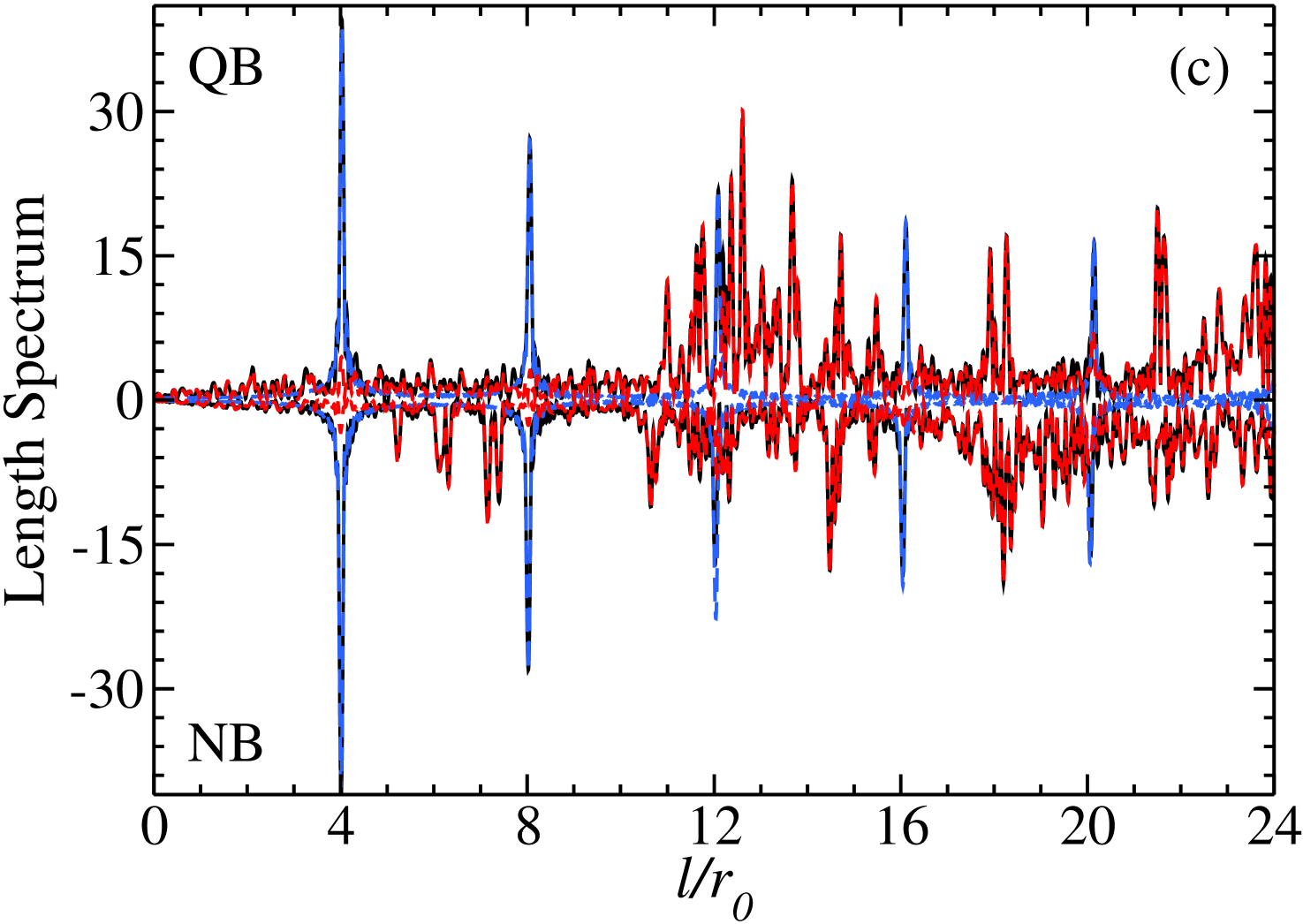}
	\caption{ (a) Fluctuating part of the integrated spectral density of the quarter-stadium QB (black dots) compared to that of the BBOs deduced from~\refeq{NbboQB} (red curve). 
        (b) Fluctuating part of the integrated spectral density of the quarter-stadium NB (black dots) compared to that of the BBOs deduced from~\refeq{NbboNB} (red curve).
	(c) Length spectrum (black) of the full stadium (QB) compared to that of the NB. The latter was multiplied by $-1$. Blue dashed lines show those of the BBOs and red dashed lines are the residual length spectra after the BBO contributions are removed from $\mathcal{N}^{fluc}(k_m)$.
        }
\label{Nfluc_FFT_Stadium}
\end{figure}
The Bunimovich stadium CB~\cite{Sinai1970,Bunimovich1979,Berry1981,Berry1983} exhibits full chaos and most of the POs are unstable and isolated and cover the whole available phase space. Furthermore, there exists a nongeneric, continuous family of neutral-stable POs of measure zero, named bouncing-ball orbits (BBOs), which bounce back and forth between the two straight-line segments. The stadium billiard became of high interest because these BBOs lead to a scarring of the wave functions of the corresponding QB~\cite{McDonald1979,Heller1984}. Bouncing-ball orbits were found experimentally in flat microwave resonators~\cite{Sridhar1991,Stein1992} in the range of microwave frequencies where the Helmholtz equation governing these systems is mathematically identical to the Schr\"odinger equation, where they are named microwave billiards. Furthermore, experiments with superconducting microwave billiards~\cite{Graef1992} revealed that BBOs, and generally scarred eigenstates cause deviations of the spectral properties from the expected GOE behavior. Due to these extraordinary features, the stadium billiard still serves as a paradigm model for the study of the effect of scarred wave functions on the properties of the eigenstates of generic quantum-chaotic systems~\cite{Selinummi2024,Graf2026,Rahkonen2025}. 

In Ref.~\cite{Sieber1993} a semiclassical trace formula is derived for the contributions of the BBOs to the level density and a procedure for the extraction of their nongeneric effects on the spectral properties is developed. Furthermore, semiclassical trace formulas are provided for other neutral-stable POs or almost POs that lead to a scarring of wave functions of the stadium billiard and whose contributions to the trace formula are of similar weight as those of the unstable ones. These nongeneric contributions manifest themselves as slow oscillations in the fluctuating part of the integrated spectral density, $N^{osc}(k_m)$. To remove their effects on the spectral properties, the eigenvalues $k_m$ are unfolded by adding these to the Weyl formula, $\epsilon_m=\mathcal{N}^{Weyl}(k_m)+N^{osc}(k_m)$. It was found in Ref.~\cite{Sieber1993}, that the BBOs are predominantly responsible for the deviations of the long-range correlations in the eigenvalue spectrum of the stadium QB from BGS predictions for typical classically chaotic systems. These semiclassical procedures were extended to relativistic NBs in Ref.~\cite{Dietz2020,Dietz2022a} and used to remove contributions from scarred eigenstates in NBs with semicircular shape, shapes with a threefold symmetry and constant-width NBs in Ref.~\cite{Zhang2021,Yupei2022,Dietz2022a}.

To reduce the numerical efforts needed to compute the eigenstates of the full stadium QB and NB, its twofold symmetry can be exploited to obtain the eigenstates belonging to the symmetry classes $l=0$ and $l=1$ separately~\cite{Zhang2021}. Another crucial advantage is that thereby degeneracies in the eigenvalue spectrum of the QB and near-degeneracies in that of the NB are avoided. In Figs.~\ref{WFs_QB_Hus_Stadium} (a) and~\ref{WFs_NB_Hus_Stadium} (a) are shown typical intensity distributions of the wave functions and local-current. The distributions exhibit enhancement along a BBO (first rows), along edge orbits, composed of whispering modes in the semicircular parts and along paths connecting diagonal corners where the straight and circular parts of the boundary are joined (second row) and isolated diameter orbits bouncing back and forth between the centers of the semicircular parts (third row). Such scarred wave functions and spinor components can be identified with help of the Husimi functions~\refeq{Hus} and on-shell momentum distributions~\refeq{Mom_Distr}~\cite{Bogomolny2006,Bogomolny2021a}. Those corresponding to the wave functions and local currents in Figs.~\ref{WFs_QB_Hus_Stadium} and~\ref{WFs_NB_Hus_Stadium} (a) are shown in (b) and (c), respectively. The Husimi functions are localized at values of $s$, where the intensity of the wave function or local current is maximal along the boundary and at values of $p$ corresponding to the directions of the paths along which they are scarred. For the QB the on-shell momentum-distributions are symmetric with respect to $\theta_k=0$, whereas those of the NB do not have mirror symmetries but exhibit a $\pi$ periodicity resulting from the twofold symmetry of the stadium. This is attributed to the unidirectionality of the local current along the boundary and chirality. Both for the QB and NB the on-shell momentum distributions are peaked around $\theta_k/\pi=\pm 1/2$ in the first row, as expected for BBOs. In the second row they exhibit peaks around $\theta_k/\pi=0\, (\pm 1)$ like for the whispering-gallery modes in a circle billiard. Furthermore, in this and the third row they exhibit sharp peaks at values of $\theta_k/\pi$ corresponding to the angles of the straight lines along which they are scarred. To quantify the localization properties of the momentum distributions the inverse of their participation numbers~\refeq{IPR}, shown in Figs.~\ref{WFs_QB_Hus_Stadium} and~\ref{WFs_NB_Hus_Stadium} (d) has been analyzed. Its largest values correspond to BBOs. For the NB also local currents which are scarred by almost BBOs that leak into the semicircular parts or by orbits from the family of the bow-tie orbit, shown as yellow crosses, have comparatively large inverse of the participation numbers, whereas for the QB also other eigenstates have similar values. For the edge orbits, marked by blue dots, the values are comparably small in the QB and NB. In both the NB and QB the largest inverse of the participation numbers correspond to BBOs with one bounce at each straight part of the boundary, and repetitions of it. It was found in~\cite{Dietz2026} that the widths of the associated on-shell momentum distributions decay with $k^{-1}$, in accordance with the feature of these inverse participation numbers that they lie on a straight line. 

Figures~\ref{Nfluc_FFT_Stadium} (a) and~\ref{Nfluc_FFT_Stadium} (b) show the fluctuating parts of the integrated spectral density, $N^{fluc}(k_m)=N(k_m)-N^{Weyl}(k_m)$ of the quarter-stadium QB and NB, respectively. The black dots correspond to the numerical results and the red curves to the analytical ones. For the QB the derivation employs Gutzwiller's trace formula~\refeq{TraceQB}~\cite{Sieber1993}, which was extended to NBs in~\cite{Dietz2020}, cf.~\refeq{rhoNB0}. Here, the relation $\rho(k)=\frac{d}{dk}N(k)$ has been used. For the QB the trace formula for BBOs is given by
        \be
        N^{QB}_{bbo}(k)=\frac{L}{2\pi}\sqrt{\frac{k}{\pi R_0}}\sum_{\tilde m=1}^\infty\tilde m^{-3/2}\cos\left(2\tilde mkR_0-\frac{3\pi}{4}\right)
        \label{NbboQB}
        \ee
        and for the NB by
        \be
        N^{NB}_{bbo}(k)=\frac{L}{2\pi}\sqrt{\frac{k}{\pi R_0}}\sum_{\tilde m=1}^\infty\tilde m^{-3/2}\cos\left(2\tilde mkR_0 -\tilde m\pi -\frac{3\pi}{4}\right),
        \label{NbboNB}
        \ee
respectively, where $L$ is the length of the straight part, $R_0=r_0$, with $r_0$ denoting the radius of the circular part, for the quarter stadium and $R_0=2r_0$ for the full stadium. The trace formula for the QB and NB just differ by an additional phase of $\tilde m\pi$ and accordingly the quantization condition for the BBOs, which is obtained from the periodicity of the cosine functions, yields just a shift of the corresponding $k$ values for the QB and NB with respect to each other. Generally, the quantization condition for a state which is scarred by an orbit (or its repetitions), requires that $k$ multiplied with its length plus a possible phase, is a multiple of $2\pi$. The additional phase depends on the BCs and for NBs also on chirality and can be read off the phases of the summands in the corresponding trace formula~\refeq{TraceQB} or~\refeq{rhoNB0}. Thus, it is linear in the wave numbers, implying that the dispersion relation is not relevant in this context. This can be seen when comparing~\reffig{Nfluc_FFT_Stadium} (a) with (b). The analytical curves describe well the slow oscillations observed in the fluctuating part of the spectral density obtained from the eigenvalues of the QB and NB. Figure~\ref{Nfluc_FFT_Stadium} (c) shows in black the length spectra of the full stadium QB and that of the NB multiplied with (-1), respectively. The blue curves show the length spectra deduced from Eqs.~(\ref{NbboQB}) and~(\ref{NbboNB}), confirming their applicability also for comparatively small values of $k_m$. To demonstrate that contributions from the BBOs are completely extracted, when subtracting $N^{XB}_{bbo}(k_m)$ with $X=Q,N$ from $N^{fluc}(k_m)$ the length spectra obtained from the Fourier transforms of $\tilde N^{fluc}(k_m)=N^{fluc}(k_m)-N^{XB}_{bbo}(k_m)$ are also shown (red dashed lines). Indeed, peaks located at the lengths of the BBOs are absent or suppressed in these length spectra. It has been demonstrated in~\cite{Dietz2026} that after removing the effect of the BBOs by employing~\refeq{NbboNB} in the associated unfolding procedure, $\epsilon_m=\mathcal{N}^{Weyl}(k_m)+N^{NB}_{bbo}(k_m)$, the spectral properties agree well with those of generic NBs whose shape generates a fully chaotic dynamics. 

\subsection{Constant-width NBs\label{CW}}
\begin{figure}[!h]
\centering
 \begin{tabular}{@{}c@{}}
        \begin{tabular}{@{}c@{}}
	\large (a)
        \includegraphics[width=0.16\linewidth]{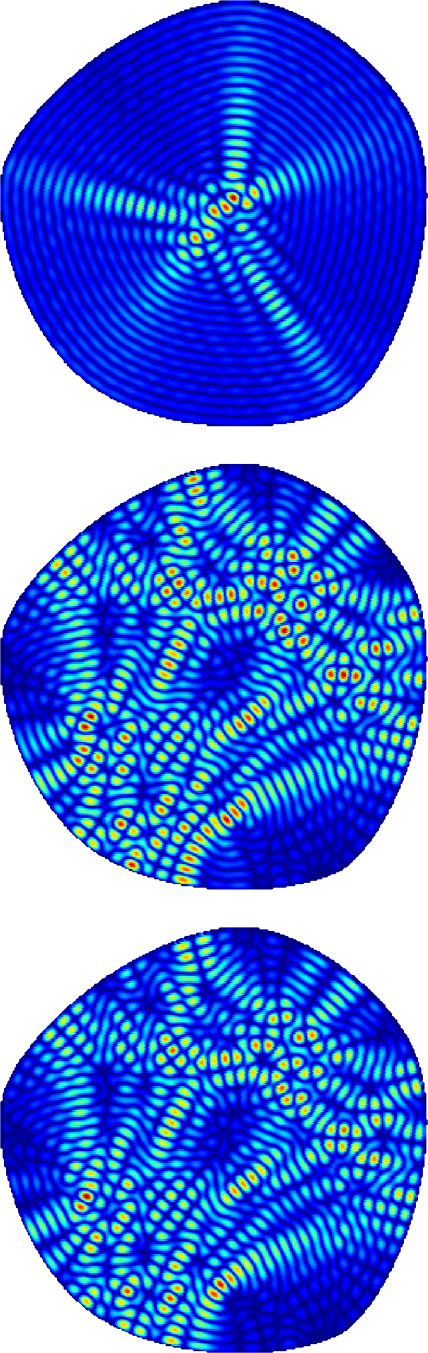}
        \includegraphics[width=.28\linewidth]{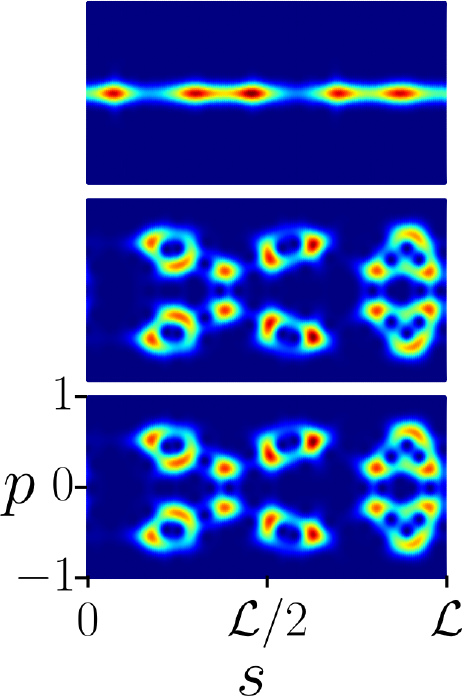}
        \end{tabular}
        \begin{tabular}{@{}c@{}}
	\large (b)
        \includegraphics[width=0.16\linewidth]{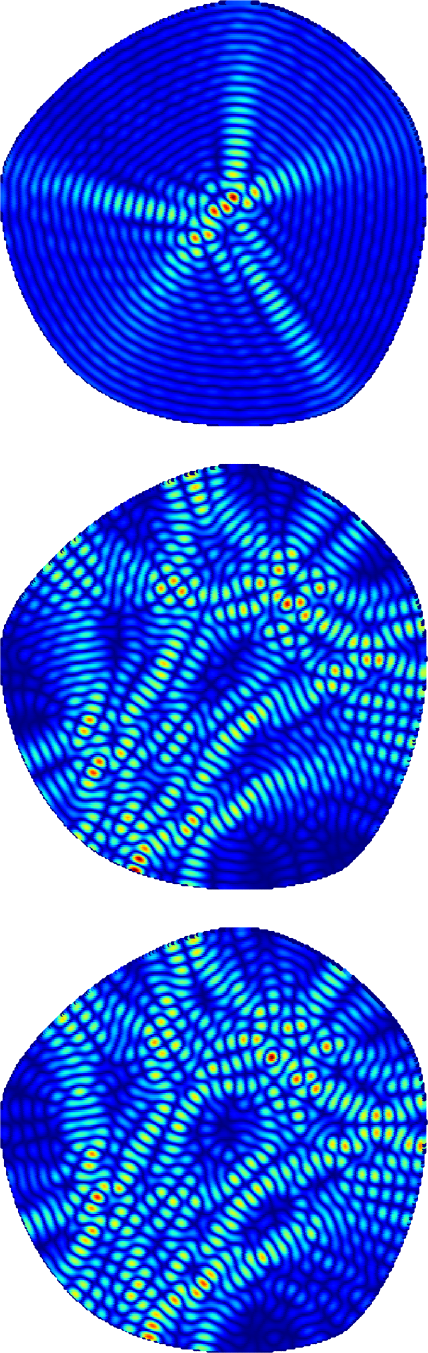}
        \includegraphics[width=.28\linewidth]{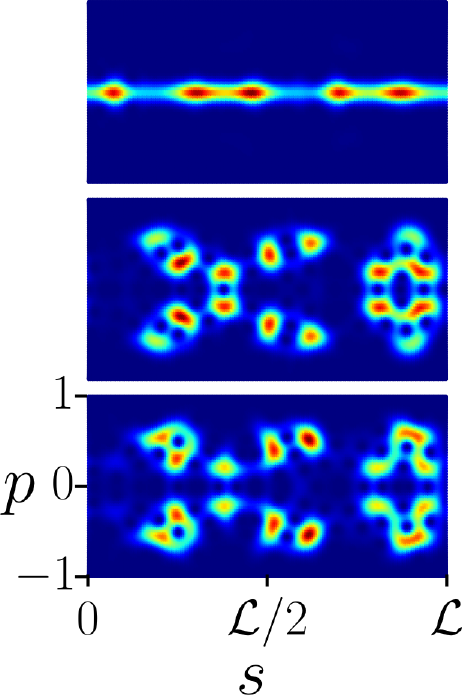}
        \end{tabular}
        \end{tabular}
        \begin{tabular}{@{}c@{}}
        \begin{tabular}{@{}c@{}}
	\large (c)
	\includegraphics[width=0.16\linewidth]{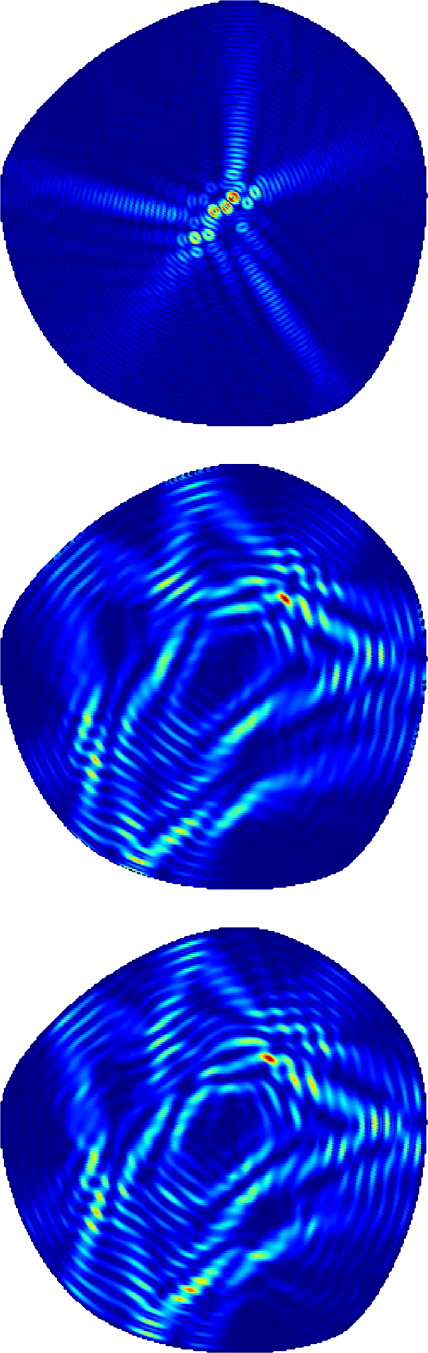}
	\includegraphics[width=.28\linewidth]{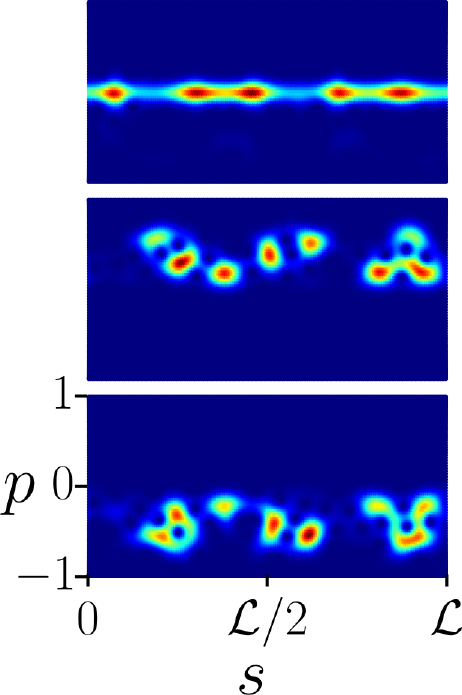}
        \end{tabular}
        \begin{tabular}{@{}c@{}}
	\large (d)
        \includegraphics[width=0.16\linewidth]{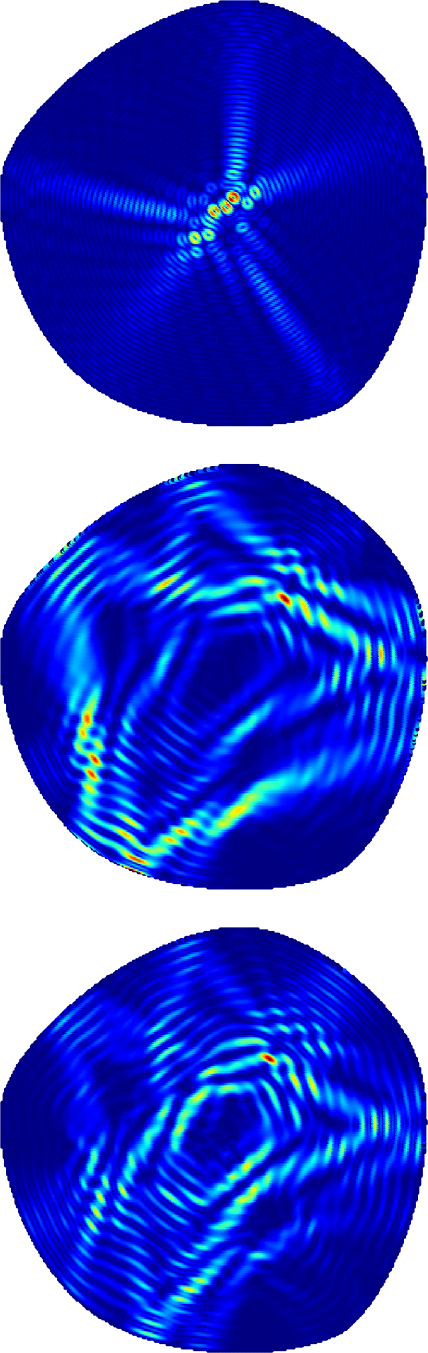}
	\includegraphics[width=.28\linewidth]{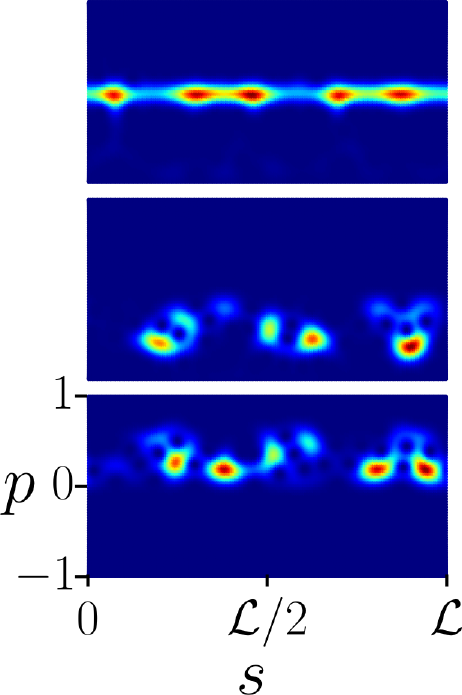}
        \end{tabular}
        \end{tabular}
	\caption{(a) Wave functions (left) and Husimi functions (right) of the constant-width QB for state numbers $m=1023,1026,1027$. The uppermost wave function is localized on the caustics formed by the diameter orbits, which are reflected at the five bending points of the billiard boundary. The corresponding Husimi function is localized around $p=0$ and $s$ values corresponding to the five bending points. Here, the starting point $s=0$ coincides with $\gamma =0$ in~\refeq{B1B2}, which is to the left near the lowest bending point. The other two wave functions are localized along the PO with 11 reflection points corresponding to the islands in the chaotic sea of the PSOS (cf.~\cite{Dietz1990,Dietz2022a}). Around these the Husimi functions are nonzero. (b)-(d) Local current and Husimi functions of the NB with $m=1026,1016,1001$. They exhibit the same localization properties as for the nonrelativistic QB. In (b) the real part of the first spinor component is inserted into~\refeq{Hus} and the Husimi functions are symmetric with respect to $p=0$ like for the QB, in (c) and (d) the complex spinor component is used and the BC $\mu=1$ and $\mu =-1$ in~\refeq{BC1}, respectively, is applied. The Husimi functions are localized in the regions $p>0$, $p<0$ and $p<0$, $p>0$, respectively.}
\label{WFs_Hus_CW}
\end{figure}
Constant-width billiards have the extraordinary property, that the classical dynamycs is unidirectional, meaning that a particle launched into the billiard never changes its rotational direction, so that only one half of the Poincar\'e surface of section (PSOS) is filled. Their name originates from the property that they can be inserted into a square box of their width, touching each of its sides independently of their orientation. In the complex plane the boundary of the family of billiards considered in~\cite{Dietz2022a} is given by~\cite{Knill1998}
\be\label{B1B2}
w(\gamma)=-R_0ie^{i\gamma}-ia_3\left[\frac{\left(e^{i4\gamma}-1\right)}{4}+\frac{\left(e^{-i2\gamma}-1\right)}{2}\right]
-ia_5\left[\frac{\left(e^{i6\gamma}-1\right)}{6}-\frac{\left(e^{-i4\gamma}-1\right)}{4}\right],\, \gamma\in [0,2\pi).
\ee
In the examples shown for the QB and massless NB the parameter values are set to $R_0=4$, $\mathcal{L}= 25.13$, and $a_3=\frac{i}{2},a_5=1$. The tangential vector and normal vector in~\refeq{BC2} are obtained from
\be
\label{Derivative}
w^\prime(\gamma)=\mathcal{R}(\gamma)e^{i\gamma},\, \mathcal{R}(\gamma)=R_0+2ia_3\sin(3\gamma)+2a_5\cos(5\gamma),\,  \mathcal{R}(\gamma)=\mathcal{R}^\ast(\gamma),
\ee
 and the curvature of the boundary at $\gamma$ equals $\kappa(\gamma)=\frac{1}{\mathcal{R}(\gamma)}$. The two halves of the PSOS with positive and negative $p$ are well separated by a barrier region of Kolmogorov-Arnold-Moser tori~\cite{Gutkin2007} around the diameter orbit. In each half of the PSOS the classical dynamics is chaotic except in the regions around the diameter orbit with $p\simeq 0$, the whispering gallery with $\vert p\vert\simeq 1$ and the tiny islands of regular motion in the chaotic sea, which for the specific choice of parameters correspond to a PO with 11 reflections and length $l_{PO}=19.21R_0$.

Switching from clockwise to counterclockwise motion is not possible for the CB, however, in the corresponding QB it is facilitated via dynamical tunneling~\cite{Keshavamurthy2011,Gutkin2007,Gutkin2009} through the barrier of KAM tori around the diameter orbit, which manifests itself in a splitting of the majority of eigenvalues into doublets of nearly degenerate ones. Namely, the Dirichlet BC $\psi (s\in\partial\Omega)=0$ implies that $\psi(s)$ is constant along the boundary and thus the tangential derivative vanishes for $s\in\partial\Omega$, $\boldsymbol{t}\cdot\vec\nabla\psi (s)=\partial_s\psi(s)=0$. Accordingly, for constant-width billiards with the property~\refeq{Derivative}, which yields $\boldsymbol{t}\equiv (\cos\gamma,\sin\gamma)$, unidirectionality leads to a separation into clockwise and counterclockwise modes, which can be associated with the modes corresponding to doublet partners~\cite{Gutkin2009}. The spectrum can be separated into three subspectra, one comprising the smaller, the other one the larger eigenvalue of the doublets, and one the singlets which are associated with the diameter orbit, that is, zero angular-momentum modes $m=0$ of the circle QB of radius $R_0$. These can be identified by comparison of the eigenvalues with the zeroes of~\refeq{ewcirc} for $m=0$, $J_0(k^0_nR_0)=0$~\cite{Berry1981}. 

In~\reffig{WFs_Hus_CW} (a) are shown wave-function intensity distributions (left) and the associated Husimi functions~\refeq{Hus} (right) for the constant-width QB. In the first row is shown an example for a singlet mode which is scarred by the caustics formed by the diameter orbit in the CB~\cite{Knill1998}. In the second and third row are shown the doublet partners of a wave function which is scarred by the PO corresponding to the tiny islands in the chaotic sea. The Husimi functions of the singlet modes are localized around $p=0$, and the ones corresponding to doublet partners resemble each other and are nonvanishing around these islands in both halves of the PSOSs. This is attributed to dynamical tunneling~\cite{Keshavamurthy2011} through the barrier of KAM tori around $p=0$, and confirms the result of Ref.~\cite{Gutkin2007}, that one doublet partner corresponds to the symmetric combination of the clockwise and counterclockwise modes, the other one to the antisymmetric one. 

The spectrum of the constant-width NB does not exhibit near degeneracies, yet comprises a sequence of eigenvalues associated with the zero angular-momentum modes of the circle NB. As in the case of the QB these can be identified by comparing its eigenvalues to the solutions of~\refeq{QCCm}, $J_{1}(kR_0)=\mathcal{K}^{-1}J_0(kR_0)$ for the massless case $m_0=0$. In~\reffig{WFs_Hus_CW} (b)-(d) are shown examples for intensity distributions of the local currents and Husimi functions of the constant-width NB corresponding to those of the QB. The Husimi functions resemble those of the QB, and reflect the structure of the PSOS. Especially, those exhibited in the second and third rows are clearly localized on the regular islands in the chaotic sea. In~\reffig{WFs_Hus_CW} (b) are shown intensity distributions of the real part of the first spinor component, $\Re[\psi_1(\boldsymbol{r})]$ and the Husimi functions deduced from them. The intensity distributions and also those of the imaginary parts of the spinor components resemble those of the corresponding QB with mixed Dirichlet-Neumann BCs~\cite{Sieber1995}. Furthermore, the Husimi functions deduced from them are localized in both halves of the PSOS like for the QB. The similarity is attributed to the fact that the BCs Eqs.~(\ref{BC2i}) turn into Robin boundary conditions~\refeq{BCRobin} when replacing $\psi_{1,2}(n,s)$ by the boundary functions, $\tilde\Phi_{1,2}(n,s)$, introduced in~\refeq{Phi12}. The Husimi functions of the complex first spinor components, shown in (c) and (d) in~\refeq{BC1}, respectively, are localized either in the upper of the PSOS in (c) and the lower part in (d) or vice versa, implying that the eigenstates exhibit unidirectionality, and chirality. The eigenvalues were obtained by imposing the BC~\refeq{BC1} with $\mu=+1$ in (c) and $\mu=-1$ in (d), respectively. They are identical, however, corresponding Husimi functions are localized in opposite halves of the PSOS. This corroborates the assumption that the feature that the related Husimi functions have nonvanishing support only in the upper or lower part of the PSOS originates from the unidrectionality of the current flow, whose rotational direction is determined by the sign of $\mu$. Like for the QB, a clear separation of the spectrum into three subspectra is also possible for the constant-width NB. Two subspectra comprise the eigenvalues corresponding to clockwise and counterclockwise modes with nonvanishing support of the Husimi functions in the upper and lower part of the PSOS, respectively, and the third one the zero modes that are associated with the diameter orbit. 

For spectral statistics in the constant-width QB only one of the two subspectra comprising doublet partners is considered. Due to the unidirectionality its spectral properties agree with GUE rather than GOE statistics for a fully classical dynamics in each half of the PSOS~\cite{Gutkin2007}. For the case considered here, the PSOS contains chaotic and regular parts and accordingly the spectral properties are between Poisson and GUE statistics. Furthermore, those of the corresponding NB are similar to those of the QB. 

\section{Summary and Outlook: The experimental realization of NBs\label{Exp}} 
\begin{figure}[!h]
\centering
 \begin{tabular}{@{}c@{}}
        \begin{tabular}{@{}c@{}}
        \large (a)
        \includegraphics[width=.42\linewidth]{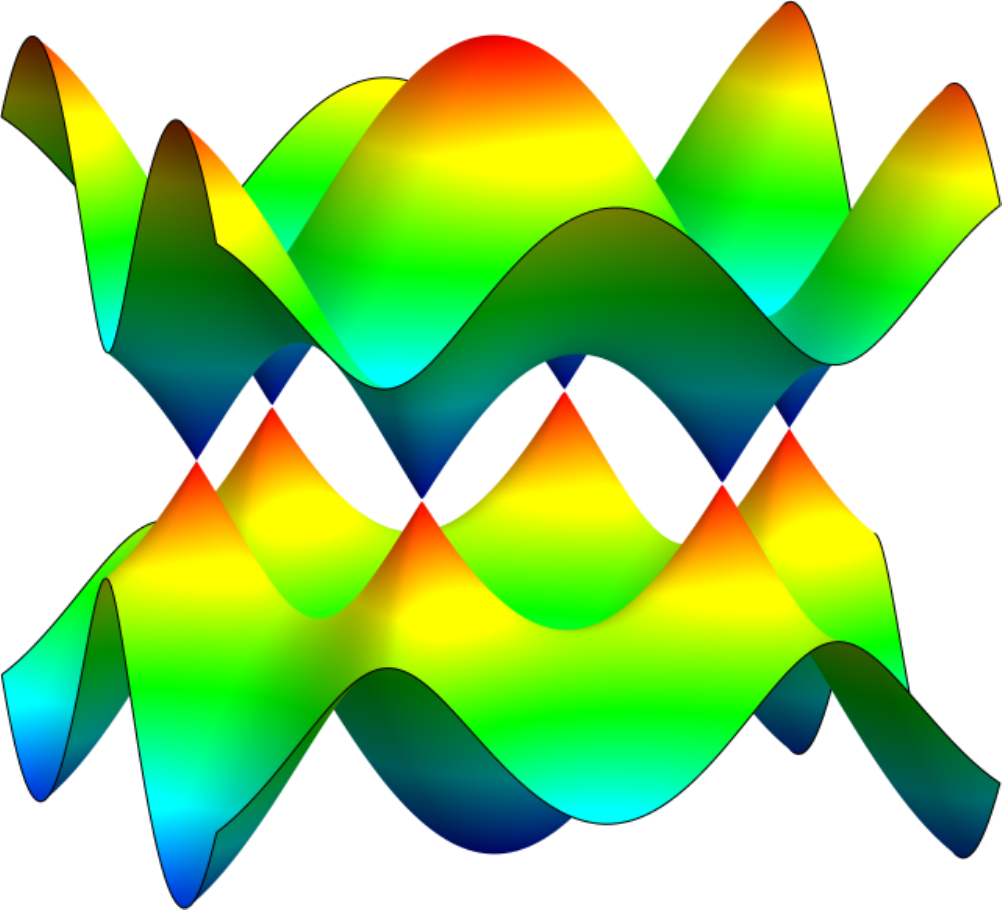}\\[\abovecaptionskip]
        \end{tabular}
        \begin{tabular}{@{}c@{}}
        \includegraphics[width=.5\linewidth]{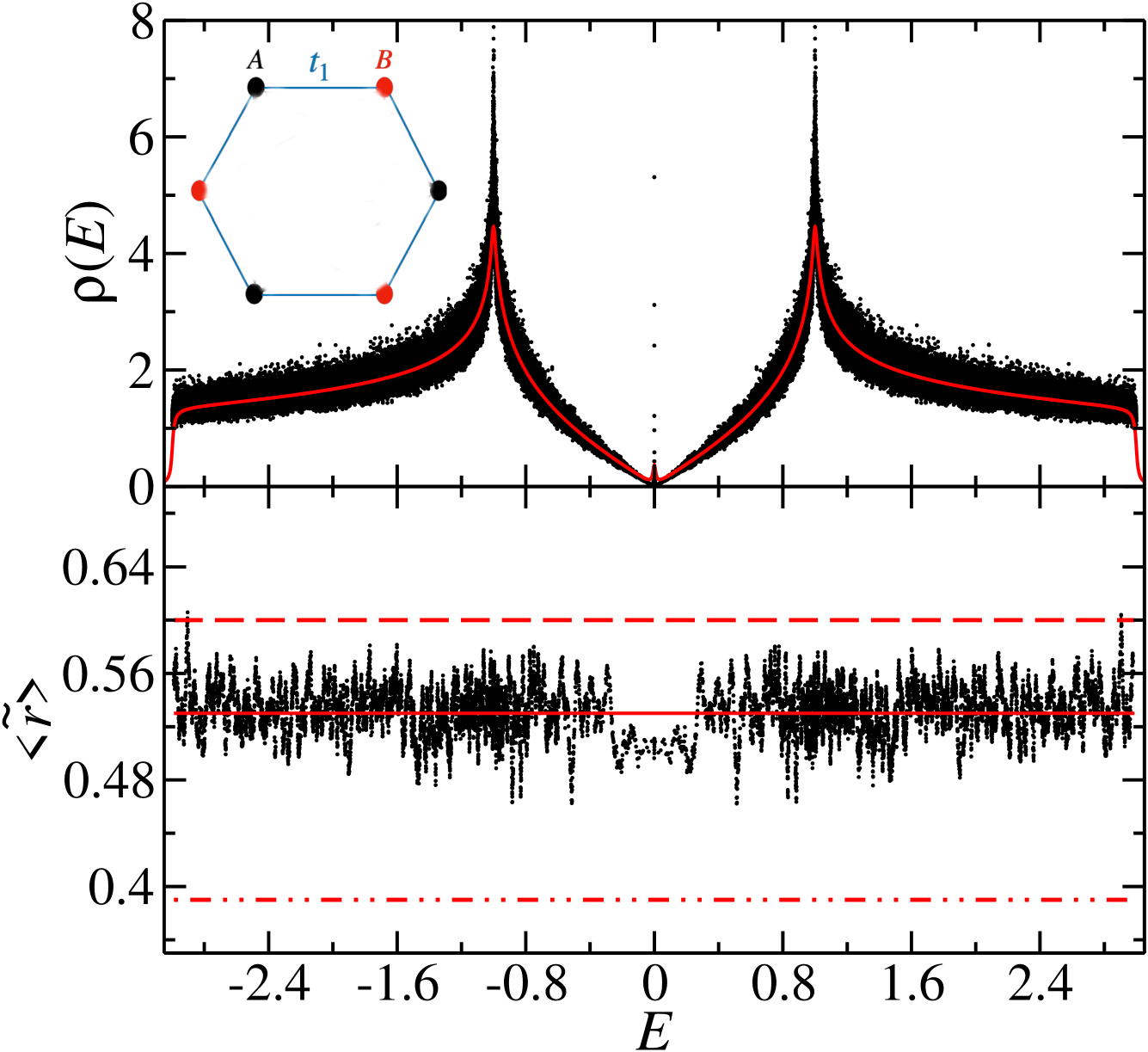}
        \end{tabular}
        \end{tabular}
	\begin{tabular}{@{}c@{}}
	\begin{tabular}{@{}c@{}}
        \large (b)
        \includegraphics[width=.42\linewidth]{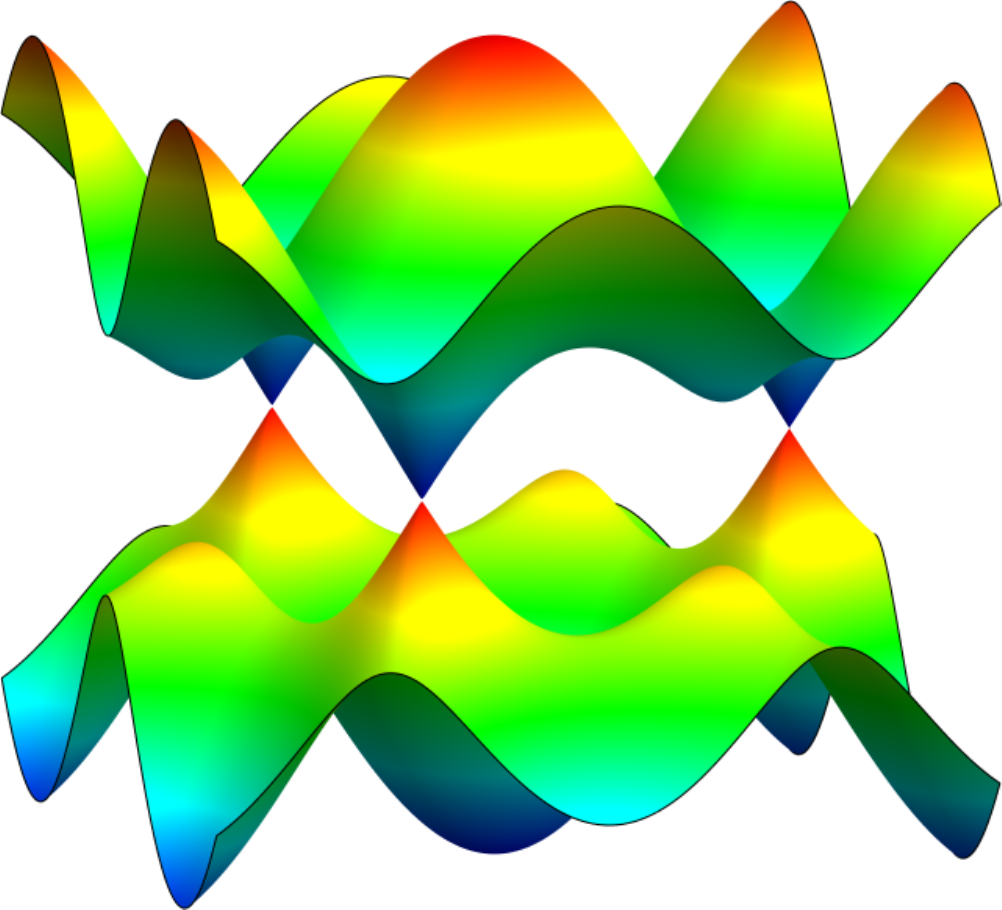}\\[\abovecaptionskip]
        \end{tabular}
        \begin{tabular}{@{}c@{}}
        \includegraphics[width=.5\linewidth]{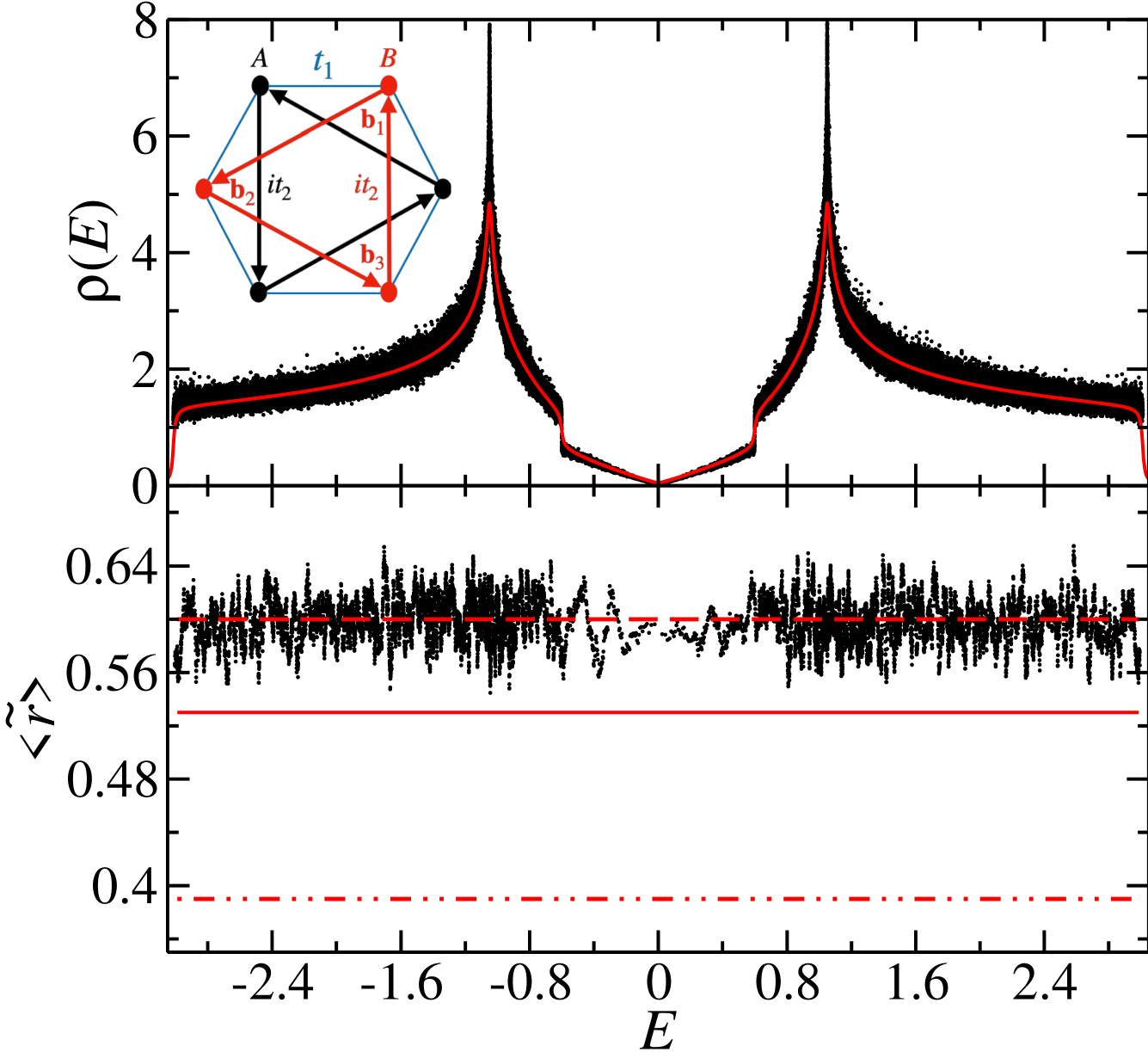}
        \end{tabular}
  \end{tabular}
	\caption{(a) Left: Conduction and valence bands of graphene. Right: The upper part shows the density of states of a GB with the shape~\refeq{AFShape} with $\omega =0.2$, the lower one the average ratios $\langle\tilde r\rangle$ as functions of energy $E$. (b) Left: Conduction and valence bands of graphene subject to the Haldane model with mass $M=0.3$. Right: The upper part shows the density of states of a Haldane graphene billiard with the shape~\refeq{AFShape} with $\omega =0.2$, the lower one the average ratios $\langle\tilde r\rangle$ as functions of energy $E$. In the panels showing average ratios the values for Poissonian are indicated as dash-dotted lines, the GOE as solid lines and GUE as dashed line at $\langle\tilde r\rangle =0.39,0.53,0.6$, respectively.}
\label{Haldane}
\end{figure}
\begin{figure}[!h]
\centering
\includegraphics[width=.8\linewidth]{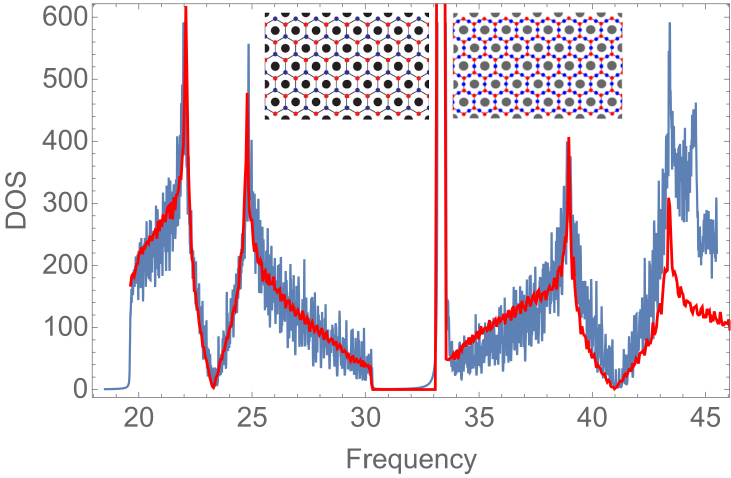}
	\caption{Comparison of an experimentally determined density of states (blue) with the TBM result for a honeycomb-Kagome lattice (red) constructed from a honeycomb and a Kagome sublattice as illustrated in the right inset. Here, up to eighth-nearest-neighbor neighbor hoppings were taken into account. The experimental results, including resonance frequencies of the two Dirac points, the band gaps, the van Hove singularities and the nearly flat band are well reproduced by the TBM. The experimental curve was measured with a superconducting rectangular DB~\cite{Dietz2015}. It comprises $\approx 4900$ resonance frequencies. Below the flat band the experimental density of states is also well described by a TBM for a honeycomb lattice shown in the left inset.}
\label{Honome}
\end{figure}
The topics of this chapter are relativistic quantum chaos and relativistic quantum billiards, namely NBs introduced in~\cite{Berry1987}. A characteristic of NBs is the chirality. It manifests itself in the symmetry properties outlined in \refsec{Syms}, in the spectral properties of sector NBs introduced in~\refsec{Sectors}, and in the length spectra as demonstrated analytically in terms of trace formulas in~\refsec{Trace}. Furthermore, examples of NBs with fully chaotic dynamics and extraordinary features, like constant-width billiards, BBOs in the stadium NB and symmetry-projected eigenstates are discussed. A central question which arose with the introduction of NBs is their experimental realization. A first attempt were finite-size graphene sheets -- referred to as graphene billiards -- in the energy region, where they exhibit relativistic phenomena.

Interest in the spectral properties of graphene billiards (GBs) came up with the pioneering fabrication of graphene, a monolayer of carbon atoms on a hexagonal lattice~\cite{Novoselov2004}, the reason being that graphene exhibits nonrelativistic and relativistic phenomena, like pseudodiffusive transport, the quantum Hall effect, Zitterbewegung, Klein tunneling and edge states~\cite{DiVincenzo1984,Novoselov2004,Geim2007,Avouris2007,Miao2007,Ponomarenko2008,Beenakker2008,Zhang2008,Neto2009,Zandbergen2010}. Graphene has the exceptional property that at the Fermi energy its valence and conduction bands shown in left part of~\reffig{Haldane} (a), touch each other conically at the corners of the hexagonal Brillouin zone. There the dispersion relation is linear and the energy excitations of graphene are effectively described by the Dirac equation for massless spin-1/2 particles~\cite{Novoselov2004,Geim2007,Beenakker2008,Neto2009}. Near the Dirac cone $\boldsymbol{K}=\left(\frac{2\pi}{3 a},\frac{2\pi}{3 \sqrt{3}a}\right)$
\begin{equation}
           \hat H^0_K(\boldsymbol{q})=\frac{3 t_1 a}{2}\boldsymbol{\hat\sigma}\cdot\boldsymbol{q},
        \end{equation}
	where $a$ is the distance between neighboring sites of the honeycomb lattice, and $\boldsymbol{q}$ denotes the quasimomentum vector with respect to the $\boldsymbol{K}$ point. Similarly, near the Dirac cone $\boldsymbol{K}'=-\boldsymbol{K}$ the effective Hamiltonian is given by
        \begin{equation}
		\hat H^0_{K^\prime}(\boldsymbol{q})=\frac{3 t_1 a}{2} \boldsymbol{\hat\sigma^\ast}\cdot\boldsymbol{q},
        \end{equation}
with $\boldsymbol{q}$ denoting the quasimomentum vector with respect to the $\boldsymbol{K}^\prime$ point. The conical shape originates from the honeycomb lattice structure, which is formed by two interpenetrating triangular sublattices. Furthermore, the occurrence of two independent Dirac cones at the touch points $\boldsymbol{K}$ and $\boldsymbol{K^\prime}$, commonly named Dirac points, is due to time-reversal symmetry, inversion symmetry, and the discrete rotational symmetry $C_3$ of the honeycomb-lattice~\cite{Beenakker2008}. Graphene billiards are constructed by cutting out of a honeycomb lattice a sheet with the shape of the billiard and their eigenstates are computed based on a TBM~\cite{Dietz2015}. The BCs to be imposed in the region around the Dirac points on the spinor components of a GB are given in Refs.~\cite{Akhmerov2008,Wurm2011}. 

The analogy to the wave equation of relativistic quantum systems in the vicinity of the Dirac points led to numerous numerical~\cite{Wurm2009,Neto2009,Polini2013} and experimental 'artificial-graphene' realizations~\cite{Polini2013,Bittner2010,Kuhl2010,Guettinger2010,Gomes2012,Rechtsmann2013}. It was expected that in that region their spectral properties are similar to those of NBs of corresponding shape and confirmed in experiments with graphene quantum dots~\cite{Ponomarenko2008}, whereas numerical~\cite{Wurm2009} and experimental studies of GBs with superconducting microwave Dirac billiards~\cite{Dietz2015,Dietz2016,Zhang2023} revealed that they comply with those of QBs of corresponding shape, that is, exhibit GOE statistics. It was shown in Ref.~\cite{Wurm2009}, that the reason is back scattering at the boundary, which leads to a mixing of valley states around the $\boldsymbol{K}$ and $\boldsymbol{K^\prime}$ points. Time-reversal invariance is violated at each Dirac cone~\cite{Beenakker2008}, and it is restored due to the occurrence of two independent Dirac cones that are mapped onto each other when applying the time-reversal operator to the eigenstates of graphene, and through the backscattering for finite-size graphene sheets.

The experiments were performed with superconducting microwave-photonic crystals~\cite{Dietz2013,Dietz2015,Dietz2016,Zhang2023}, consisting of a flat microwave resonator containing metallic cylinders arranged on a triangular grid~\cite{Bittner2012} and have been named Dirac billiards (DBs). The microwave frequency was restricted to the range where only the lowest transverse magnetic modes are excited. There the electric-field strength is perpendicular to the top and bottom plate of the resonator and it is described by the scalar Helmholtz equation with Dirichlet BCs for the electric field strength at the walls of the resonator and cylinders, which is mathematically equivalent to the Schr\"odinger equation of a quantum billiard of corresponding shape with scatterers at the positions of the cylinders. The honeycomb structure is formed by the electric-field strength which in the microwave-frequency range below the flat band is localized in the voids at the centers of three neighboring cylinders. The density of states obtained from measurements with a rectangular DB~\cite{Dietz2015} is shown in~\reffig{Honome}. It exhibits two Dirac points framed by van Hove singularities. These are separated by a flat band of extraordinarily high spectral density. It was demonstrated that in the bands of propagating modes framing the lower Dirac point, the density of states and properties of the resonance frequencies and electric-field distributions, that is, wave functions, agree well with those of the eigenstates of the corresponding GB when taking into account in the TBM up to third-nearest neighbor hoppings and wave-function overlaps. There the electric-field strength is localized at the voids as illustrated in the left inset of~\reffig{Honome}. However, the occurrence of the flat band and of the upper Dirac point can not be explained with the honeycomb-lattice based TBM. Indeed, it was demonstrated in~\cite{Maimaiti2020}, that the density of states and spectral properties of the DBs are well captured by a TBM for a lattice structure that is formed by a combination of a honeycomb and kagome sublattice~\cite{Jacqmin2014}, shown as red curve in~\reffig{Honome}. Indeed, for resonance frequencies around and above that of the flat band the electric-field strength is localized at the voids and also at the centers between two adjacent cylinders which form a kagome lattice; cf. right inset in~\reffig{Honome}. The spectral properties of DBs agree well with those of the GB and thus with those of the nonrelativistic QB of corresponding shape. In the right part of~\reffig{Haldane} (a) are shown the density of states (upper part) of a GB with the shape of an Africa billiard and its average ratios $\langle\tilde r\rangle$ as a function of energy. These agree well with the value $\langle\tilde r\rangle=0.53$ for the GOE

Recently, GBs subject to the Haldane model, called Haldane graphene billiards (HGB) were proposed as a suitable graphene system to emulate the properties of NBs~\cite{Nguyen2024}. Here a gap is introduced at one of the Dirac points by applying the Haldane model~\cite{Haldane1988,KaneMele2005} so that within the energy range of that gap the eigenstates are confined to the valley region around the other one. This is achieved by appropriately tuning the Semenoff mass, generated by opposite constant onsite potentials on the triangular sublattices of the honeycomb lattice, thereby breaking inversion symmetry, and the Haldane mass arising from purely imaginary next-nearest-neighbor hoppings. In~\cite{Nguyen2024} a nonzero purely imaginary next-to-nearest neighbor tunneling parameter, $i t_2$~\cite{Haldane1988}, was added, as illustrated in the inset of the right part of~\reffig{Haldane} (b), such that, even though time-reversal invariance is broken by complex tunneling, the total magnetic flux through a unit cell is zero. Furthermore, onsite potentials $M$ were introduced on all sites of sublattice $A$ and $-M$ on all sites of sublattice $B$, yielding in the quasimomentum space around the $\boldsymbol{K}$ and $\boldsymbol{K^\prime}$ points the effective Hamiltonian
\begin{align}
\hat H_K(\mathbf{q})  &=\frac{3 t_1 a}{2}\boldsymbol{\hat\sigma}\cdot\mathbf{q} + \left(M- 3 \sqrt{3}t_2\right)\hat\sigma_z \\
\hat H_{K'}(\mathbf{q})  &=\frac{3 t_1 a}{2}\boldsymbol{\hat\sigma^\ast}\cdot\mathbf{q} + \left(M+ 3 \sqrt{3}t_2\right)\hat\sigma_z .
\end{align}
Thus, for example at the critical point $t_2=\frac{M}{3\sqrt{3}}$ only the Dirac cone $\boldsymbol{K}$ survives and the other one at $\boldsymbol{K^\prime}$ is gapped with the effective mass $2M$, implying that in the low energy limit $|E|< M$, there is only one Dirac cone. Accordingly the conduction and valence bands, exhibited in the left part of~\reffig{Haldane} (b) touch each other only at one of the Dirac points, while the other valley is gapped. The purely imaginary next-to-nearest neighbor tunneling term induces time-reversal invariance violation in the region of linear dispersion relation, however not in the region of the band edges where the lattice structure is not discernible. It was found in~\cite{Nguyen2024} that in the region of the Dirac point the spectral properties and properties of the eigenstates are consistent with those of the NB, whereas around the band edges the eigenstates coincide with those of the corresponding GB and thus spectral properties agree with those of the QB. In the upper and lower parts of~\reffig{Haldane} (b) are shown the density of states and the average ratios $\langle\tilde r\rangle$ for the Africa-shaped HGB. The latter agrees with the value for the GUE, $\langle\tilde r\rangle =0.6$. 

Similarly, it has been demonstrated recently~\cite{Dietz2026} that, contrary to HGBs and to the general assumption~\cite{Huang2009,Xu2013,Ge2024,Dietz2025} the quantum scars observed in stadium GBs in the region of the Dirac point do not exhibit the characteristics of relativistic ones, even though they are governed by the relativistic Dirac equation. Indeed, it has been demonstrated for a GB and HGB with the shape of a full stadium that the slow oscillations of their spectral densities are described by~\refeq{NbboQB} and~\refeq{NbboNB}, respectively. Furthermore, the eigenstates of the GB maybe be separated into 4 mirror symmetry classes like the corresponding QB, whereas for the HGB they are separated into the two symmetry classes for two-fold symmetry, like the corresponding NB. Also, the $k$ values of the BBOs of the GB and HGB agree with those of the QB and NB, respectively. These findings corroborate the analogy of HGBs and NBs. Recently it was proposed to emulate the Haldane model with photonic crystals~\cite{Jotzu2014,Liu2021,Lannebere2019}, which would open the possibility to realize energy spectra exhibiting the particular features of NBs experimentally. Indeed, the Haldane model has been realized experimentally in different setups, however the measuement of energy spectra comprising many levels is still challenging.

\section{Conclusions}
In summary, the spectral properties of NBs whose shape does not have a geometric symmetry are consistent with the BGS conjecture for quantum systems with a chaotic classical dynamics and violated time-reversal invariance. On the other hand, NBs with shapes of QBs with an integrable classical dynamics that exhibit Poisson statistics do not necessarily follow the BT conjecture. Generally NBs exhibit in their length spectra peaks at the length of POs of the corresponding CB, where in the ultrarelativistic limit those with an odd number of reflections are absent. Momentum distributions, their IPR values and Husimi functions serve as suitable tools to identify scarred or localized wave functions. Furthermore, contrary to the general belief~\cite{Huang2009,Ge2024} the quantum scars observed in GBs do not exhibit chirality and other characteristic features of relativistic QBs~\cite{Zhang2021,Dietz2026} even though they are effectively described by the Dirac equation of relativistic spin-1/2 particles. Consequently, even though GBs show at the Dirac points relativistic phenomena, they do not provide a model for the study of aspects of relativistic quantum chaos in the region of linear dispersion around the Dirac points. The recently proposed HGBs exhibit the characteristics of relativistic QBs and therefore constitute a suitable candidate for the experimental realization of NBs. Yet, currently it is not possible to measure complete spectra of HGBs comprising several hundreds of eigenvalues.  

\begin{ack}[Acknowledgments]
tThis work was partially funded by the Deutsche Forschungsgemeinschaft (DFG, German Research Foundation) Project No. 290128388. We acknowledge financial support from the Institute for Basic Science in Korea through the project IBS-R024-D1.
\end{ack}
\bibliography{References_Chapter}
\end{document}